\documentclass[usenatbib, onecolumn]{mn2e}
\usepackage{graphicx}
\usepackage{amssymb}
\usepackage{amsbsy}
\usepackage{color}

\usepackage{hyperref}
\definecolor{grey}{rgb}{0.75,0.75,0.75}
\definecolor{Orange}{rgb}{1.0,0.5,0.15}
\definecolor{brown}{rgb}{0.7,0.25,0.0}
\definecolor{pink}{rgb}{1.0,0.5,0.5}
\definecolor{darkerred}{rgb}{0.8,0,0}
\definecolor{darkerblue}{rgb}{0,0,0.8}
\definecolor{Blue}{rgb}{0,0.08,0.65}
\definecolor{Red}{rgb}{0.65,0.08,0.05}
\definecolor{Green}{rgb}{0.15,0.45,0.25}

\begin{document}
\title[Polarization transfer in  relativistic magnetized plasmas]{Polarization transfer in 
relativistic magnetized plasmas}

\author[J. Heyvaerts, C. Pichon, S. Prunet, J. Thi\'ebaut]{Jean~Heyvaerts$^{1,2}$, Christophe~Pichon$^{1,3}$, 
Simon~Prunet$^{1}$, J\'er\^ome~Thi\'ebaut$^{1}$ \\
$^{1}$ Institut d'Astrophysique de Paris, UMR 7095, CNRS, UPMC Univ. Paris VI, 98 bis boulevard Arago, 75014 Paris, France\\
$^{2}$ Observatoire Astronomique, Universit\'e de Strasbourg, CNRS,
UMR 7550, 11 rue de l'Universit\'e, 67000 Strasbourg, France\\
$^{3}$ CEA Saclay, DSM/IPhT, B\^atiment 774, 91191 Gif-sur-Yvette, France
}

%\offprints{J. Heyvaerts, \\ \email{jean.heyvaerts@astro.unistra.fr}}
\date{\small{Accepted 2013 January 18. Received 2013 January 17; in original form 2012 November 30\\
Reference: MNRAS 2013; doi:10/1093/mnras/stt135}}
\maketitle
\begin{abstract}
The polarization transfer coefficients of a relativistic magnetized plasma are derived.
These results apply to any momentum distribution function of the particles, isotropic or anisotropic. 
Particles interact with the radiation either in a non-resonant mode
when the frequency of the radiation exceeds
their characteristic synchrotron emission frequency,
or quasi-resonantly otherwise. These two classes of particles contribute
differently to the polarization transfer coefficients. For a given frequency, 
this dichotomy corresponds to a regime change in the dependence of the transfer coefficients 
on the parameters of the particle's population, since these parameters control the relative weight of 
the contribution of each class of particles.
Our results apply to either regimes as well as  the intermediate one.
The derivation of the transfer coefficients involves 
an exact expression of the conductivity tensor of the relativistic magnetized plasma that 
has not been used hitherto in this context. 
Suitable expansions 
valid at frequencies much larger than the cyclotron frequency
allow us to analytically perform
the summation over all resonances at high harmonics of the 
relativistic gyrofrequency. 

The transfer coefficients are represented in the form of two-variable integrals that can
be conveniently computed for any set of parameters by using Olver's expansion of high-order Bessel functions.
We particularize our results to a number of distribution functions, isotropic, thermal or power-law,
with different multipolar anisotropies of low order, or strongly beamed.
Specifically, earlier exact results
for thermal distributions are recovered. 
For isotropic distributions, the Faraday coefficients are expressed in the form 
of a one-variable quadrature over energy, for which we provide the kernels in the
high-frequency limit and in the asymptotic low-frequency limit. An interpolation 
formula extending over the full energy range is proposed for these kernels. 
A similar reduction to a one-variable quadrature over energy is derived at high frequency 
for a large class of anisotropic distribution functions that may form a basis 
on which any smoothly anisotropic distribution could be expanded.
\end{abstract}

\setcounter{tocdepth}{3}
\pagenumbering{arabic}

%\tableofcontents

%%%%%%%%%%%%%%%%%%%%%%%%%%%%%%%%%%%%%%%%%%%%%%%%%%%%%%%%%%
\section{Introduction and motivation}
\label{introduction}

The transfer of the intensity and polarization  
of radiation propagating in a relativistic magnetized plasma has been a long standing problem which
still remains incompletely solved.
Its potential diagnostic applications
have long been known \citep{JonesODell, Cioffi} and it has
recently attracted renewed interest \citep{Brentjens,Thiebautetal} in connection with
the study of the magnetic field structure 
of synchrotron emitting objects where 
the high-energy emitting particles may constitute the bulk of the plasma. Polarization data
constrain the structure of geometrically complex sources and of their magnetospheres.
Astrophysical environments such as
the vicinity of the central black hole in the Milky Way \citep{Agol, Beckert, CerbakovSagA},
jet-sources, galactic \citep{SS433} or extragalactic \citep{3C84}, pulsars \citep{Kanbach, Dyks, PetriPolar, MacDo, Yuen} or
disks around black holes \citep{Dovciak} have been studied by these methods.
The perspectives of using polarization data to constrain magnetic fields have improved strongly by the
deployment of the low frequency array, LOFAR, \citep{Beck2009}, and will blossom with the advent of 
square kilometer array, SKA, \citep{Dewdney,SKA-Carilli}, 
ranging from the largest scales \citep{SKA-Feretti, SKA-Gaensler} to the smallest ones \citep{SKA-Bicknell}.
  
Using polarization data for the reconstruction of the magnetic field distribution in volume
from multiwavelengths observations of relativistically hot astrophysical media rests
on the availability of good polarization tranfer codes  and
on the knowledge of the transfer coefficients that would modify the intensity and the polarization of light
as it propagates in the medium.
Synchrotron radiation plays an important role in the emission of magnetized high-energy sources.
The polarization properties of its emission and
absorption caused by the high-energy population are well known.
The non-dissipative transfer coefficients of Faraday rotation and Faraday conversion from circular
to linear polarization have been more difficult to derive.
Substancial progress has nevertheless been achieved 
in time \citep{Trubnikov, Zelezniakoff, Sazonov69, MelroseJplPhyspremier, Cerbakov2008, HuangCerbakov}.
However, the polarization transfer coefficients for high harmonic numbers have not been generally expressed yet
in a manner similar to the dissipative transfer coefficients that involve, for any given distribution function,
the integral of the product of some derivative of it
with a well-defined energy- and direction-dependent kernel. Such an expression 
would be most useful for use in data inversion codes.

In this paper, we obtain
such expressions for the polarization transfer coefficients in a relativistic
homogeneous magnetized plasma for any distribution function and provide efficient means 
to calculate the general two-variable kernel. These expressions are reduced to simple
one-variable quadratures for isotropic distributions and 
for a large class of non-isotropic ones as well.
The derivation of these results parallels that 
of dissipative transfer coefficients. It is based on an expression of the high
frequency conductivity tensor of a magnetized relativistic plasma
that has not been hitherto used in this context and 
involves, at some late stage of the derivation, well-justified approximations that allow us
to explicitly sum series of principal parts in the limit of large harmonics numbers. 
Our results are shown to
exactly coincide with previously known exact results in the case of thermal distributions.

Section~\ref{derivation} summarizes the basics of polarized radiation transfer in
a uniformly magnetized medium.
Section~\ref{SecResults} presents in anticipation the end results of our derivation 
of the  polarization transfer coefficients as a double regular quadrature and illustrates 
it for isotropic and anisotropic distributions, where it can sometimes be reduced to a 
single quadrature.  This unconventional presentation has been chosen because the derivation of these results
from those in Section~\ref{derivation}, which is carried in Section~\ref{Secderivation} and Section~\ref{sec:NR-QR},
is mathematical in some respects, as many previous contributions to this subject \citep{Westfold, 
Sazonov69, MelrosePhysRev, Swanson}. The reader who would like to closely follow the 
derivations presented in Sections \ref{Secderivation} and \ref{sec:NR-QR} is invited to consult the series of 
appendices that 
are available in the on-line version of this paper only.
Sections \ref{Secderivation} and \ref{sec:NR-QR} can be skipped on a first read.
Section \ref{seclowfrequencylimit} focuses on analytical results applicable to 
isotropic distribution functions in the low-frequency (LF) limit.
Section~\ref{SecDiscussion} wraps up.

%%%%%%%%%%%%%%%%%%%%%%%%%%%%%%%%%%%%%%%%%%%%%%%%%%%%%%%%%%
\section{Formulation }
\label{derivation}

This section briefly establishes ab initio some known results concerning the description of 
polarized radiation transfer in a uniformly magnetized medium
and the associated derivation of the high-frequency electrical conductivity 
from which the elements of the transfer matrix are obtained. Meanwhile, we set our
notations and specify the nature of the large harmonic number approximation on which this work is based.

\subsection{Description of polarization transfer}
The polarization properties 
of transverse electromagnetic radiation 
are represented by a column vector of four Stokes parameters, ${\mathbf{I}} = \, ^{T}\!(I, Q, U, V)$,
$^T\! {\bf{\sf M}}$ denoting the transpose of the matrix $\bf{\sf M}$. 
The Stokes vector ${\mathbf{I}}$ depends on the frequency and direction of propagation
of the radiation, as well as on the point 
in space and on the time at which it is observed. 
In the limit where the frequency is higher than any gyrofrequency,
the Stokes vector satisfies a radiative transfer equation in which the generalized emission coefficient is
a four-component column vector ${\mathbf{W}} = \, ^{T}\!(W_I,W_Q, W_U, W_V)$  
and the generalized absorption coefficient is a 4$\times$4 transfer matrix. This 
transfer equation can be written, for radiation of frequency $\omega$ and group velocity 
(based on an isotropic mean conductivity) ${\mathbf{v}}_g$ as:
\begin{equation} 
\left(\partial_t + {\mathbf{v}}_g \cdot {\boldsymbol{\nabla}}\right) 
\left( \begin{array}{c}
I\\
Q\\
U\\
V
\end{array} \right)
= 
\left( \begin{array}{c}
W_I\\
W_Q\\
W_U\\
W_V
\end{array} \right)
-
\left( \begin{array}{cccc}
K_{II}&K_{IQ}&K_{IU}&K_{IV}\\
K_{IQ}&K_{II}&K_{QU}&K_{QV}\\
K_{IU}&-K_{QU}&K_{II}&K_{UV}\\
K_{IV}&-K_{QV}&-K_{UV}&K_{II}
\end{array} \right)  
\left( \begin{array}{c}
I\\
Q\\
U\\
V
\end{array} \right)
\,.\label{Lequationdetransfert}
\end{equation}
The four matrix elements $K_{II}$, $K_{IQ}$, $K_{IU}$, $K_{IV}$ and the emission column vector $\mathbf{W}$ result from
dissipative effects. The emission column vector $\mathbf{W}$ represents the emission rate and polarization
of the spontaneously emitted radiation. The elements 
$K_{QU}$, $K_{QV}$, $K_{UV}$ of the transfer matrix result from non-dissipative 
effects caused by the slowly varying  phase difference between different transverse components
of the electric vector of the radiation. 
The elements of the matrix in equation~(\ref{Lequationdetransfert})  can be expressed in terms of the components of
the high-frequency conductivity tensor $\boldsymbol{\sigma}({\mathbf{k}}, \omega)$. 
Equation (\ref{Lequationdetransfert}) has been established 
by a number of authors and by different methods \citep{SasonovTsytovitch, HeyvaertsASpSc, Zelezniakoffetal}.
A direct derivation of the absorption-like term is presented below. The derivation of 
the emission term is less straightforward \citep{SasonovTsytovitch}. 

Since the wavelengths of the plasma eigenmodes of a given frequency $\omega$ do not differ 
by much from each other at high frequency,
their phase difference evolves slowly in space and time. 
As a result, the correlations between these phases,
which determine the state of polarization,
are maintained over a length much longer than the wavelength, justifying a radiative transfer approach. 
When the plasma frequency $\omega_p$ exceeds by much the gyrofrequencies $\Omega_A$ of all particle species $A$, 
the wavelengths of the eigenmodes still differ by a little amount when $\omega \gg \Omega_A$
although the dispersion relation
may depart from the vacuum one when $\omega$ is of order $\omega_p$. 
In any other case, the difference of the phases of the eigenmodes 
evolves on a wavelength scale
and the radiative transfer approach is inappropriate.
Assuming there are only two eigenmodes with transverse components, 
if their phases would be correlated at some point and time, 
the rapid growth of their difference would reduce this correlation to nothing
in only a few wavelengths. 
Thus, the polarization state of radiation supported by modes of
largely different wavelengths for a given frequency reduces to that of the sum of two mutually incoherent eigenmodes.
Equation~(\ref{Lequationdetransfert}) is then only relevant
in the high-frequency limit, $\omega \gg \Omega_A$. It can be written in this form in the 
weakly anisotropic medium approximation, in which 
the wave dispersion is approximated by neglecting the longitudinal component
of partially transverse eigenmodes \citep{Zelezniakoffetal}. 

We now show how the elements of the absorption matrix relate to those
of the conductivity tensor. We start from
the propagation equation of the electric field ${\mathbf{E}}({\mathbf{r}}, t)$,
deduced from Maxwell's equations:
\begin{equation}
\frac{1}{c^2} \frac{\partial^2 {\mathbf{E}} }{\partial t^2} - \Delta {\mathbf{E}} + \mu_0 \frac{\partial {\mathbf{J}} }{\partial t}
= 0
\,.\label{propagcrue}
\end{equation}
Our equations are written in the MKSA system of units, where $\varepsilon_0$
is the dielectric permittivity of vacuum and
$\mu_0$ its magnetic permeability. The velocity of light in vacuo is $c = (\varepsilon_0 \mu_0)^{-1/2}$. In equation (\ref{propagcrue}),
${\mathbf{E}}$ is the electric field and ${\mathbf{J}}$ the current density.
The electric field is
assumed to be transverse, so that
${\mathrm{div}}\, {\mathbf{E}} = 0$, which is a good approximation at high frequencies.
The microscopic electric field is then Fourier expanded,
its complex vectorial amplitude ${\hat{\mathbf{E}}}_{{\mathbf{k}}, \omega}$ being regarded as 
slowly depending on space and time, which is equivalent to coarse graining in Fourier space. 
The electric current density ${\mathbf{J}}({\mathbf{r}}, t)$ is similarly expanded.
The phase, and possibly the amplitude, 
of the microscopic electric field is a random variable. Omitting the indices ${\mathbf{k}}, \omega$ for simplicity,
each term of the Fourier expansion can be written as
\begin{equation}
{\mathbf{E}}({\mathbf{r}}, t) = 
{\hat{\mathbf{E}}}({\mathbf{r}}, t) \ e^{i ({\mathbf{k}} \cdot {\mathbf{r}} - \omega t)} 
\,.\label{quasimono}
\end{equation}
The current in equation~(\ref{propagcrue}) is the sum of an induced current ${\mathbf{J}}'$, 
which results from the electric field ${\mathbf{E}}({\mathbf{r}}, t)$
and a spontaneous current ${\mathbf{J}}^{{'}\!{'}}$, the microcurrent created by the individual plasma particles,
which gives rise to spontaneous emission, while
the induced current gives rise to the absorption-like term in equation~(\ref{Lequationdetransfert}).
This latter part of the current is given in terms of the electric field by a linear non-local 
and causal conductivity tensor operator:
\begin{equation}
{\mathbf{J}}'({\mathbf{r}}, t)  
=  \int_0^\infty \! d\tau \!\!\int\!\!\!\! \int\!\!\!\!\int\! d^3\!R \ \  
{\mathbf{\Sigma}}( {\mathbf{R}}, \tau) \cdot {\mathbf{E}}({\mathbf{r}}-{\mathbf{R}} , t-\tau)
\,.\label{conductoperator}
\end{equation}
In the weakly anisotropic medium approximation,
the conductivity operator can be split into a scalar isotropic part, ${{\mathbf{\Sigma}}}_{\rm s}$, and 
a tensorial anisotropic part ${{\mathbf{\Sigma}}}_1$. The isotropic part may, for example, represent 
the dispersive properties of a cold population of relatively large density, 
neglecting the magnetization of these particles.
An isotropic conductivity
has no effect on the polarization but affects the propagation properties of electromagnetic waves. 
No component of the conductivity need to be
separated out when the propagation is 
represented well enough by the vacuum dispersion relation.
The tensorial anisotropic part consists of the anisotropic residual of the conductivity,
not included in ${{\mathbf{\Sigma}}}_{\rm s}$, and of the conductivity of populations
which negligibly contribute to the dispersive properties of the plasma waves. 
The frequency $\omega$  supposedly being much larger than all gyrofrequencies, 
the component ${{\mathbf{\Sigma}}}_1$ must be small. 
Its contribution to the dispersion properties 
may be neglected but it nevertheless contributes to the polarization transfer matrix. 

The form of the electric field in equation~(\ref{quasimono}) is 
inserted in equations (\ref{propagcrue}) and (\ref{conductoperator}).
When acting on the imaginary exponential phase factor the operators $\partial_t$ and $\nabla$  
are of order $\omega$ and $k$ respectively,
but when acting on the slowly varying amplitude ${\hat{\mathbf{E}}}({\mathbf{r}},t)$ 
they are of order $1/T$ and $1/L$, 
$T$ and $L$ being the time and length scales
on which ${\hat{\mathbf{E}}}$ varies, which are much longer than the wave period and wavelength. 
Only the terms of first order in
$1/(\omega T)$ and $1/(kL)$ are retained. For consistency, this requires that 
${\hat{\mathbf{E}}}({\mathbf{r}}-{\mathbf{R}} , t-\tau)$ 
in equation~(\ref{conductoperator}) be expanded as:
\begin{equation}
{\hat{\mathbf{E}}}({\mathbf{r}}-{\mathbf{R}} , t-\tau) 
= {\hat{\mathbf{E}}}({\mathbf{r}}, t) - \tau \ \partial_t {\hat{\mathbf{E}}}
-({\mathbf{R}}\cdot {\boldsymbol{\nabla}})\,  {\hat{\mathbf{E}}} \,.
\end{equation}
The time and space integrations in equation~(\ref{conductoperator}) then involve 
the Fourier transforms of ${\mathbf{\Sigma}}_{\rm s}$ and ${\mathbf{\Sigma}}_1$, 
$\sigma_{\rm s}({\mathbf{k}}, \omega)$ and ${\boldsymbol{\sigma}} ({\mathbf{k}}, \omega)$ (no index $1$), 
the latter being a tensor, 
and the derivatives of $\sigma_{\rm s}$ with respect to $\omega$
and $\mathbf{k}$. This results in:
\begin{equation}
\left(\omega^2 - c^2 k^2 + \frac{i\omega}{\varepsilon_0} \, \sigma_{\rm s}\right) {\hat{\mathbf{E}}}
+ \left(2 i \omega - \frac{1}{\varepsilon_0} \frac{\partial\, (\omega \sigma_{\rm s})}{\partial \omega} \right)  
\, \partial_t {\hat{\mathbf{E}}}
+ \left(\left(2 i c^2 {\mathbf{k}} 
+ \frac{\omega}{\varepsilon_0} 
{\boldsymbol{\nabla}}_{\mathbf{k}} \sigma_{\rm s}\right) \cdot {\boldsymbol{\nabla}}\right) {\hat{\mathbf{E}}} 
= - \frac{i \omega}{\varepsilon_0}  \, 
{\boldsymbol{\sigma}}({{\mathbf{k}}, \omega}) \cdot {\hat{\mathbf{E}}} 
\,.\label{propagQMbrut1}
\end{equation}
The term of dominant order in equation~(\ref{propagQMbrut1}) is the first one. 
The two other terms on the left are 
of first order in $1/(\omega T)$ and $1/(kL)$ and the term on the right is also regarded as small.
Equation~(\ref{propagQMbrut1}) is satisfied at the dominant order when the 
factor of ${\hat{\mathbf{E}}}$ vanishes, which means that the electric field 
consists of fluctations, the frequency of which
is $\omega_k$, the solution of the isotropic dispersion relation:
\begin{equation}
\omega^2 - c^2 k^2 + \frac{i\omega}{\varepsilon_0} \, \sigma_{\rm s}(k, \omega) =0
\,.\label{disprelationisotr}
\end{equation}
The corresponding group velocity, ${\mathbf{v}}_g$, is 
obtained by differentiating equation~(\ref{disprelationisotr}) with respect to $\omega$ and $k$.
When $\sigma_{\rm s}$ represents the dispersion due to
a cold unmagnetized plasma, which we assume for simplicity,  
$\omega \sigma_{\rm s}$ does not depend on $\omega$ nor on $k$.
With these simplifications, the three subdominant terms in equation~(\ref{propagQMbrut1}) reduce to:
\begin{equation}
\left(\partial_t  + ({\mathbf{v}}_g \cdot {\boldsymbol{\nabla}})\right) \,  {\hat{\mathbf{E}}} = 
- \, \frac{{\boldsymbol{\sigma}}({\mathbf{k}}, \omega_k) \cdot {\hat{\mathbf{E}}}
}{
2 \varepsilon_0}  
\,,\label{propagcharcute}
\end{equation}
where the amplitude ${\hat{\mathbf{E}}}$, which slowly varies with position and time, 
is only non-vanishing when the wave vector $\mathbf{k}$
and the frequency $\omega_k$ are linked by the dispersion relation in equation~(\ref{disprelationisotr}).
An equation for the correlation tensor of the electric field of this radiation, 
$\langle {\hat{\mathbf{E}}}\otimes {\hat{\mathbf{E}}}^*\rangle$
can be derived from equation~(\ref{propagcharcute}) by tensorial multiplication by ${\hat{\mathbf{E}}}^*$,
the superscript~$^*$ denoting complex conjugation, and averaging over the statistical 
distribution of the phases (and possibly of the amplitudes) of the components of ${\hat{\mathbf{E}}}$. 
This operation is represented by brackets.
The electric field supposedly being transverse, 
it pertains the plane perpendicular to the propagation direction $\mathbf{n}$ and may be represented 
by its components on two abitrarily chosen unit basis vectors 
${\mathbf{e}}_1$ and ${\mathbf{e}}_2$ in this plane. An intensity tensor with components $I_{ij}$
relative to this basis that is proportional 
to the two-dimensional electric field correlation tensor must be defined such that it has
the dimension of a specific intensity. 
We define the Stokes parameters of the radiation as being related to this intensity tensor by 
\begin{equation}
I_{ij} \equiv  v_g \, \frac{k^2 dk}{d\omega_k} 
\left( \begin{array}{cc}
\varepsilon_0 \langle{\hat{E}}_1 {\hat{E}}_1^*\rangle&\varepsilon_0\langle{\hat{E}}_1 {\hat{E}}_2^*\rangle\\
\varepsilon_0 \langle{\hat{E}}_2 {\hat{E}}_1^*\rangle& \varepsilon_0 \langle{\hat{E}}_2{\hat{E}}_2^*\rangle
\end{array} \right)  
= 
\left( \begin{array}{cc}
I+Q & U + iV \\
U - iV & I -Q
\end{array} \right)  
\,.\label{Stokesdecorrtensor}
\end{equation}
The transfer equation for the components $I_{ij}$ 
can be deduced from equation~(\ref{propagcharcute})  which writes, using the dummy index rule
\begin{equation}
\left(\partial_t  + ({\mathbf{v}}_g \cdot {\boldsymbol{\nabla}})\right) \, I_{ij} =
- \, \frac{1}{2 \varepsilon_0} \left( \sigma_{ip} \delta_{jq} + \delta_{ip} {\sigma}^*_{jq} \right) I_{pq}
\,.\label{eqproppolarinduit}
\end{equation}
By equation~(\ref{Stokesdecorrtensor}), the term on the right of equation~(\ref{eqproppolarinduit})
can be be converted into the transfer matrix term 
in equation~(\ref{Lequationdetransfert}), some elements of which turn out to
involve the Hermitian (dissipative) part of the conductivity while others
involve its anti-Hermitian (non-dissipative) part. These parts are respectively defined by:
\begin{equation}
\sigma^H_{ij} = \frac{1}{2} \left( \sigma_{ij} + \sigma^*_{ji} \right) \,,\qquad \qquad \qquad 
\sigma^A_{ij} = \frac{1}{2} \left( \sigma_{ij} - \sigma^*_{ji} \right) \,.
\end{equation}
The elements of the transfer matrix in equation~(\ref{Lequationdetransfert}) are given in terms of
these parts of the conductivity by:
\begin{eqnarray}
&&K_{II} = \frac{{\cal{R}}e\left(\sigma^H_{11} + \sigma^H_{22} \right)}{2\varepsilon_0} \,,\qquad \qquad 
K_{IQ} =   \frac{{\cal{R}}e\left(\sigma^H_{11}- \sigma^H_{22} \right)}{2\varepsilon_0}   \,,\qquad \qquad 
K_{IU} =   \frac{{\cal{R}}e\left( \sigma^H_{12} \right)}{\varepsilon_0}  \,,\qquad \qquad  
K_{IV} =   \frac{{\cal{I}}m\left(\sigma^H_{12}\right)}{\varepsilon_0} \,, \nonumber \\
&&K_{QU} = \frac{{\cal{R}}e\left(\sigma^A_{12}\right)}{\varepsilon_0}  \,,\quad \qquad\qquad \quad
K_{UV} =   \frac{{\cal{I}}m\left(\sigma^A_{22} - \sigma^A_{11} \right)}{2 \varepsilon_0}   \,,\qquad \qquad
K_{QV} =   \frac{ {\cal{I}}m\left(\sigma^A_{12}\right)}{\varepsilon_0}  
\,.\label{coeffdesigmaHA}
\end{eqnarray}
These matrix elements refer to the components of the local conductivity for
the wave vector ${\mathbf{k}}$ and  frequency $\omega_{\mathbf{k}}$ related
to it by the dispersion relation in equation~(\ref{disprelationisotr}). The relations in equation (\ref{coeffdesigmaHA}) 
between the transfer coefficients and the elements of the conductivity matrix agree with those in \citet{Cerbakoff2011}
given the relation between conductivity and dielectric tensors for perturbations varying as in equation (\ref{quasimono})
and accounting for the fact, mentioned above, that their definition of the Stokes parameter $V$ differs from ours by a sign.

The polarization properties of the synchrotron emission are well known, both in 
vacuo \citep{Westfold} and in a cold plasma \citep{Tsytovitch, Razin, Ramaty}.
The dissipative absorption coefficients $K_{II}$, $K_{IQ}$, $K_{IU}$, $K_{IV}$
given in equation~(\ref{coeffdesigmaHA}) in terms of the components of the Hermitian part of the conductivity
have also been presented in the literature \citep{GinzburgSirovatski, SazonovSA}. 
These components can be calculated using the same well-known approximations
that also  provide the emission coefficient. The dissipative absorption 
can be derived from the emission by Einstein's coefficients method,
ignoring dispersion \citep{WildAA}, or not \citep{Zelezniakoff}. 

The non-dissipative coefficients $K_{QU}$, $K_{UV}$, $K_{QV}$ have proved more difficult to calculate.
This paper concentrates on their calculation in the case of ultrarelativistic
plasma particles immersed in a static uniform
magnetic field ${\mathbf{B}}_0$ and for frequencies of large harmonic number. 
The motion of the particles is described in the Vlasov approximation.
We neglect the dispersive properties of a cold population that might be present, so that
the dispersion relation (\ref{disprelationisotr}) supposedly reduces to $\omega = ck$.

\subsection{Formal solution for the high-frequency conductivity tensor}
\label{Solformelle}

The conductivity tensor is found by calculating the high-frequency current 
${\mathbf{J}}$ 
resulting from a high-frequency electric field 
${\mathbf{E}}$
by solving the linearized Vlasov equation for the perturbation 
$f({\mathbf{r}}, {\mathbf{p}})$ of the distribution function,
\begin{equation}
\partial_t f + {\mathbf{v}} \cdot {\boldsymbol{\nabla}} f + (q {\mathbf{v}} \times {\mathbf{B}}_0)
\cdot {\boldsymbol{\nabla}}_{\mathbf{p}} f =
- q ({\mathbf{E}} +  {\mathbf{v}} \times {\mathbf{B}}) \cdot {\boldsymbol{\nabla}}_{\mathbf{p}} f_0 
\,,\label{Vllineaire}
\end{equation}
where $q$  is the charge (with its sign) of the particle species considered.
In equation~(\ref{Vllineaire}), $f_0({\mathbf{p}})$ is the unperturbed 
homogeneous distribution function of the species considered, that only depends on the
component $p_\parallel$ of the particle's momentum 
along the magnetic field and on the modulus $p_\perp$ of its component perpendicular to it, normalized to the density
of particles, so that 
\begin{equation}
\int_0^\infty \!\!\!\int_{-\infty}^{+\infty} 2\pi p_\perp dp_\perp dp_\parallel f_0(p_\perp, p_\parallel) = n\,,
\end{equation}
where $n$ is the volume density of this species of particles.
The gradient with respect to position is denoted as $\boldsymbol{\nabla}$ and the gradient
with respect to some other vectorial variable ${\mathbf{u}}$, such as the momentum ${\mathbf{p}}$, is denoted 
by $\boldsymbol{\nabla}_{\mathbf{u}}$.
The first three terms of equation~(\ref{Vllineaire}) represent the time-derivative of $f$ following the
unperturbed particle's motion, described by the position and momentum ${\mathbf{r}}(t')$ 
and ${\mathbf{p}}(t')$ of this particle at time $t'$, following the unperturbed motion.
At time $t$, the particle is at ${\mathbf{r}}$ with momentum $\mathbf{p}$. 
A standard procedure to obtain the
perturbation $f$ is by then integrating in time the right-hand side term of equation~(\ref{Vllineaire}) 
following the unperturbed particle's motion, as shown for example in the textbooks
by \citet{Tidmanbook} or \citet{Ichimaru}. This gives
\begin{equation}
f({\mathbf{r}}, {\mathbf{p}}, t)  = - q \, \int_{-\infty}^t \! dt' \ \left({\mathbf{E}}({\mathbf{r}}(t'), t')
+  {\mathbf{v}}(t') \times {\mathbf{B}}({\mathbf{r}}(t'), t')\right) \cdot {\boldsymbol{\nabla}}_{\mathbf{p}} f_0({\mathbf{p}}(t'))
\,.\label{soltrajnonperturb}
\end{equation}
The integration is over all times $t'$ earlier than $t$. The unperturbed motion 
is easily expressed in a frame where ${\mathbf{B}}_0$ is along the $Z$ axis so that
${\mathbf{B}}_0 = B_0 \, {\mathbf{e}}_Z$ and the two unit vectors ${\mathbf{e}}_X$ and ${\mathbf{e}}_Y$ 
are perpendicular to it, but otherwise unspecified. The velocity of a particle can be written as
\begin{equation} 
{\mathbf{v}} = v_\perp \, \cos \phi \ {\mathbf{e}}_X +
v_\perp \, \sin \phi \ {\mathbf{e}}_Y + v_\parallel \  {\mathbf{e}}_Z
\,,\label{vent}
\end{equation}
where $v_\perp$ and $v_\parallel$ are conserved by the unperturbed motion, as are also
the modulus $p$ of the momentum, the associated Lorentz factor $\gamma$ and the particle's pitch angle $\vartheta$. 
The angle $\phi$, the gyration angle of the particle, 
rotates in time at the synchrotron frequency $\Omega_*$, which for non-relativistic particles reduces 
to the cyclotron frequency $\Omega$.  These frequencies, which have the sign opposite to that of the charge,
$s_q$,  depend on the rest mass $m$ of the particles and are given by
\begin{equation} 
\Omega_* = - \, \frac{q \, B_0}{\gamma \, m} \,,\qquad \qquad \qquad \qquad  \Omega = - \, \frac{q \, B_0}{m}\,,
\qquad \qquad \qquad \qquad {\mathrm{and}} \qquad \qquad s_q = {\mathrm{sign}}(q)
\,.\label{omega*}
\end{equation}
At a time $t' = t - \tau$ earlier than $t$,
the components of the velocity of a freely-moving particle 
having at time $t$ a gyration angle $\phi$ and a position $X$, $Y$, $Z$ were
\begin{equation}
v_X(t-\tau) =  v_\perp(t) \cos (\phi - \Omega_* \tau) \,,\qquad \qquad \qquad
v_Y(t -\tau) =  v_\perp(t) \sin (\phi - \Omega_* \tau) \,,\qquad \qquad \qquad
v_Z(t -\tau) = v_\parallel(t) \,.
\end{equation}
The position of that particle at time $t - \tau$  was
\begin{equation}
X(t-\tau) = X - \frac{v_\perp}{\Omega_*} \ (\sin \phi - \sin (\phi - \Omega_* \tau))\,, \qquad \,\, 
Y(t-\tau) = Y + \frac{v_\perp}{\Omega_*} \ (\cos \phi - \cos (\phi - \Omega_* \tau))\,, \qquad \,\,
Z(t-\tau) = Z - v_\parallel \ \tau\,.
\end{equation}
The magnetic perturbation is related to the electric one by the Faraday equation, ${\mathrm{curl}} \, {\mathbf{E}} =
- \partial_t {\mathbf{B}}$. The perturbation in the distribution function which develops as a result
of an electric field perturbation such as that in equation (\ref{quasimono})
can be written as
$ f({\mathbf{r}}, {\mathbf{p}}, t) = {\hat{f}}({\mathbf{p}}) \exp(i({\mathbf{k}}\cdot {\mathbf{r}} - \omega t)$.
It depends linearly on the perturbation field by equation~(\ref{soltrajnonperturb}) and
can be expressed in terms of operators acting on the unperturbed distribution function, such as
the anisotropy operator $D$, 
defined by:
\begin{equation}
D(f_0)  = \left( v_\perp \frac{\partial f_0}{\partial p_\parallel} - v_\parallel \frac{\partial f_0}{\partial p_\perp}\right)
\,.\label{defDanisotrop}
\end{equation}
$D(f_0)$ vanishes for an isotropic distribution function. 
The wave vector $\mathbf{k}$ is taken to be in the $X-Z$ plane and has components $k_X = k \sin \alpha$ and
$k_Z = k \cos \alpha$, the propagation angle $\alpha$ being comprised between $0$ and $\pi$.
When fully expanded, the solution for ${\hat{f}}$ can be written from equation~(\ref{soltrajnonperturb}) as
\begin{eqnarray}
{\hat{f}}({\mathbf{p}}) &=& - \frac{q}{\omega}\,  \int_0^\infty  d\tau 
\exp\Big( i\, \Big(\, (\omega - k_\parallel v_\parallel) \, \tau - \frac{k_\perp v_\perp}{\Omega_*}
(\sin \phi - \sin (\phi - \Omega_* \tau)\, 
\Big)\Big) \times
\nonumber \\
&&\Big[  \Big({\hat{E}}_X  \cos (\phi - \Omega_* \tau) 
+ {\hat{E}}_Y \sin (\phi - \Omega_* \tau)\Big) \Big(\omega \frac{\partial f_0}{\partial p_\perp} + k_\parallel 
D(f_0)\Big)
+ {\hat{E}}_Z \Big(\omega \frac{\partial f_0}{\partial p_\parallel} - \cos (\phi - \Omega_* \tau) \, k_\perp 
D(f_0)\Big) \Big]
\,.\label{F1}
\end{eqnarray}
The associated Fourier coefficient of the current density is 
the average value of $q {\mathbf{v}}$ wheighted by ${\hat{f}}({\mathbf{p}})$.
Using the solution for ${\hat{f}}$ in equation~(\ref{F1}), each component of the current 
appears to be a linear function
of the electric field components ${\hat{E}}_X$, ${\hat{E}}_Y$
and ${\hat{E}}_Z$ from which the components of the conductivity tensor may be identified. Each one involves
a fourfold integral on the variables $p_\parallel$, $p_\perp$, $\phi$ of the particle momentum and
on the delay time $\tau$. 
As can be seen in equation (\ref{F1}), integration over the delay time involves
imaginary exponentials with an argument linear and 
trigonometric in $\tau$. We refer to these integrals as phase integrals. 
Expanding the trigonometric functions
of $(\phi - \Omega_* \tau)$ on the second line of equation~(\ref{F1}) in imaginary exponentials, 
three phase integrals appear, that can be written as 
\begin{equation}
P_\varepsilon(\omega, k_\perp, k_\parallel, {\mathbf{v}} )
= \int_0^\infty \!\!\! d\tau \
\exp\Big(i \Big((\omega - k_\parallel v_\parallel) \tau - \frac{k_\perp v_\perp}{\Omega_*} (\sin \phi - \sin (\phi - \Omega_* \tau))
+ \varepsilon (\phi - \Omega_* \tau) \Big) \Big)
\,.\label{defintphase}
\end{equation}
The index $\varepsilon$ may take the three values ($-1$, $0$, $+1$), or ($-$, $0$, $+$). 
At this point, it is necessary to take care of the sign of the charge, and  to introduce dimensionless 
variables, simpler than those in equation~(\ref{defintphase}), by:
\begin{equation}
x = \frac{k_\perp v_\perp}{\mid \! \Omega_*\!\mid}\,,
\qquad  \qquad \sigma = \frac{\omega - k_\parallel v_\parallel}{\mid \! \Omega_*\!\mid}\,,
\qquad  \qquad u = \frac{\omega}{\mid \! \Omega\!\mid} \,,\qquad  \qquad 
\sigma_0 = u \sin \alpha \,,\qquad  \qquad
\varpi = \frac{c k_\parallel - v_\parallel k}{\mid \! \Omega_*\!\mid}
\,.\label{defxsigma}
\end{equation}
The variables $x$ (not a coordinate), $\sigma$ and $u$ are all positive and usually large
since $\omega = ck \gg \mid \! \Omega\!\mid$. The variable $\varpi$, which
plays the role of an angular variable,
may be positive or negative. The modulus of $\varpi$ assumes values of the same order as $\sigma$. 
The properties of the variables $\varpi$ 
and $\sigma$, which may replace $p_\parallel$ and $p_\perp$ as dynamical variables of
freely moving particles, are detailed in Appendix~\ref{Apprhosigma}. 
From equation~(\ref{F1}), the current density can be written in terms of the three phase integrals as:
\begin{eqnarray}
{\hat{\mathbf{J}}} &=&
- \frac{2\pi q^2}{\omega}
\int_{-\infty}^{+\infty} \!\! dp_\parallel\!\! \int_0^\infty\!\! p_\perp  dp_\perp \!\!
\int_0^{2\pi}\!\! \frac{d\phi}{2\pi} \ \ \, \Big[v_\perp \cos \phi \, {\mathbf{e}}_X + v_\perp \sin \phi \, {\mathbf{e}}_Y
+ v_\parallel \, {\mathbf{e}}_Z\Big] \times
\nonumber \\
&&\Big[
\frac{P_+ \!+\! P_-}{2}\, \Big(\omega \frac{\partial f_0}{\partial p_\perp} + k_\parallel D(f_0) \Big) \, {\hat{E}}_X
+ \frac{P_+ \!-  \! P_-}{2i} \, \Big(\omega \frac{\partial f_0}{\partial p_\perp} + k_\parallel D(f_0) \Big) \, {\hat{E}}_Y
+ \Big(P_0 \, \omega \frac{\partial f_0}{\partial p_\parallel} - \frac{P_+ \!+\! P_-}{2} \ k_\perp D(f_0) \Big)\, {\hat{E}}_Z
\Big]
\,.\label{courantJphasegeneral} 
\end{eqnarray}

\section{Results }
\label{SecResults}

We derive in Section~\ref{Secderivation} the components of the conductivity tensor for a relativistic plasma
at frequencies much higher than the gyrofrequency and carry out two
out of the four integrals involved in equations (\ref{defintphase}) and  (\ref{courantJphasegeneral}).
In particular cases, the corresponding quadrature may
be pursued one step farther. 
In this section we anticipate the  
polarization transfer coefficients and illustrate
them for isotropic and anisotropic distributions.

\subsection{Dissipation-less polarization transfer coefficients}
\label{sectionexpressions}

\subsubsection{Quasi-exact expressions}

We obtain in Section~\ref{sectresonancesnegl} a general expression for the 
Faraday rotation and conversion coefficients $f$ and $h$, 
defined $f = K_{QU}/c$ and $h= K_{UV}/c$, where $ K_{QU}$ and $K_{UV}$ are given by
equation (\ref{coeffdesigmaHA}). These coefficients can be explicitely written as in 
equations (\ref{fpremierjus})--(\ref{hpremierjus}), which we reproduce here
neglecting the principal value terms that are shown in Section~\ref{sectresonancesnegl}
to be safely negligible when $\omega \gg \Omega$. These equations are therefore almost exact in this limit.
We refer to them as being nearly exact or quasi-exact. They can be written as
\begin{eqnarray}
f &=& - \, \frac{2\pi^2 s_q}{c} \ \frac{\omega_{\rm pr}^2 \Omega^2}{\omega^3}
\int\!\!\!\!\int \!\! \frac{m^3\!c^3}{\sin^2\!\! \alpha} \, d\varpi \, d\sigma
\ \ \varpi \, x \ \frac{\partial F_0}{\partial \sigma} \
\left[
J'_\sigma(x) N_\sigma(x) + \frac{1}{\pi x}\right]
\,,\label{fexacteanticip}
\\
h &=& \frac{\pi^2}{c} \frac{\omega_{\rm pr}^2 \Omega^2}{\omega^3}
\int\!\!\!\!\int \!\! \frac{m^3\!c^3}{\sin^2\!\! \alpha} \, d\varpi d\sigma
\left[
\frac{\partial F_0}{\partial \sigma} 
\left(x^2 J'_\sigma(x) N'_\sigma(x) - \varpi^2 J_\sigma(x) N_\sigma(x)\right)
+ \frac{1}{\pi} \left(\varpi \frac{\partial F_0}{\partial \varpi} - \sigma \frac{\partial F_0}{\partial \sigma}\right)
\right]
\,.\label{hexactanticip}
\end{eqnarray}
The particle's dynamical variables $\varpi$ and $\sigma$, defined in equation (\ref{defxsigma}), incorporate 
properties of the radiation of interest. Much of the dependence of $f$ and $h$ on the
propagation angle $\alpha$ and on the frequency is in fact hidden in these variables.
The distribution function $F_0$ is normalized to unity [in the sense of equation (\ref{definiGrandF0}) below], 
the particle density $n$ being implicit in 
the square of the specie's plasma frequency $\omega_{\rm pr}^2 = n q^2/m\varepsilon_0$.
The integrals in equations (\ref{fexacteanticip}) and (\ref{hexactanticip}) involve 
two-variable kernels in which one of the variables, $\sigma$,
appears as a continuous Bessel function index. This index 
is larger than their argument $x$, as can be seen
from equation (\ref{defxsigma}), but possibly not much larger, 
which complicates the search for suitable approximations. 

There are essentially 
two very different regimes, defined in Section~\ref{sec:NR-QR}, for the wave-particle interaction 
which we refer to as non-resonant (NR) and quasi-resonant (QR). Different particles 
make very different contributions to the transfer coefficients
depending on whether they interact in the NR or in the QR regime. Equations
(\ref{fexacteanticip}) and (\ref{hexactanticip}) however 
encompass all regimes, being  written in a form that does not require
any regioning of the $\varpi$-$\sigma$ domain for their evaluation. 
The domain of validity of specific approximations, NR or QR,
to the functions in equations (\ref{fexacteanticip}) and (\ref{hexactanticip}) are
different and complementary (Fig. \ref{figBQR}).  

\subsubsection{HF limit}

It is shown in Section~\ref{contribQR} that when 
the inequality
\begin{equation}
\frac{3\, \gamma^2 \sin \alpha \mid\!\Omega\!\mid}{\omega} < 1
\label{regimechange-maintext}
\end{equation}
holds true, only the NR domain, defined in that section, contributes to the transfer coefficients.
For a given value of the Lorentz factor $\gamma$,
the inequality (\ref{regimechange-maintext}) is satisfied when the radiation's frequency $\omega$
exceeds the characteristic synchrotron frequency $\omega_c(\gamma, \alpha)$ 
for particles of this energy 
travelling in the direction of the radiation  (equation (\ref{defomegacritsynch})).
We therefore refer to this situation as the high-frequency (HF) limit.
When all particles in the distribution interact with the wave in this regime, the integration on the NR domain
actually extends over the full $\varpi$-$\sigma$ domain.
In this case, the Faraday coefficients $f$ and $h$ result from equations (\ref{fNR}) and (\ref{hNRreduit}),
expanded to the first non-vanishing order in $\Omega/\omega$:
\begin{eqnarray}
&& f_{\rm HF} = 2\pi s_q \, \frac{\omega_{\rm pr}^2 \Omega^2}{c \, \omega^3} \!\!
\int\!\!\!\!\int  \frac{m^3\!c^3}{\sin^2\!\! \alpha}  d\varpi \, d\sigma
\ \,
\left(\frac{\varpi \, x^2}{2 \ (\sigma^2 -x^2)^{3/2} }
\right) \ \frac{\partial F_0}{\partial \sigma}
\,.\label{ffinalordr1pisigma}
\\
&&h_{\rm HF} =-\, \pi \, \frac{\omega_{\rm pr}^2 \! \Omega^2}{c \, \omega^3} \!
\int\!\!\!\!\int \frac{m^3\!c^3}{\sin^2\!\! \alpha} \, d\varpi  \, d\sigma \ \,  
\left(\frac{2 x^4 (\sigma^2 - x^2) + \sigma_0^2 \, x^2 (4 \sigma^2 + x^2) }{8\, (\sigma^2 - x^2)^{7/2}}\right)
\ \frac{\partial F_0}{\partial \sigma} \,.
\label{hfinalrecappisigma}
\end{eqnarray}
Equations (\ref{ffinalordr1pisigma}) and (\ref{hfinalrecappisigma})
do not assume isotropy of the distribution function $F_0$ but are only valid  when
the QR contribution can be neglected, which makes them
less general than equations (\ref{fexacteanticip})--(\ref{hexactanticip}).

\subsubsection{Medium-frequency ({MF}): cutting through the QR domain }

The integration over the
NR domain cannot be extended to the full domain when the inequality
in equation (\ref{regimechange-maintext}) is not satisfied for a non-negligible number
of particles in the distribution.
This arises in particular when the parameter on the left of equation (\ref{regimechange-maintext})
is of order unity, which we refer to as a medium-frequency (MF) situation.
A slightly more sophisticated approach is then needed, which we now describe.
The double quadratures in equations (\ref{ffinalordr1pisigma})--(\ref{hfinalrecappisigma}) should
in this case be extended over the NR domain only and the contribution of the
QR domain should be added to it. When it is not dominant, the integral over the QR domain may
be approximately evaluated.
Our proposed approximation rests on separating the NR from the QR domain,
and expressing the integral over the QR one by using a very simple approximation. The boundary
${\cal{B}}_{\rm QR}$ between the NR and QR regions
is where the variable $g$ defined in equation (\ref{gtexte}) below is near unity.
Since $x \approx \sigma$ in the vicinity of this limit, ${\cal{B}}_{\rm QR}$ 
is represented in the $\varpi$--$\sigma$ plane by the equation
\begin{equation}
B_{\rm QR}(\varpi, \sigma) \equiv \varpi^2  - 3^{2/3} \sigma^{4/3} + \sigma_0^2 = 0 \, .
\label{bordNRQR}
\end{equation}
The sign of $B_{\rm QR}(\varpi, \sigma)$ tells whether the point $\varpi$--$\sigma$ is
in the NR domain ($B_{\rm QR} > 0$) or in the QR one ($B_{\rm QR} < 0$).
The QR domain is then cut out of the $\varpi$-$\sigma$ domain, the remainder constituting the
properly-defined NR domain.
The kernels of the integrands in equations (\ref{ffinalordr1pisigma})--(\ref{hfinalrecappisigma}) 
are kept at their NR values
when in the NR domain. For perfect accuracy, the kernels in the QR region
should be taken to be those in equations (\ref{fQR})--(\ref{hQR}).
When however the QR region does not contribute predominantly,
these complicated kernels 
can be replaced by simple approximations.
Their effect essentially being to limit
the growth of the modulus of the NR kernels when the QR region is reached, we
replace them by linear interpolations in $\varpi$, at fixed $\sigma$, between the
values of the NR kernels evaluated on the border ${\cal{B}}_{\rm QR}$ for that value of $\sigma$.
Let $\Theta_{\rm H}(y)$ denote the Heaviside function, equal to unity for
positive argument $y$ and zero otherwise. The above-described approximations to $f$ and $h$ can be written as
\begin{eqnarray}
&&  \!\!\!\!\! \!\!\!\!\! \!\!
f_{\rm MF} = 2\pi s_q \, \frac{\omega_{\rm pr}^2 \Omega^2}{c \, \omega^3} \!\!
\int\!\!\!\!\int \!\! \frac{m^3c^3}{\sin^2 \alpha}  d\varpi \, d\sigma
\ \varpi  \left[ \Theta_{\rm H}^+
 \, \frac{x^2}{2 \, (\sigma^2 -x^2)^{3/2} }  + \Theta_{\rm H}^-
\, \frac{x_{\rm QR}^2}{2\, (\sigma^2 -x_{\rm QR}^2)^{3/2} } \right]
\frac{\partial F_0}{\partial \sigma}
\, ,\label{fcoupure}
\\
&& \!\!\!\!\! \!\!\!\!\! \!\!
h_{\rm MF} =-\! \pi \, \frac{\omega_{\rm pr}^2 \! \Omega^2}{c \, \omega^3} \!\!
\int\!\!\!\!\int \!\! \frac{m^3c^3}{\sin^2\! \alpha} \, d\varpi  \, d\sigma 
\Big[ \Theta_{\rm H}^+
\, \frac{2 x^4 (\sigma^2 - x^2) + \sigma_0^2 \, x^2 (4 \sigma^2 + x^2) }{ 8\  (\sigma^2 - x^2)^{7/2}}  + \Theta_{\rm H}^-
\, \frac{ 2 x_{\rm QR}^4 (\sigma^2 - x_{\rm QR}^2) + \sigma_0^2 \, x_{\rm QR}^2 (4 \sigma^2 + x_{\rm QR}^2) 
}{ 
8\ (\sigma^2 - x_{\rm QR}^2)^{7/2} }
\Big]
\frac{\partial F_0}{\partial \sigma}\cdot 
\label{hcoupure}
\end{eqnarray}
where $\Theta_{\rm H}^\pm=\Theta_{\rm H}(\pm B_{\rm QR}(\varpi, \sigma))$
and $\varpi_{\rm QR}(\sigma)$ and $x_{\rm QR}(\sigma)$ are the values of $\mid\!\!\varpi\!\!\mid$ and $x$ on
the boundary ${\cal{B}}_{\rm QR}$ for the given value of $\sigma$ (equation~(\ref{bordNRQR})). 
Equations (\ref{fcoupure}) and (\ref{hcoupure}) provide approximate values of the transfer coefficients
$f$ and $h$ valid for isotropic as well as for non-isotropic distribution functions.
The approximations in equations (\ref{fcoupure}) and (\ref{hcoupure})
will be illustrated
on the example of a thermal distribution function
in Section~\ref{sec:thermal}. 
\subsubsection{LF limit}
When the reverse inequality to equation (\ref{regimechange-maintext}) applies in a strong sense,
both the NR and QR domains contribute to the transfer coefficients.
For a given frequency and a given propagation angle, $\alpha$,
this essentially happens when the dynamical variable $\sigma$ of the particle is above the limit
defined by equation (\ref{bordNRQR}) for $\varpi = 0$ or when its Lorentz factor $\gamma$
largely exceeds the threshold defined by the reverse of inequality (\ref{regimechange-maintext}),
which happens when the
frequency $\omega$ is much less than the critical synchrotron frequency
for a particle of Lorentz factor 
$\gamma$ travelling in the direction of the considered radiation (equation (\ref{defomegacritsynch})).
We refer to this situation as the low-frequency (LF) limit. The LF contribution
to the Faraday coefficients comprises both NR and QR parts, which we 
calculte for isotropic distributions in Sections \ref{isotropicfLF} and \ref{kernelHlowfreqanalytiq}.
%
%%%%%%%%%%%%%%%%%%%%%%%%%%%%%%%%%%%%%%%%
\subsection{Kernels of polarization transfer coefficients for isotropic distribution functions}
\label{SecIsotrope}
Let us illustrate how to use equations~(\ref{fexacteanticip}) and~(\ref{hexactanticip})
in isotropic models of the distribution function. When the distribution function is isotropic, 
it depends only on the Lorentz factor $\gamma$ of the particles,
but nevertheless depends on both variables $\varpi$ and $\sigma$.
When dealing with isotropic distribution functions, it is therefore wise to
change from the variables $\varpi$, $\sigma$ to the variables $\varpi$, $\gamma$. Integrating over $\varpi$, 
which then does not involve the distribution function, equations (\ref{fexacteanticip}) and~(\ref{hexactanticip})
can be written in the general form
\begin{equation}
f = \frac{\omega_{pr}^2 \Omega^2}{c \, \omega^3} \, m^3c^3 \int_1^\infty\!d \gamma \,
\frac{dF_0}{d\gamma} \, F^{\rm iso}(\gamma)\,,
\qquad \qquad \qquad 
h= \frac{\omega_{pr}^2 \Omega^2}{c \, \omega^3} \, m^3c^3 \int_1^\infty\!d \gamma \,
\frac{dF_0}{d\gamma} \, H^{\rm iso}(\gamma)\,,
\label{defnoyauxisotropes}
\end{equation}
where $F^{\rm iso}(\gamma)$ and $H^{\rm iso}(\gamma)$ are one-variable kernels for isotropic distribution functions.
\subsubsection{Isotropic distributions in the HF limit}
\label{isotropichighfreqresult}
In the HF limit, the integration over $\varpi$ 
in equations (\ref{ffinalordr1pisigma})--(\ref{hfinalrecappisigma}),
which defines these kernels, 
extends, for a given value of $\gamma$, over the full domain.
The kernels $F^{\rm iso}$ and $H^{\rm iso}$ 
can then be written in this case as
\begin{eqnarray}
&& F^{\rm iso}_{\rm HF} (\gamma)= 
\frac{2\pi s_q}{\sin^2 \alpha} 
\int_{\varpi_-}^{\varpi_+}\, d\varpi \, \ \left(\frac{\varpi  \, x^2}{2\, (\sigma^2 -x^2)^{3/2}} \right)
\,.\label{ffinalordr1pigamma}
\\
&&H^{\rm iso}_{\rm HF} (\gamma)=-\, \frac{\pi}{\sin^2\! \alpha} \, 
\int_{\varpi_-}^{\varpi_+}\, d\varpi \   \,  
\left(\frac{2 x^4 (\sigma^2 - x^2) + \sigma_0^2 \, x^2 (4 \sigma^2 + x^2) }{8\ (\sigma^2 - x^2)^{7/2}  }\right)
\,.\label{hfinalpigamma}
\end{eqnarray}
The variables $\sigma$, $x$ and $(\sigma^2\! -\! x^2)$ can 
be expressed from equation (\ref{pperpderhoq}) in terms of 
$\gamma$, $\varpi$ and parameters in equation (\ref{defxsigma}) as
\begin{equation}
\sigma = \gamma \, u \sin^2\!\! \alpha + \varpi \, \cos \alpha \,,
\qquad \qquad 
x^2 = (u^2 (\gamma^2 \sin^2\!\! \alpha - 1) + 2\gamma u \, \varpi \cos \alpha  -\varpi^2)\sin^2\alpha \, , \qquad \qquad 
\sigma^2 - x^2 = \varpi^2 + u^2 \sin^2\!\! \alpha \, .
\label{xgammapi}
\end{equation}
The boundaries $\varpi_-(\gamma)$ and $\varpi_+(\gamma)$ of the integration  over $\varpi$ at a given $\gamma$ are
the abcissas of the intersection of the line of constant $\gamma$ with the boundary of the 
physical domain in the $\varpi$-$\sigma$ plane (Fig. \ref{figpisigma}). 
Since this boundary is where $x = 0$, equation (\ref{xgammapi}) implies that 
\begin{equation}
\varpi_{\pm}(\gamma) = u \left( \gamma \cos \alpha \pm \sqrt{\gamma^2 - 1} \right)\, ,
\label{varpiplusoumoins}
\end{equation}
The integration over $\varpi$ in equations (\ref{ffinalordr1pigamma}) and (\ref{hfinalpigamma}) 
plays the role of an angular integration.
It can be performed analytically, the results, valid at high frequency, eventually being
\begin{eqnarray}
&& F^{\rm iso}_{\rm HF}(\gamma) =  4 \, \pi s_q \, \frac{\omega \cos \alpha}{\mid \!\!\Omega\!\!\mid}
\left(\gamma {\cal L}(\gamma)
- \sqrt{\gamma^2 -1} \right) \,,\,\,\, \quad \quad
{\rm where}\qquad  {\cal L}(\gamma )= {\cosh}^{-1}\!(\gamma)=
\ln \left(\sqrt{\gamma ^2-1}+\gamma \right)
\! ,\label{fisotfinal}
\\
&&H^{\rm iso}_{\rm HF}(\gamma) =-\, \pi \, 
\frac{\sin^2 \alpha}{2} \left(\gamma \, \sqrt{\gamma^2 -1} \, \left(2 \gamma^2 - 3\right) + {\cal{L}}(\gamma) 
\right) 
\,.\label{hisotfinal}
\end{eqnarray}
\subsubsection{Isotropic distributions in the LF limit}
\label{isotropiclowfreqresult}
It comes out of the discussion in Section~\ref{comparQRNR} 
that the coefficient of Faraday rotation $f$ does not depart much 
from its HF approximation, although corrections are 
necessary for a quantitative agreement in LF situations. 
We find in Section~\ref{seclowfrequencylimit} that the isotropic kernel for $f$ 
can be written in the asymptotic LF limit as
\begin{equation}
F^{\rm iso}_{\rm LF}(\gamma) = \pi s_q \, \frac{\omega \cos \alpha}{\mid \!\!\Omega\!\!\mid} \, \gamma
\left(\frac{4}{3} \, \ln\left(\frac{\gamma u}{\sin \alpha}\right) - 1.260\, 724\, 39\right)
\,. \label{FisoLF}
\end{equation}
By contrast, the Faraday
conversion coefficient $h$ drastically differs from its 
HF expression when there is a QR contribution. 
The kernel of this coefficient is derived in Section~\ref{isotropicfLF}
in the asymptotic LF limit and for an isotropic distribution function. It can then be written as:
\begin{equation} 
H^{\rm iso}_{\rm LF}(\gamma) =  \frac{\pi}{8}
\left(\frac{\omega^2 \sin\!\alpha}{\Omega^2}\right)^{2/3}
\left(4  - \frac{1}{3^{4/3}}\right)
\ \gamma^{4/3} \,.
\label{hinlowfreqResu}
\end{equation}

\subsubsection{Exact isotropic kernels and composite approximations to them}

The results in Sections \ref{isotropichighfreqresult} and \ref{isotropiclowfreqresult} apply in different regions of 
the energy space. For distributions with only a very few particles in the LF regime, the HF 
approximation is sufficient. When the distribution is more
extended in energy than the threshold $\gamma_{\rm QR}$ defined in equation (\ref{sigmagammaqr}), 
the kernels should account for the QR contribution in the LF regime. They may then be 
calculated by integrating over $\varpi$ at fixed $\gamma$  the right-hand sides 
of equations (\ref{fexacteanticip})--(\ref{hexactanticip}). This however requires a double numerical integration. 
Section~\ref{kernelsinterpol} proposes an approximate formula interpolating 
between the HF expression of these kernels, valid when 
$\gamma < \gamma_{\rm QR}$, and their asymptotic LF expressions, valid for $\gamma \gg \gamma_{\rm QR}$. 
\citet{HuangCerbakov} adopt a similar approach by providing fits to their numerically computed kernels and
claim similar accuracy.

For extended energy distributions, the region at $\gamma \sim \gamma_{\rm QR}$ should not dominantly contribute 
to integrals such as those in equation (\ref{defnoyauxisotropes}). These interpolation formulae then
provide about 10 per cent accuracy on $h$ for power-law distributions and slightly better accuracy on $f$. 
When on the contrary the energy range close
to $\gamma_{\rm QR}$ is determinant, the MF  approximations in equations (\ref{fcoupure})--(\ref{hcoupure})
could be used instead or, if accurate results are needed, the transfer coefficients 
would be numerically calculated from equations  
(\ref{fexacteanticip}) and (\ref{hexactanticip}) by double integrating over the variables $\varpi$ and $\sigma$.

\subsubsection{Illustrations and comparison to known results}\label{sec:illustre}
\begin{figure}
\begin{center}
\includegraphics[width=0.45\hsize]{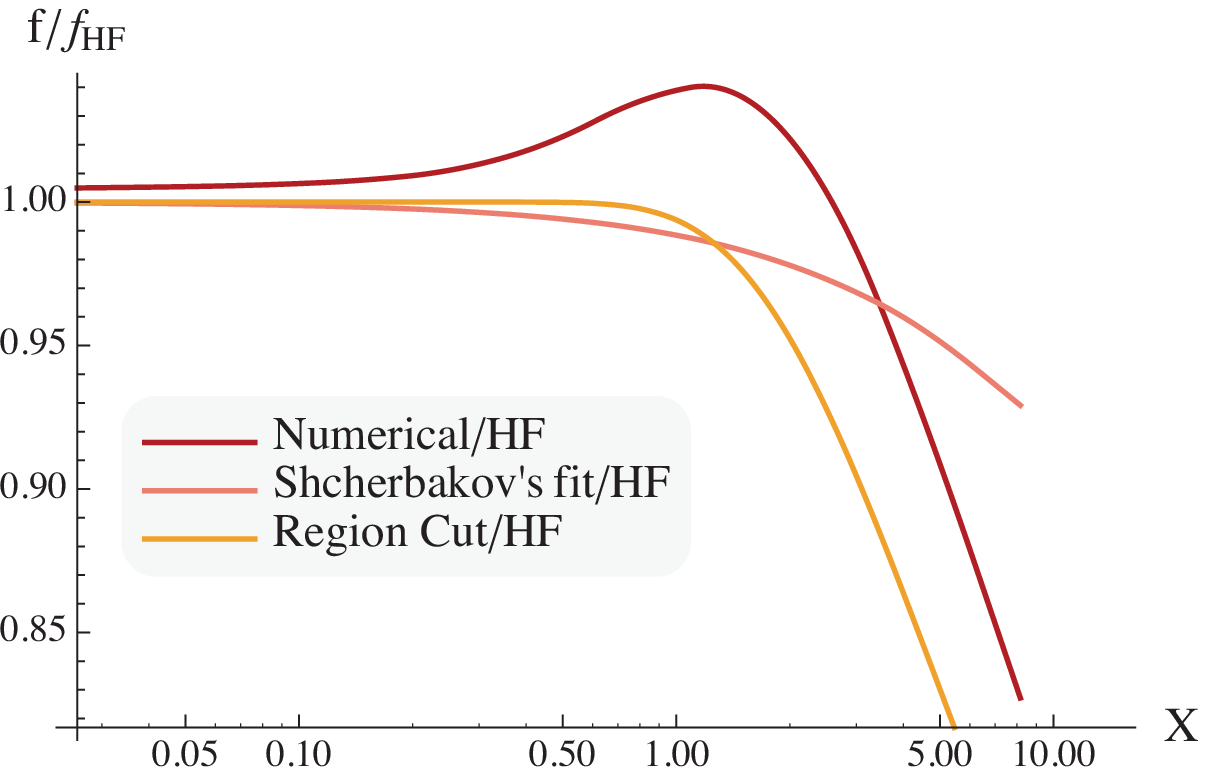}
\includegraphics[width=0.45\hsize]{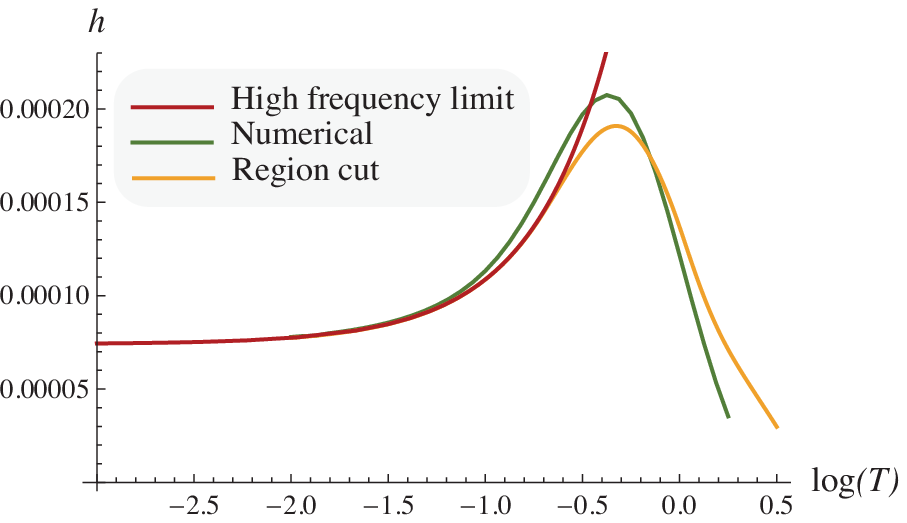}
\caption{ Left: ratio to the high-frequency expression $f_{HF}^{\rm th}$ in equation (\ref{cerbakov-thermal})
of, respectively, the numerically computed value of the 
thermal Faraday rotation coefficient $f^{\rm th}$ for the isothermal model of equation~(\ref{defFthermal}),
based on equation~(\ref{fexacteanticip}), the fit formula provided for it by \protect\cite{Cerbakov2008}
and the region-cut approximation in equation (\ref{fcoupure}), as labelled. 
These ratios are displayed as a function of 
the scaling parameter $X$ introduced by \protect\cite{Cerbakov2008}, defined in equation (\ref{XCherbak}).
Right: the HF limit of the Faraday conversion coefficient
$h^{\rm th}$ from equation (\ref{cerbakov-thermal}), the region-cut formula from equation (\ref{hcoupure})
and the almost exact value 
numerically computed from equation~(\ref{fexacteanticip}), as a function of the temperature, 
$T$ (as labelled). In both cases, the region-cut procedure around the resonant particles 
seems to improve significantly the domain of validity of  the match to the 
exact transfer coefficients.
}
\label{ratio}
\end{center}
\end{figure}
\noindent
{\it{Previous approaches to polarization transfer coefficients.}}$\qquad$
The polarization transfer coefficients for a thermal distribution have been previously derived exactly,
by a method entirely different from ours,
in a form that eventually requires only one numerical quadrature \citep{Trubnikov}.
It is therefore of interest to test different approximate results 
against this exact result, or against results proposed for the thermal case by different authors. 
\citet{Trubnikov} considers a J\"uttner distribution, 
as in equation (\ref{fddthermique}) below. The conductivity is derived
by integrating the changes of the perturbed distribution function over the delay time along unperturbed trajectories,
just as shown in Section~\ref{derivation}, and then taking suitable moments of the components of momentum.
The exponential dependence of the J\"uttner distribution on energy
can be associated with the complex exponential that arises from
the evolution over time of the phase functions, as in equation (\ref{intdephasefactorsigmaeps}).
The integration over
the particle's energy can then be carried out, 
yielding a result that involves a modified Bessel function of the second kind of
a complex argument depending on the delay time \citep{MelroseJplPhyspremier, Cerbakoff2011}.
The integration over the delay time then remains to be carried out. This
can be done by different numerical strategies \citep{HuangCerbakov, Cerbakoff2011},
yielding exact results. Else, it can be carried out analytically  in an approximate way
in different limits \citep{MelroseJplPhys, Cerbakov2008}
or by using the method of stationary phase \citep{MelrosePhysRev}. 
\citet{Swanson} calculates the dielectric tensor of a moderately relativistic thermal plasma
from the familiar multiple resonance series \citep{Bekefi} which he sums in a form similar 
to our equation (\ref{lesGabBesselsigma}) below, yielding expressions that involve 
Bessel functions depending on a continuous argument. This approach is analogous to those used by
\citet{Qin}, \citet{theseJean}, \citet{theseJerome} and in this paper, where we furthermore 
take full advantage of the high harmonic limit. \citet{Swanson} then uses his result to perform the
integration over the directions of motion of the particles in terms of hypergeometric functions for which
he presents approximations suitable for weakly relativistic plasmas.
\citet{HuangCerbakov} have
extended Trubnikov's approach to non-thermal isotropic distribution functions, the integration with respect to delay time
being performed first for a given value of the Lorentz factor. These authors
numerically calculate transfer coefficients for Dirac distributions in energy and
provide fitting formulae for the kernels of Faraday rotation and conversion.
The coefficients for arbitrary isotropic distributions 
may be obtained by a further integration over energy. Examples are provided 
for power-law distributions covering a finite energy interval and for thermal distributions.
In the following sections we compare our results for a thermal distribution to results obtained by these approaches.\\

\noindent
{\it{Polarization transfer coefficients for a thermal 
distribution in the HF limit.}}\label{sec:thermal}$\quad$
Following \cite{Cerbakov2008} we express the thermal 
J\"uttner distribution as
\begin{equation}
F_0(\gamma)=\frac{\exp\left({-{\gamma }/{T}}\right)}{4 \pi  m^3 c^3 T K_2\left(T^{-1}\right)}\,,
\quad \mbox{normalized so that} \quad \int 4 \pi  \gamma  \sqrt{\gamma ^2-1} \ m^3\! c^3{F_0}(\gamma ) d\gamma=1\,. 
\label{defFthermal}
\label{fddthermique}
\end{equation}
where $T$ is the ratio of 
the thermal energy to the rest mass energy of the particles.
By Trubnikov's method, \citet{Cerbakov2008} finds the following 
polarization transfer coefficients in the HF limit
\begin{equation}
f^{\rm th}_{\rm HF} =\frac{\Omega  \cos \alpha \, 
\omega_{\rm{pr}}^2 K_0\!\left({T^{-1}}\right)}{c \,\omega ^2 K_2\!\left({T^{-1}}\right)}\,,
\qquad \quad {\rm and} \qquad \quad
h^{\rm th}_{\rm HF} =\frac{\Omega ^2 \sin^2\!\! \alpha \,  
\omega_{\rm{pr}}^2 \left(K_1\!\left(T^{-1}\right)+6 T K_2\!\left(T^{-1}\right)\right)}{2 c \omega ^3 K_2\!\left(T^{-1}\right)}\,,
\label{cerbakov-thermal}
\end{equation}
where $K_1$ and $K_2$ are modified Bessel functions of the second kind.
In this HF limit his results are identical to ours. To compare them, we evaluate $f$ and $h$ 
from equations (\ref{fisotfinal})--(\ref{hisotfinal}), also valid in this limit,
considering the case of an electronic population, so that $s_q=-1$.  
The $f$ coefficient, derived from equations (\ref{defnoyauxisotropes}) and (\ref{fisotfinal}), becomes in this case 
\begin{equation}
f^{\rm th}_{\rm HF} =  \frac{\Omega \, \cos \alpha\, \omega_{\rm pr}^2}{c \, \omega^2}  
\frac{1}{T^2 K_2\!\left(T^{-1}\right)} \ \, 
\int_1^\infty \!\!  d\gamma  \  
e^{-{\gamma }/{T}} \left(\gamma\, {\cal L}(\gamma) -  \sqrt{\gamma^2 -1} \right) \, . 
\label{fisot-gammapart}
\end{equation}
Keeping the integrand part in $\sqrt{\gamma^2-1}$ untouched, the term $\gamma\, {\cal L}(\gamma) \exp(-\gamma/T)$
is integrated by parts, giving successively
\begin{equation}
\int_1^\infty\!\! d\gamma \ \gamma \ {\cal L}(\gamma)\, e^{-\gamma/T} 
= T^2\int_1^\infty \!\! d\gamma\ {e^{-\gamma/T} \over \sqrt{\gamma^2 -1}} + T\, \int_1^\infty \!\!
d\gamma \ {\gamma \over \sqrt{\gamma^2-1}} \ e^{-\gamma/T} 
= T^2\, K_0(T^{-1}) + \int_1^\infty \!\! d\gamma \ \sqrt{\gamma^2-1}\ e^{-\gamma/T}\,,
\end{equation} 
where the last equality has been obtained again by integrating by parts the second integral of the middle part. 
Substituting this into equation~(\ref{fisot-gammapart}), we recover $f^{\rm th}$ in equation~(\ref{cerbakov-thermal}).
Starting  from equations~(\ref{defnoyauxisotropes}) and (\ref{hisotfinal}) for $h^{\rm iso}_{\rm HF}$ we similarly 
recover $h^{\rm th}_{\rm HF}$ in equation~(\ref{cerbakov-thermal}) by noting that 
$\int d\gamma\,  {\cal L}(\gamma) \exp(-\gamma/T) =TK_0(T^{-1})$ and that
\begin{equation}
\int_1^\infty\!\!  d\gamma\ \gamma\, (2\gamma^2-3)\sqrt{\gamma^2-1}\ e^{-\gamma/T} 
= T\, \left(-K_2(T^{-1})+6\, T\, K_3(T^{-1}) \right)
= -T\, K_0(T^{-1}) +4T^2\left(K_1(T^{-1})+6\, T\, K_2(T^{-1})\right)\,.
\label{eqHFpourhthermal}
\end{equation}

\begin{figure}
\begin{center}
\resizebox{0.45 \hsize}{!}{\includegraphics{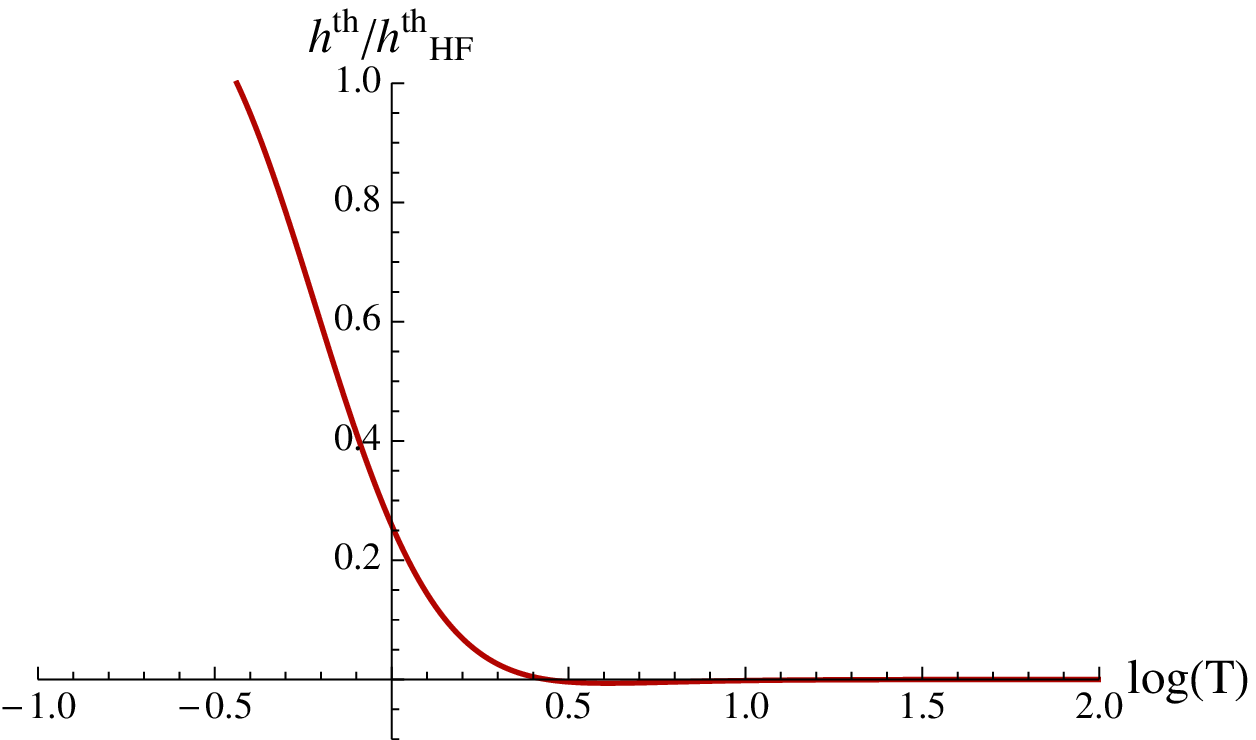}}
\resizebox{0.45 \hsize}{!}{\includegraphics{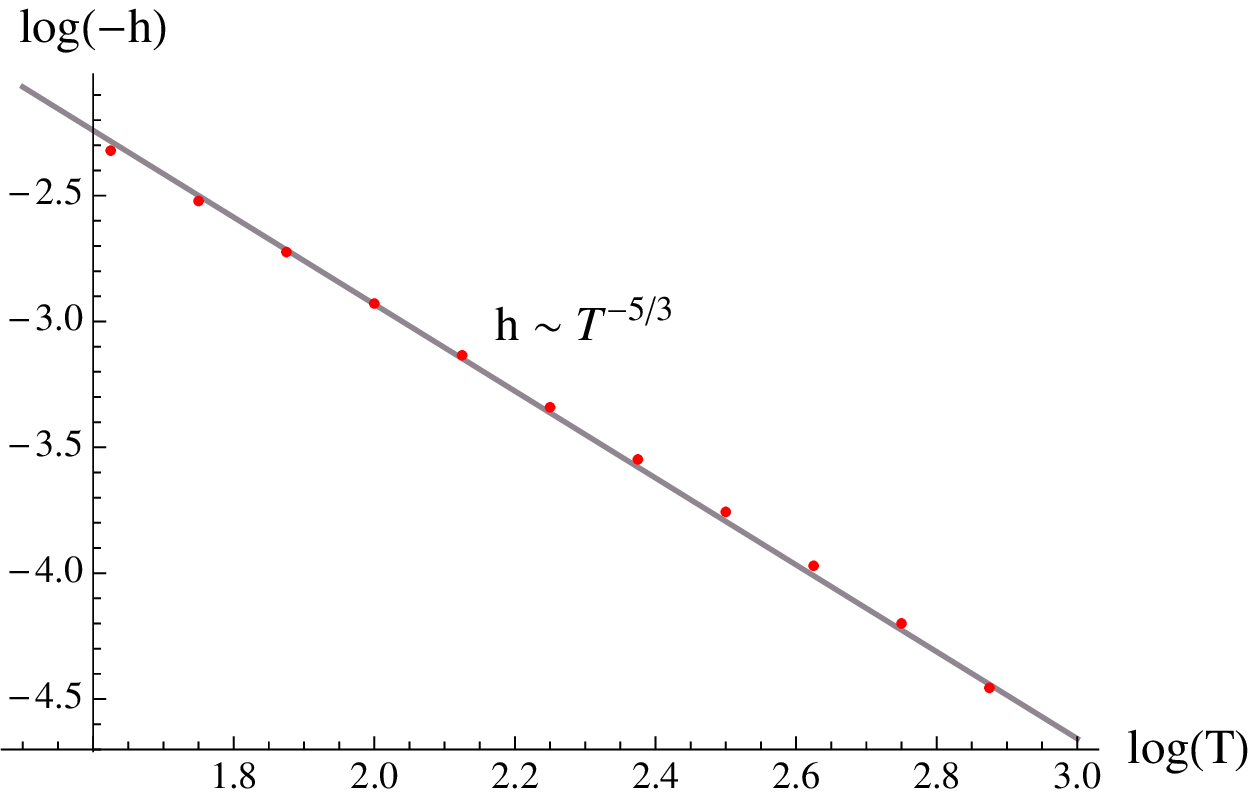}}
\caption{ 
Left: The evolution with temperature $T$ of the ratio of $h^{\rm th}$ to its high-frequency approximation $h^{\rm th}_{\rm HF}$.
Right: The evolution of $\log |h|$ versus $\log T$. The $T^{-5/3}$ asymptote at large $T$ is a 
direct consequence of the $\gamma^{4/3}$ 
asymptotic behaviour of the isotropic kernel of the Faraday coefficient $h$ at large $\gamma$.
}
\label{figisothermalhasymptotes}
\end{center}
\end{figure}
\smallskip
\noindent
{\it{Polarization transfer coefficients for a thermal distribution in the MF regime.}} \label{secthermalinterm}$\quad$
Stepping out of the domain of validity of the HF approximation, 
Fig.~\ref{ratio} comparatively illustrates the behaviour of different approximations
to the polarization coefficients by 
representing their HF approximations~(\ref{cerbakov-thermal}) and contrasting
them  to  the numerical estimation of their exact counterparts (equations~(\ref{fexacteanticip}) and~(\ref{hexactanticip})) 
and to the fits provided by \cite{Cerbakov2008} to describe this regime. 
The difference between the approximations to $f^{\rm th}$  being small, 
the left panel of  Fig.~\ref{ratio} represents their ratio to their HF approximation. 
These ratios are displayed
as a function of the parameter 
\begin{equation}
X=T \left(10^3\, \sqrt{2}\, \sin \alpha /u\right)^{1/2}\,,
\label{XCherbak}
\end{equation}
defined in \citet{Cerbakov2008},
which is proportional to the square root of the left-hand side of the inequality in equation (\ref{regimechange-maintext}). 
The right-hand panel of Fig.~\ref{ratio}  
shows the corresponding approximations to $h^{\rm th}$ as a function of temperature,
the quantity represented in this case being the coefficient itself.
This figure uses the following fiducial parameters $m=c=1$, $\alpha=\pi/4$, $u=15$. 
It is found that the region-cut approximation 
to the intermediate regime discussed in Section~\ref{sectionexpressions} and
represented by the expressions in equations (\ref{fcoupure}) and (\ref{hcoupure})
provides a good extension of these coefficients to higher temperatures where
the HF limit becomes insufficient.\\ 

\noindent
{\it{Polarization transfer coefficients for a thermal  distribution in the LF limit.}}
\label{isotropiclowfreqresultthermal}$\quad$
As an illustration of equation~(\ref{hinlowfreqResu}) let us consider again 
the isothermal distribution, equation~(\ref{defFthermal}).
It follows that when most particles interact with the radiation in the LF regime we expect
\begin{equation}
h^{\rm th}_{\rm LF}\propto-\frac{1}{8 \pi}\left(\frac{1}{T}\right)^{5/3} \Gamma \left(\frac{7}{3}\right)\,,
\end{equation}
which corresponds indeed to the asymptotic behaviour that can be computed numerically (using the Olver uniform expansion, 
see Section~\ref{subsubthermallowfreq} and Appendix~\ref{app:olver}), as shown on figure~\ref{figisothermalhasymptotes}.\\

\noindent
{\it{Polarization transfer coefficients for an isotropic power-law distribution function.}}
$\quad$
Let us also consider a power-law distribution scaling like $\gamma^{-n}$ with a low-energy cutoff $\gamma_{\rm m}$, such as:
\begin{equation}
F_0(\gamma)= \frac{\Theta_{\rm H}(\gamma-\gamma_{\rm m})}{N(\gamma_{\rm m}, n) \, \gamma^n}\,.
\label{definePL}
\end{equation}
For an isotropic distribution, assuming $n> 3$ for convergence,
the normalization factor $N(\gamma_{\rm m}, n)$ results from equation~(\ref{defFthermal})
\begin{equation}
\frac{4\pi m^3c^3}{N(\gamma_{\rm m}, n)} 
\int_{\gamma_{\rm m}}^\infty \frac{\sqrt{\gamma^2 -1} \ \gamma \, d\gamma}{\gamma^n} = 1\,.
\label{NormalisePL}
\end{equation}
The Faraday coefficients 
can be expressed for such distribution functions 
in terms of a $\gamma$-dependent kernel as in equations (\ref{defnoyauxisotropes}).
The low-energy cutoff introduces a Dirac-type singularity in the derivative $dF_0/d\gamma$ 
at $\gamma_{\rm m}$, that should be taken care of when
equations (\ref{defnoyauxisotropes}) is used.
Alternatively, these relations may be integrated by parts,
the integrated term vanishing because $F_0$ vanishes at infinity and the kernels do at $\gamma =1$.

\begin{figure}
\begin{center}
\resizebox{0.45 \hsize}{!}{\includegraphics{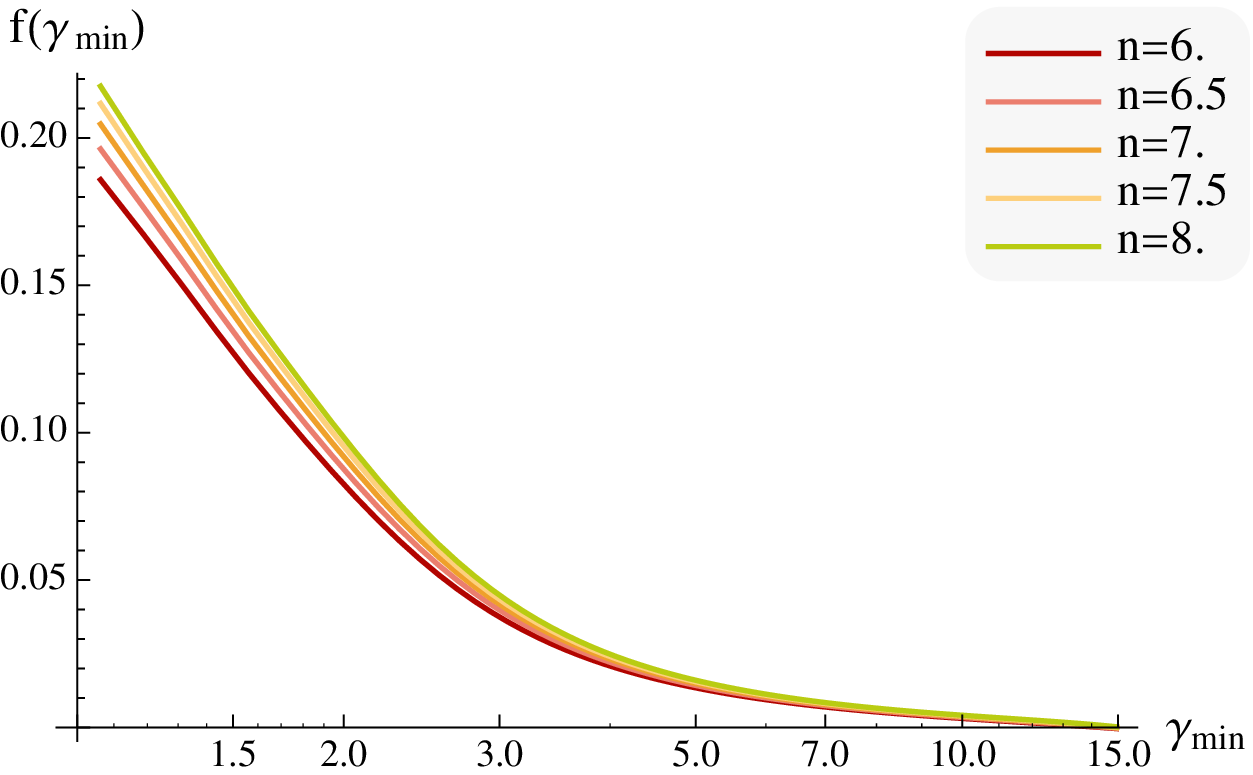}}
\resizebox{0.45 \hsize}{!}{\includegraphics{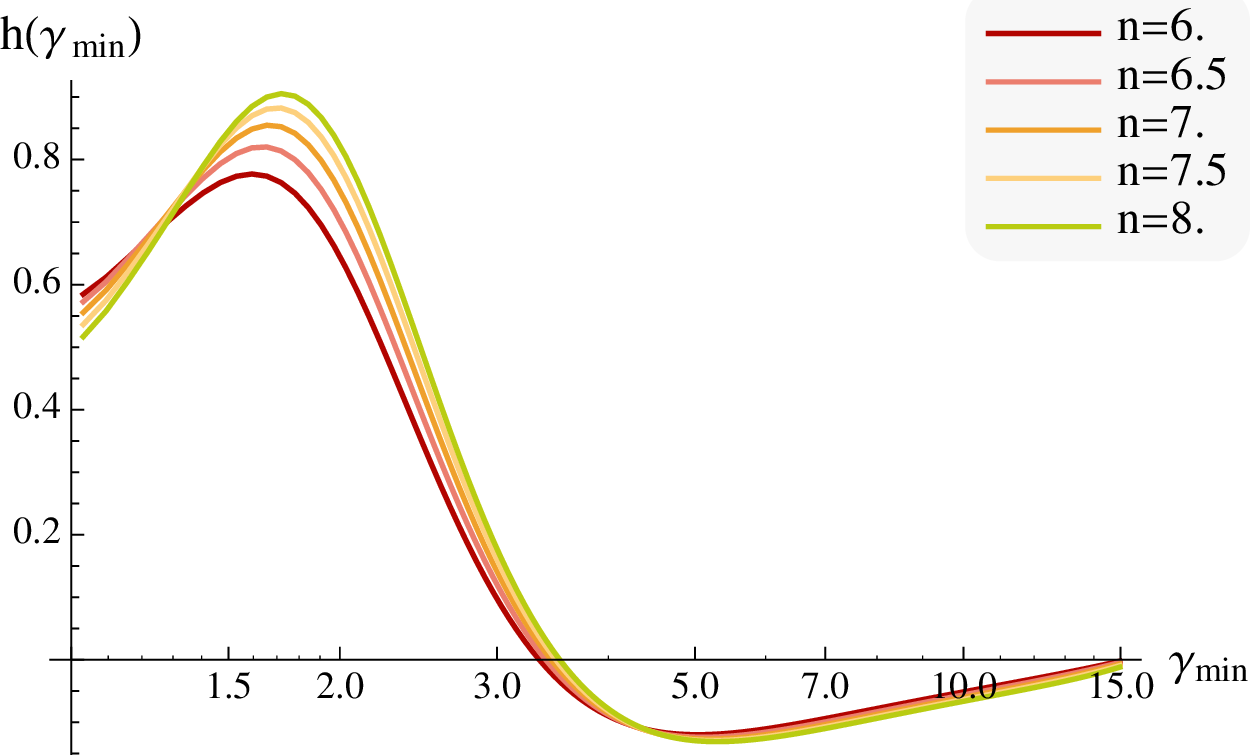}}
\caption{Polarization transfer coefficients ( left: $f^{\rm pl}_{\rm iso}$ and right: $h^{\rm pl}_{\rm iso}$)  
for an isotropic power-law distribution function as a function of the low energy cutoff $\gamma_{\rm min}$ for different values of the 
power-law index $n$, as labelled. The coefficient $f$ is normalized to 
$-\pi s_q \cos \alpha (\omega_{\rm pr}^2\!\mid\!\!\Omega\!\!\mid\!/c \omega^2)$
and $h$ to $\omega_{\rm pr}^2 \Omega^2\!/c\omega^3$. This figure has been produced by 
numerically integrating equations (\ref{fexacteanticip})--(\ref{hexactanticip}) over $\varpi$ and $\sigma$ 
for the distribution (\ref{definePL}). 
}
\label{fig:powerlaw}
\end{center}
\end{figure}

Fig. \ref{fig:powerlaw} shows the polarization transfer coefficients
for power-law distributions of various exponents
as a function of the low-energy cutoff $\gamma_{\rm min}$.
These coefficients have been 
numerically calculated from equations (\ref{fexacteanticip}) and (\ref{hexactanticip}) by integrating 
over $\varpi$ and $\sigma$ for the distribution (\ref{definePL}), using equation 
(\ref{relapisigmagammappar}) to express $\gamma$ in terms of $\sigma$ and $\varpi$.

Approximations to the one-variable kernels could have been used instead.
We derive in Section~\ref{kernelsinterpol} approximations to $F^{\rm iso}(\gamma)$ and $H^{\rm iso}(\gamma)$
obtained by interpolating between their
HF expressions and their asymptotic LF limits.
They provide results of reasonable, though limited, accuracy, but conveniently
reduce the calculation to a simple one-variable quadrature. Fig. \ref{figcomparPLexactinterpol}
in Appendix~\ref{AppinterpolKernels} compares 
the results in Fig. \ref{fig:powerlaw} with those obtained from these
approximations. Equations (\ref{defnoyauxisotropes}),(\ref{NormalisePL}) and (\ref{FisoLF}) 
indicate that $f^{\, \rm pl}$ scales for large $\gamma_{\rm m}$ as
$\ln(\gamma_{\rm m})/\gamma_{\rm m}^2$.

In a distribution that is extended in energy,
particles of Lorentz factor $\gamma$ less than the upper limit $\gamma_{\rm QR}$ 
defined by equation (\ref{regimechange-maintext})
contribute in the HF regime, as in equations (\ref{fisotfinal}) and (\ref{hisotfinal}), 
while particles of a larger Lorentz factor bring a 
QR contribution that should be added to their NR one.
For $\gamma\gg \gamma_{\rm QR}$, these two contributions add 
to form the asymptotic LF contribution 
to the coefficients in equations  (\ref{FisoLF}) and (\ref{hinlowfreqResu}).  
The QR contribution to the coefficient $f$ is small but does not vanish, being
the integral of an almost odd function over an interval
almost symmetrical with respect to zero (Appendix~\ref{Applowfrequencyapprox}). The isotropic 
LF kernel that results for $f$ has been derived in Section~\ref{seclowfrequencylimit} 
and is shown in equation (\ref{FLFdetail}).

\subsection{Polarization transfer coefficients for anisotropic distribution functions}
\label{SecAnisotrope}

The HF results in equations (\ref{ffinalordr1pisigma}) and (\ref{hfinalrecappisigma})
make no assumption about the isotropy, or otherwise, of the distribution function $F_0$.
But for the work of \citet{Melroseanis} on anisotropic thermal plasmas,
there does not seem to be any other result published so far in the literature on general
anisotropic distributions.
\citet{HuangCerbakov} give results and fits for a number of
isotropic distribution functions, including monoenergetic ones.
The form in which our results are obtained allows to calculate transfer coefficients
in the HF approximation not only for isotropic distribution functions but also for a large class
of anisotropic ones, providing the first terms of an expansion of the anisotropy in multipoles. 
Strong anisotropies still demand a two-variable quadrature, as exemplified by
the beam model below.
Stepping out of the HF limit in this case would provide one-variable quadratures that are no simpler than 
the two-variable ones from which they originate. We therefore restrict in this section to HF results 
for different anisotropic distribution functions. 

\subsubsection{Quadrupolar and higher multipolar anisotropies at high frequency}

Let us parametrize quadrupolar anisotropic distribution functions as the product of a function of the Lorentz factor $\gamma$ 
of the particles by a second-order Legendre function of the cosine of the particle's pitch angle 
$\vartheta$, defined such that $p_\parallel= p \cos \vartheta$: 
\begin{equation}
F(\gamma,\vartheta)=F_2(\gamma)P_2(\cos \vartheta). \label{eq:ansaztFaniso}
\end{equation}
For such distributions it is best to switch to the spherical coordinates $\gamma$ and $\vartheta$.
The $\sigma$- and $\varpi$-derivatives of the distribution function
in equations (\ref{fexacteanticip})--(\ref{hexactanticip}) or in equations
(\ref{ffinalordr1pisigma})--(\ref{hfinalrecappisigma})  always separate into the sum of two terms, 
one of which is proportional to ${\rm d} F_2/{\rm d} \gamma$ and the other is proportional to $F_2$. 
Generically, owing to the factorization property of the distribution function as in equation (\ref{eq:ansaztFaniso}),
the transfer coefficient could be written, for example in the case of the Faraday rotation coefficient $f$,  as 
\begin{equation}
f^{\rm aniso}_{\rm HF} = A_f \int_1^\infty \!\! \gamma \, d\gamma  \ \left(\frac{dF_2}{d\gamma} \ D'_{f}(\gamma,\mu) + F_2(\gamma)
D_{f}(\gamma,\mu)\right)\,,
\label{DetDprim}
\end{equation}
where $A_f$ is a constant factor, still dependent on $\omega$ and $\alpha$, 
and the functions, respectively, multiplying $F'_2(\gamma)$ and $F_2(\gamma)$,
$D'_{f}(\gamma,\mu)$ and $D_{f}(\gamma,\mu)$, depend on $\gamma$ and on the parameter $\mu = \cos \alpha$, 
$\alpha$ being the propagation angle of the radiation (Fig. \ref{axesJeanJerome}).
Similar quantities referring  to the coefficient $h$ can be defined. 
From equations~(\ref{ffinalordr1pisigma}), (\ref{hfinalrecappisigma}), using equations~(\ref{xgammapi}) 
and~(\ref{relapisigmagammappar}), it follows after integration over $\vartheta$ that,
for the quadrupolar model in equation (\ref{eq:ansaztFaniso})
\begin{eqnarray}
f^{\rm aniso}_{\rm HF}\!\! \!\! &=& \!\! \pi s_q \, \frac{\omega_{\rm pr}^2 \mid\!\!\Omega\!\!\mid}{c \, \omega^2} \, m^3\!c^3\!
\int\!\! \gamma  d\gamma
\frac{d{F_2}}{d\gamma } \left( \scriptstyle{
\frac{4  {\cal L}(\gamma) \left(  3 \left(2 \gamma ^2+3\right) P_3(\mu )
-\left(\gamma ^2-1\right) \mu\right)}{5 \left(\gamma ^2-1\right)}
+\frac{4 \left(\left(\gamma ^2-1\right) \mu -\left(11 \gamma ^2+4\right) P_3(\mu )\right)}{5 \gamma  \left(\gamma ^2-1\right)^{1/2}}}\right)
\nonumber \\
&&\hskip -0.cm+\, \pi s_q \, \frac{\omega_{\rm pr}^2 \mid\!\!\Omega\!\!\mid}{c \, \omega^2} \, m^3\!c^3\!
\int\!\! \gamma d\gamma{F_2}(\gamma ) \left(
\scriptstyle{
\frac{6 {\cal L}(\gamma) \left( 2 \left(\gamma ^4-1\right) \mu-    \left(2 \gamma ^4+15 \gamma ^2+\scriptscriptstyle{3}\right) P_3(\mu ) \right)}{5 \gamma \left(\gamma ^2-1\right)^2}+\frac{\left(34 \gamma ^2+86\right) P_3(\mu )-24 \left(\gamma ^2-1\right) \mu }{5  \left(\gamma ^2-1\right)^{3/2}}
}
\right),
\label{fanisotfinal}
\end{eqnarray}
and
\begin{eqnarray}
h^{\rm aniso}_{\rm HF}\!\! \!\!\!  &=&  \!\!\!\!\! \pi \frac{\omega_{\rm pr}^2 \Omega^2}{c \, \omega^3} \, \!m^3\!c^3\!\!\! 
\int\!\! \gamma   d\gamma
{F_2}(\gamma ) \! \left(\scriptstyle{
\frac{{\cal L}(\gamma) \left(
\left(24 \gamma ^2+5\right) P_2(\mu )-12 \left(2 \gamma ^2+1\right) P_4(\mu )+7\right)}{7 \left(\gamma ^2-1\right)^2}-\frac{7 \left(-6 \gamma ^4+17 \gamma ^2+4\right)+5 \left(18 \gamma ^4+65 \gamma ^2+4\right) P_2(\mu )-12 \left(4 \gamma ^4+37 \gamma ^2+4\right) P_4(\mu )}{105 \gamma  \left(\gamma ^2-1\right)^{3/2}}
}
\right)\nonumber \\ && \hskip -1.25cm  + \pi \frac{\omega_{\rm pr}^2 \Omega^2}{c \, \omega^3} \, m^3\!c^3\!\!\!
\int\!\! \gamma  d\gamma\frac{d{F_2}}{d\gamma } 
\left(\scriptstyle{
\frac{{\cal L}(\gamma) \left(14 \gamma ^2-108 \gamma ^2 P_4(\mu )+\left(94 \gamma ^2-7\right) P_2(\mu )+7\right)}{42 \gamma  \left(\gamma ^2-1\right)}+\frac{7 \left(4 \gamma ^4-8 \gamma ^2+19\right)+5 \left(-20 \gamma ^4+76 \gamma ^2+31\right) P_2(\mu )+36 \left(2 \gamma ^4-9 \gamma ^2-8\right) P_4(\mu )}{210 \left(\gamma ^2-1\right)^{1/2}}
}\right).
\label{hanisotfinal}
\end{eqnarray}
Equations~(\ref{fanisotfinal}) and~(\ref{hanisotfinal}) are analytical expressions, valid
in the HF limit, for the 
Faraday coefficients $f$ and $h$ for an anisotropic distribution 
function of the form~(\ref{eq:ansaztFaniso}). Appendix~\ref{app:Anisotrotropic} gives 
the corresponding coefficients $D_f'$ and $D_f$ for higher multipoles of the form $F_n(\gamma)P_n(\mu)$ for
$n=0,1\cdots 6$.  
Any axisymmetric distribution $F(\gamma, \vartheta)$ can be represented as a 
linear combination of such functions, as $F(\gamma,\vartheta)=\sum_n F_n(\gamma)P_n(\cos \vartheta)$. Since the conductivity depends linearly
on the unperturbed distribution function, the conductivity of a sum of functions is the sum of the corresponding conductivities.
In each individual term, 
neither the multipole 'distribution function' $F_n(\gamma)$ nor the angular factor $P_n(\cos \vartheta)$ needs to be positive. 
The only constraint is that the sum over $n$ be positive for any $\gamma$ and $\vartheta$.

As expected, the coefficients $f$ and $h$  depend on the direction of propagation $\alpha$ in a different manner
than those associated with an isotropic distribution function given in equations (\ref{fisotfinal})--(\ref{hisotfinal}).
For a quadrupolar distribution, 
$f$ varies with $\alpha$ as a combination of $P_1(\cos \alpha)$ and $P_3(\cos \alpha)$ while
$h$ varies as a combination of $P_2(\cos \alpha)$ and $P_4(\cos \alpha)$. 

\subsubsection{Beam model at high frequency}
\label{sec:beam}
\begin{figure}
\begin{center}
\resizebox{0.45 \hsize}{!}{\includegraphics{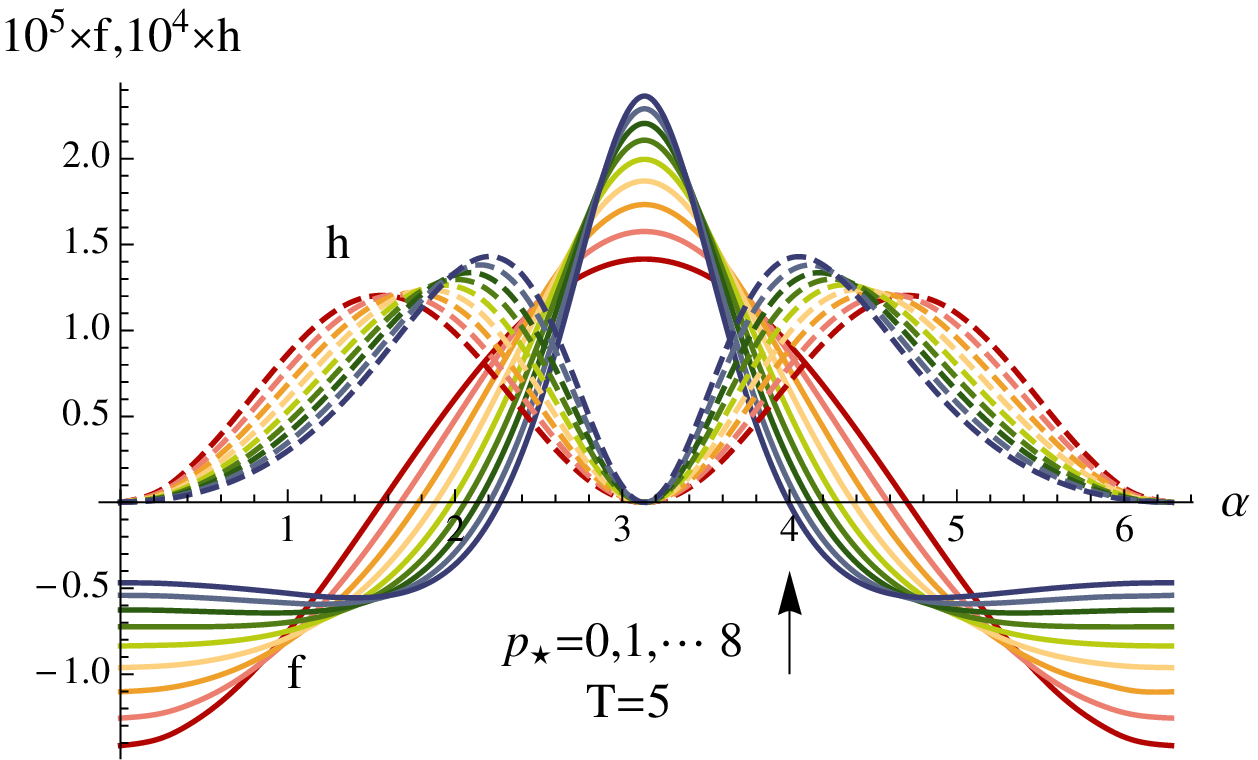}}
\resizebox{0.45 \hsize}{!}{\includegraphics{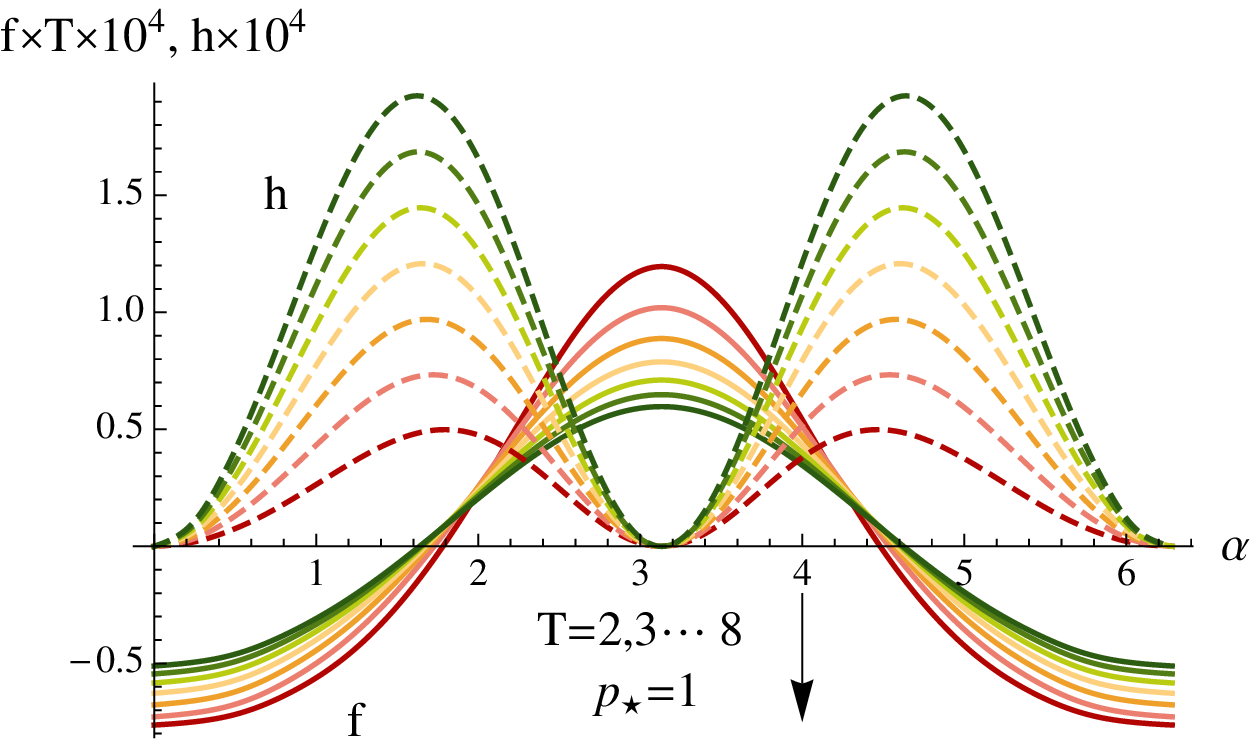}}
\caption{ Polarization transfer coefficients ($f,h$) versus the angle of propagation $\alpha$,
represented here on the interval $[0, \, 2 \pi]$ for the 'beam' model of Section~\ref{sec:beam} 
as a function of the beam momentum parameter $p_\star$ (left-hand panel) and the temperature $T$ (right-hand panel), as labelled. 
The vertical arrows indicate in which sense the parameter varies at $\alpha = \pi$ as $p_\star$ or $T$ vary as indicated.
The Faraday response becomes more collimated along the magnetic field direction 
as the distribution becomes more anisotropic or colder. 
}
\label{beam}
\end{center}
\end{figure}

Let us consider now an anisotropic distribution representing a beam. Since theory is constructed 
for particles, the unperturbed motion of which is ruled only by the static magnetic field
in the chosen frame of reference, 
the beam must be assumed to propagate parallel to the magnetic field. Otherwise, for example for winds propagating at an angle to
the magnetic field, a static convection electric field would also be present in the observer's frame
and the calculation in Section~\ref{Solformelle} would then have to be revisited or the Stokes parameters transformed
from the beam proper frame to the observer's frame.
We form the beam distribution $F_\star$ by boosting 
an isothermal distribution, so that
\begin{equation}
F_\star(\gamma,\theta)\propto \exp\left({-{\gamma_\star(p,\theta) }/{T}}\right)\,, \label{defFthermalboost}
\quad \mbox{where \ $\gamma_\star(p,\theta)$ \ obeys}\quad
c^2 m^2 \gamma_\star^2(p,\vartheta) = {{c^2 m^2+\left(p_\star+p \cos (\vartheta )\right){}^2+p^2 \sin ^2(\vartheta )}}\,.
\end{equation}
For such a distribution, the derivative of $F_\star$ with respect to $\sigma$ 
entering equations~(\ref{ffinalordr1pisigma}) and (\ref{hfinalrecappisigma}) is given by
\begin{equation}
 \frac{\partial F_\star}{\partial \sigma} =-F_\star \
\frac{\mid\! \Omega\!\mid }{T\, \omega \, \gamma _{\star }\sin^2 \alpha} \  
\left(\sqrt{c^2 m^2+p^2}+\cos (\alpha )\, p_{\star }\right)\,. \label{eq:dFdsigbeam}
\end{equation}
For the purpose of numerical integration,  
equations~(\ref{ffinalordr1pisigma}), (\ref{hfinalrecappisigma}) and~(\ref{eq:dFdsigbeam})  
are expressed as a function of $p_\perp$ and $p_\parallel$. The result of this integration, valid in the HF limit, 
is shown on Fig.~\ref{beam} 
for our beam model and for different values of $p_\star$ and $T$ as labelled, 
with the same fiducial parameters as above. As expected, for $p_\star=0$ we recover 
the thermal solution of Section~\ref{sec:thermal}, but as the 
beam becomes more anisotropic (i.e. $p_\star$ increases at fixed $T$, 
or $T$ decrases  at fixed  $p_\star\neq 0$), the Faraday coefficients are more focused near the axis of symmetry.
This may be understood from the fact that the values of the kernels depending on $\varpi$ and $\sigma$ in equations 
(\ref{ffinalordr1pisigma})--(\ref{hfinalrecappisigma}) are the largest where the difference $\sigma^2 - x^2$ is the smallest, 
that is where the pitch angle $\vartheta$ of the particle is closest to the direction of propagation $\alpha$ of the radiation,
causing a larger response when the angle $\alpha$ is in the particle's beam.

The next two sections formally carry out two of the four integrals involved in equation~(\ref{courantJphasegeneral})
in order to write equations~(\ref{fexacteanticip}) and (\ref{hexactanticip}), and investigate the corresponding  resonant and 
non-resonant regimes.  A detailed analysis of the various frequency regimes allows us to write simple one-dimensional quadrature. 
These two sections can be skipped on a first read.

\section{Derivation of the polarization transfer coefficients} \label{Secderivation}
%%%%%%%%%%%%%%%%%%%%%%%

We now describe our derivation of the polarization transfer coefficients,
which amounts to calculating the elements of the Hermitian and anti-Hermitian parts of the conductivity 
deduced from equation (\ref{courantJphasegeneral}), that is, of the phase integrals in equation (\ref{defintphase}).
The usual approach to the phase integrals is to 
integrate over the time delay $\tau$ by expanding 
$\exp(i x \sin \phi)$, or any similar expression, in discrete Fourier series such as for example
$ \exp(i x \sin \phi) = \sum_{n} J_n(x) \ \exp(i \, n\, \phi)$ where $J_n(x)$ is a 
Bessel function of relative integer order.
An integration over times $t'$ earlier than $t$ (that is over  positive delay times $\tau = t - t'$)
is then performed that yields resonant denominators $\omega - k_\parallel v_\parallel - n\Omega_*$.
The integrand being proportional to $\exp(+i \omega \tau)$, and
it being understood that $\omega$ really is a complex Laplace
variable, the resonant denominators should be regarded as having an infinitesimal positive imaginary part, ensuring
the convergence of integrals over $\tau$ at $\tau = +\infty$. 
These denominators 
are then to be understood as  complex numbers with an infinitesimal 
positive imaginary part, the real part of which is the Cauchy principal value, 
the imaginary part being $-i\pi \delta_{\rm D}(\omega - k_\parallel v_\parallel - n\Omega_*)$.
This gives an exact expression of the conductivity \citep{Bekefi} 
in the form of a series, each term of which involves one such
resonant denominator. 

However, since  any approximation to the sum of these series 
has, in the present context, to be carried out up to 
very large values of $n$, of order $\omega/\mid\! \Omega_*\!\mid$ at least, 
this representation of the conductivity
is not suitable for our purpose. As when deriving the synchrotron emission spectrum, it 
would be preferable to somehow substitute to the discrete series a representation in which
the discrete summation is replaced by an integration over a continuous variable. \citet{Sazonov69} attempted 
this by changing the discrete sums into integrals, considering the index $n$
of the Bessel functions as being a continuous variable which can be identified with the
variable $\sigma$ defined in equation~(\ref{defxsigma}). This approach is known to be successful when evaluating the 
dissipative part of the conductivity, 
in which the many resonant Dirac functions form a sum similar to a Riemann one
that can be approached by the integral of the so-defined interpolating function, $F(\sigma)$ say.
But this is an unfortunate approach for calculating the non-dissipative part, 
in which principal values are involved.
A continuous approximation to the derivative of $F(\sigma)$ would have been preferable in this case, 
although any a-priori choice of such an interpolation would be tainted with arbitrariness.

\subsection{An exact continuous-spectrum-type representation of the conductivity} 
\label{secPrinceton}

We  improve here over Sazonov's approach by deriving exact expressions for the phase integrals 
in equation~(\ref{defintphase}) as functions of the 
continuous variable $\sigma$ defined in equation~(\ref{defxsigma}). This variable will eventually appear as
a continuous index of some Bessel functions. This will achieve, without any arbitrariness, 
the desired continuous-spectrum-type representation. 
\citet{Qin} have derived this transformation by using the invariance associated to the periodicity of the 
unperturbed motion while \citet{theseJean} and \citet{theseJerome} 
derived it by exactly turning the familiar discrete series representation
into a continuous one by means of an integral representation of the principle values involved in the series.
To show here how this is achieved, we follow a method similar to that of \citet{Qin}. 
The phase integrals in equation~(\ref{defintphase}) 
are first expressed in terms of the variables in equation~(\ref{defxsigma}) 
and the delay time $\tau$ is replaced by a delay angle $\psi$, so that
\begin{equation}
P_\varepsilon(\sigma, x, \phi) =
\exp({ - i\, s_q  (\sigma \phi - x \sin \phi)})
\int_{s_q\phi}^{+\infty}   \frac{d\psi}{\mid\! \Omega_*\!\mid} \ \
\exp({i\, ( (\sigma + s_q \varepsilon) \, \psi 
-  x \, \sin \psi )})\,,
\qquad {\mathrm{with}}  \qquad \psi = s_q \phi \, + \, \mid \! \Omega_*\!\mid \, \tau
\,.\label{intdephasefactorsigmaeps}
\end{equation}
To account for the periodicity of the unperturbed motion, the integration range over the angle
$\psi$ is separated into segments of length
$2\pi$, and the integration on the $n^{\mathrm{th}}$ segment is carried out
by changing from $\psi$ to $w = \psi -(s_q \phi + 2 n \pi)$.
Then,
\begin{equation}
P_\varepsilon(\sigma, x, \phi)  = 
\frac{\exp({i\, (\varepsilon \phi + s_q x \sin \phi)}) }{\mid\! \Omega_*\!\mid}  
\left(\sum_{n = 0}^{\infty} e^{2 i\pi n\,(\sigma + s_q \varepsilon)}\right)
\int_0^{2\pi} \!\! dw \ 
\exp(i ((\sigma + s_q \varepsilon) w - x \, \sin (w + s_q \phi)))
\,.\label{Pepsavecretours}
\end{equation}
The fact that the periodicity of the motion is fully taken into account 
in this exact transformation could be considered superfluous
because, on the one hand, the period is very long 
in the ultrarelativistic limit and because, on the other hand,
the integral over $\psi$ in equation (\ref{intdephasefactorsigmaeps})
could conceivably be evaluated by the method of stationary phase applied to a unique interval of quasi-stationarity. 
It will however become clear
below that the anti-Hermitian  part of 
the conductivity is  in fact not determined only by the properties of the motion near the quasi-stationary phase, 
a situation that may be compared to that of a classical plasma, in which
the cyclotron resonances determine the rates of emission and absorption
but where NR particles make the most of the contribution to the medium's dispersive properties.
From equation~(\ref{Pepsavecretours}) we calculate the combinations of the phase integrals 
that appear in equation~(\ref{courantJphasegeneral}), namely
\begin{equation}
\left( \begin{array}{c}
P_0\\
(P_+ - P_-)/\, 2i\\
(P_+ + P_-)/\, 2 
\end{array} \right)
= \frac{e^{i s_q x \sin \phi} }{\mid \!\Omega_*\!\mid} \,  \left(\sum_{n=0}^\infty e^{2 i \pi n \sigma} \right)
\int_0^{2\pi} \!\! dw \, e^{i(\sigma w - x \sin(w + s_q \phi))} 
\left( \begin{array}{c}
1\\
\sin(\phi + s_q w)\\
\cos(\phi + s_q w)
\end{array} \right)
\,.\label{intphasesexplicit}
\end{equation}
The sum over $n$ of the imaginary exponentials is not absolutely convergent. It should be 
regarded as a distribution rather than as a proper function because it enters an integration over the variable $\sigma$ 
that will have to be performed at a later stage. 
We define three functions
$T_1$, $T_2$, $T_3$ by
\begin{equation}
T_1  = 1 \,,\qquad \qquad \qquad T_2 = \sin \,,\qquad \qquad \qquad T_3 = \cos   \,.
\label{lesfonctionsT}
\end{equation}
The expressions in equation~(\ref{intphasesexplicit}) are then inserted
in equation~(\ref{courantJphasegeneral}) which gives the components of the current.
Then gathering the terms wich depend on the gyration angle $\phi$ and integrating over it,
the components of the conductivity tensor can be written as 
\begin{equation}
\sigma_{ij} = -\frac{2\pi q^2}{\omega} 
\int_{-\infty}^{+ \infty} \!\!\! dp_\parallel \!\int_0^\infty \!\!\!p_\perp dp_\perp \ \, 
M_{ij}(p_\perp, p_\parallel, k_\perp, k_\parallel)
\,,\label{conductdesMij}
\end{equation}
where the matrix $M_{ij}$ is, with $i$ a line index and $j$ a column one:
\begin{equation}
\left( \begin{array}{ccc} \displaystyle
v_\perp (\omega \frac{\partial f_0}{\partial p_\perp} + k_\parallel D(f_0) ) \, Q_{33}(\sigma, x)& \displaystyle
v_\perp (\omega \frac{\partial f_0}{\partial p_\perp} + k_\parallel D(f_0) ) \, Q_{32}(\sigma, x)&\displaystyle
(v_\perp \omega \frac{\partial f_0}{\partial p_\parallel} \, Q_{31}(\sigma, x) 
- k_\perp v_\perp D(f_0) \, Q_{33}(\sigma, x)) \\
v_\perp (\omega \frac{\partial f_0}{\partial p_\perp} + k_\parallel D(f_0) ) \, Q_{23}(\sigma, x)&\displaystyle
v_\perp (\omega \frac{\partial f_0}{\partial p_\perp} + k_\parallel D(f_0) ) \, Q_{22}(\sigma, x)&\displaystyle
(v_\perp \omega \frac{\partial f_0}{\partial p_\parallel} \, Q_{21}(\sigma, x)
- k_\perp v_\perp D(f_0) \, Q_{23}(\sigma, x))\\
v_\parallel (\omega \frac{\partial f_0}{\partial p_\perp} + k_\parallel D(f_0) )\, Q_{13}(\sigma, x)&\displaystyle
v_\parallel (\omega \frac{\partial f_0}{\partial p_\perp} + k_\parallel D(f_0) )\, Q_{12}(\sigma, x)&\displaystyle
(v_\parallel \omega \frac{\partial f_0}{\partial p_\parallel}  \, Q_{11}(\sigma, x) 
- k_\perp v_\parallel  D(f_0) \, Q_{13}(\sigma, x))
\end{array} \right)
\,.\label{lesMIJ}
\end{equation}
The coefficients $Q_{ab}$ that appear in equation (\ref{lesMIJ}), $a$ and $b$ varying between $1$ and $3$, 
encompass the results of the integrations over
the delay angle and over the gyration angle.
They depend on the sign $s_q$ of the electric charge of the particle species and may
eventually be written as
\begin{equation}
Q_{ab}(\sigma, x) =  (-s_q)^{a+b} \, \left(2\pi\!\sum_{n=0}^\infty e^{2 i \pi n \sigma} \right) \! G_{ab}(\sigma, x) \,, \ \ {\mathrm{with}} \ \ \,
G_{ab} = \!
\int_0^{2\pi}\!\! \frac{d\phi}{2\pi} \!\! \int_0^{2\pi} \!\!\! \frac{dw}{2\pi}
\, T_a(\phi) T_b(\phi - w)
e^{i \big(\sigma w + x (\sin (\phi - w)-\sin \phi)\big)} 
\,.\label{Qabforme2}
\end{equation}
The parenthesis in the first part of equation (\ref{Qabforme2}) can be more explicitly written as
\begin{equation}
2\pi\!\sum_{n=0}^\infty e^{2 i \pi n \sigma} = i\pi \ \frac{e^{-i\pi \sigma} }{\sin \sigma \pi} \ \left(\lim_{N\rightarrow \infty} 
\left(1 - e^{2i\pi N \sigma} \right)\right)
\,.\label{Lasomme}
\end{equation}
The calculation of the elements $G_{ab}$ involves integrations over 
the  angles $\phi$ and $w$. 
\citet{Qin} perform these integrations  along the lines described in Appendix~\ref{LesGQin}.
The elements $G_{ab}$  may
finally be expressed in terms of Bessel functions of the first kind, with
indices $\sigma$ or $-\sigma$ and argument $x$, as listed below:
\begin{equation}
\begin{array}{ccc}
G_{11}(\sigma, x) = e^{i \sigma \pi}  J_{\sigma}(x) J_{-\sigma}(x) \qquad \hfill
&G_{12}(\sigma, x) =  - \frac{i}{2} \, e^{i \sigma \pi} \frac{\partial}{\partial x}
\!\left(J_{\sigma}(x) J_{-\sigma}(x)\right) \,,\hfill
& \\
& &\\
G_{13}(\sigma, x) = e^{i \sigma \pi} \left(- \frac{\sin \sigma \pi}{\pi x} 
+ \frac{\sigma}{x} \  J_{\sigma}(x) J_{-\sigma}(x)\right) \qquad \hfill 
&G_{22}(\sigma, x) = \frac{\sin \sigma \pi}{\pi}  \ e^{i \sigma \pi} 
\left(\frac{\pi}{\sin \sigma \pi} \ J'_{\sigma}(x) J'_{-\sigma}(x) + \frac{\sigma}{x^2} \right)\,,\hfill
&\\
& &\\
G_{23}(\sigma, x) = \frac{i \sigma}{2 x} \ e^{i \sigma \pi} \frac{\partial}{\partial x}
\left(J_{\sigma}(x) J_{-\sigma}(x)\right) \,,\qquad  \hfill
&G_{33}(\sigma, x) = - \frac{\sigma}{\pi x^2} \ \sin \sigma \pi \ \, e^{i \sigma \pi} 
\, \left(1 - \frac{\pi \sigma}{\sin \sigma \pi} \ J_{\sigma}(x) J_{-\sigma}(x)\right)\,, \hfill 
&\\
& &\\
G_{21} = - \,  G_{12} \,, \qquad \hfill  &G_{31} = + \, G_{13}\,,  \hfill & \hskip -3cm G_{32} = - \, G_{23} \,. \hfill
\end{array}
\label{lesGabBesselsigma}
\end{equation}
The elements $Q_{ab}$, and then the matrix elements $M_{ij}$,
are given in terms of these coefficients in equation~(\ref{Qabforme2}).
The variable $\sigma$ being positive, it is useful to eliminate the Bessel functions of negative indices
for other Bessel functions with positive indices. This is possible since the Bessel function of the first kind 
and negative index  $J_{-\sigma}(x)$  can be expressed in terms of Bessel functions of the first and second kind
with positive indices, $J_{\sigma}(x)$ and $N_{\sigma}(x)$ by the left identity in equation~(\ref{passesigmapositif})
below \citep{AbramStegun}.  When derivatives are involved, 
the second relation in equation~(\ref{passesigmapositif}), that gives the wronskian of the functions 
of the first and second kind \citep{AbramStegun}, may be used to eliminate $N'_{\sigma}$ if needed. We then base 
the transformation to positive $\sigma$ values on the relations:
\begin{equation}
J_{-\sigma}(x) = \cos \sigma \pi \, J_\sigma(x) - \sin \sigma \pi N_\sigma(x)\,,
\qquad \qquad \qquad \qquad \qquad \qquad  J_\sigma(x) N'_\sigma(x) - J'_\sigma(x) N_\sigma(x) = \frac{2}{\pi x}
\,.\label{passesigmapositif}
\end{equation}
Using equation (\ref{passesigmapositif}), the elements $Q_{ab}$ in equation~(\ref{Qabforme2})
can all be expressed in terms of $J_\sigma(x)$, $N_\sigma(x)$
and their derivatives with respect to $x$. Some components of these expressions turn out to be  regular at integer values of $\sigma$, while some other
keep a singular denominator $\sin \sigma \pi$ which generates expressions that must, again, 
be understood in the sense of distributions. The matrix elements $M_ {ij}$ in equations (\ref{conductdesMij}) and (\ref{lesMIJ}) 
can then be explicitly written as
\begin{equation}
\begin{array}{ccc} 
M_{XX} =\displaystyle  (+i) \
\lim_{N\rightarrow \infty} \left(1 - e^{2i\pi N\sigma}\right) 
\hfill & 
\displaystyle
\frac{v_\perp}{\mid \Omega_*\!\mid} \, 
(\omega \frac{\partial f_0}{\partial p_\perp} + k_\parallel D(f_0) )
\hfill 
& \displaystyle
\pi \,\frac{\sigma^2}{x^2}  \left( \, \frac{\cos \sigma \pi}{\sin \sigma \pi}
J^2_{\sigma}(x) - J_{\sigma}(x) N_{\sigma}(x) - \frac{1}{\sigma \pi} \right) \, ,
\hfill 
\\
M_{XY} =\displaystyle   
(-s_q)  \lim_{N\rightarrow \infty} \left(1 - e^{2i\pi N\sigma}\right) 
\hfill & \displaystyle
 \frac{v_\perp}{\mid \Omega_*\!\mid} \, 
(\omega \frac{\partial f_0}{\partial p_\perp} + k_\parallel D(f_0) )
\hfill & \displaystyle
\pi \, \frac{\sigma}{x} \left( \frac{\cos \sigma \pi}{\sin \sigma \pi} J_{\sigma}(x) J'_{\sigma}(x)
- J'_{\sigma}(x) N_{\sigma}(x)  - \frac{1}{\pi x} \right)\,,
\hfill 
\\
M_{XZ} =\displaystyle (+i) \
\lim_{N\rightarrow \infty} \left(1 - e^{2i\pi N\sigma}\right) 
\ 
\hfill & \displaystyle
\frac{v_\perp}{\mid \Omega_*\!\mid}  
(\omega \frac{\partial f_0}{\partial p_\parallel} 
- k_\perp D(f_0) \frac{\sigma}{x} )
\hfill
&\pi \,\frac{\sigma}{x} \left(\frac{\cos \sigma \pi}{\sin \sigma \pi} J^2_\sigma(x) - J_\sigma(x) N_\sigma(x) - \frac{1}{\sigma \pi} \right)\,,
\hfill 
\\
M_{YX} =\displaystyle  (+s_q) 
\lim_{N\rightarrow \infty} \left(1 - e^{2i\pi N\sigma}\right) 
\hfill & \displaystyle
\frac{v_\perp}{\mid \Omega_*\!\mid} \, 
(\omega \frac{\partial f_0}{\partial p_\perp} + k_\parallel D(f_0) )
\hfill & \displaystyle 
\pi \,\frac{\sigma}{x}  \left( \frac{\cos \sigma \pi}{\sin \sigma \pi} J_{\sigma}(x) J'_{\sigma}(x)
- J'_{\sigma}(x) N_{\sigma}(x)  - \frac{1}{\pi x} \right)\,,
\hfill 
\\
\end{array}
\end{equation}
\begin{equation}
\begin{array}{ccc}
M_{YY} =\displaystyle  (+i) \ \lim_{N\rightarrow \infty} \left(1 - e^{2i\pi N\sigma}\right) 
\hfill & \displaystyle
\frac{v_\perp}{\mid \Omega_*\!\mid} \,
(\omega \frac{\partial f_0}{\partial p_\perp} + k_\parallel D(f_0) )
\hfill & \displaystyle
\pi \left( \frac{\cos \sigma \pi}{\sin \sigma \pi} {J'}^2_\sigma(x) - J'_\sigma(x) N'_\sigma(x) + \frac{\sigma}{\pi x^2} \right)\,,
\hfill
\\
M_{YZ} =\displaystyle  (+s_q) 
 \lim_{N\rightarrow \infty} \left(1 - e^{2i\pi N\sigma}\right) 
\hfill & \displaystyle
\frac{v_\perp}{\mid \Omega_*\!\mid}  
( \omega \frac{\partial f_0}{\partial p_\parallel} \, 
- k_\perp D(f_0)\,
\frac{\sigma}{ x}) 
\hfill & \displaystyle
\pi \left( \frac{\cos \sigma \pi}{\sin \sigma \pi} J_{\sigma}(x) J'_{\sigma}(x)
- J'_{\sigma}(x) N_{\sigma}(x)  - \frac{1}{\pi x} \right)\,,
\hfill 
\\
M_{ZX} =\displaystyle  (+i) \ \lim_{N\rightarrow \infty} \left(1 - e^{2i\pi N\sigma}\right) 
\hfill & \displaystyle
\frac{v_\parallel}{\mid \Omega_*\!\mid} \, 
(\omega \frac{\partial f_0}{\partial p_\perp} + k_\parallel D(f_0) ) 
\hfill & \displaystyle
\pi \, \frac{\sigma}{x} 
\left( \frac{\cos \sigma \pi}{\sin \sigma \pi} J^2_{\sigma}(x) - J_{\sigma}(x) N_{\sigma}(x) 
-\frac{1}{\pi \sigma} \right)\,,
\hfill 
\\
M_{ZY} =\displaystyle  (-s_q) 
\lim_{N\rightarrow \infty} \left(1 - e^{2i\pi N\sigma}\right) 
\hfill & \displaystyle
\frac{v_\parallel}{\mid \Omega_*\!\mid} \, 
(\omega \frac{\partial f_0}{\partial p_\perp} + k_\parallel D(f_0) )
\hfill & \displaystyle
\pi \left( \frac{\cos \sigma \pi}{\sin \sigma \pi} J_{\sigma}(x) J'_{\sigma}(x)
- J'_{\sigma}(x) N_{\sigma}(x)  - \frac{1}{\pi x} \right)\,,
\hfill 
\\
M_{ZZ} =\displaystyle  (+i) \ \lim_{N\rightarrow \infty} \left(1 - e^{2i\pi N\sigma}\right) 
\hfill & \displaystyle  
\frac{v_\parallel}{\mid \Omega_* \! \mid} \quad
\big((\omega \frac{\partial f_0}{\partial p_\parallel}  - k_\perp D(f_0) \frac{\sigma}{x})
& \displaystyle
\pi \left(\frac{\cos \sigma \pi}{\sin \sigma \pi} J^2_{\sigma}(x) - J_{\sigma}(x) N_{\sigma}(x)\right)
\ + \ \frac{1}{x} \  k_\perp D(f_0) \, \big) \,.
\end{array}
\label{MzzJN}
\end{equation}
Let us now specify the meaning of $\lim \left(1 - e^{2i\pi N\sigma}\right)$, which 
enters with other factors in integrals over $\sigma$ implied by the integration over
momenta in equation~(\ref{conductdesMij}).
When this factor multiplies a regular function of $\sigma$, its limit is unity since, 
when $N$ diverges, $\exp(2i\pi N\sigma)$ oscillates infinitely rapidly leaving
in the limit a vanishing integral when multiplied by any regular function. The case when a factor $(\cos \sigma/\sin \sigma \pi)$ 
multiplies $\left(1 - \exp(2i\pi N\sigma)\right)$ deserves closer scrutiny, since both $\sin \sigma \pi$
and $(1 - \exp(2i\pi N\sigma))$ vanish at integer values.
It is shown in Appendix~\ref{Appdistrib} that the limit of their product is the distribution 
\begin{equation}
\lim_{N\rightarrow \infty} \left(1 - e^{2i\pi N\sigma}\right) \ \frac{\cos \sigma \pi}{\sin \sigma \pi}  \ \equiv 
\ {\cal{D}}\!\left( \frac{\cos \sigma \pi}{\sin \sigma \pi}\right) \ = \
\left( \, {\cal{P}}\!\left(\frac{\cos \sigma \pi}{\sin \sigma \pi}\right) - i \sum_{n \in {\mathbb{N}}} \delta_{\rm D}(\sigma - n)\, \right)
\,.\label{lafameusedistrib}
\end{equation}
where ${\mathbb{N}}$ is the set of positive integers, $\delta_{\rm D}$ a Dirac
distribution and the notation $\cal{P}$ 
means that the function which follows in the parenthesis should be taken as 
a Cauchy principal value near each of its singularities, 
which in this case are the integer values of $\sigma$. 
The effect of the multiple resonances at all integer
values of $\sigma$ that are involved in the matrix elements of the conductivity are concentrated on those terms 
in equations (\ref{MzzJN}) in which factors $(\cos \sigma \pi/\sin \sigma \pi)$ 
subsist, although it could have been expected that resonances should be present in all terms since equation~(\ref{Lasomme}) 
exhibits such a singular factor. 
In changing according to equation~(\ref{passesigmapositif}) for Bessel functions with positive indices only, some of these 
singular factors have been disposed of, having been regularized owing to the favourably phased 
term proportional to $\sin \sigma \pi\, N_\sigma(x)$ in equation (\ref{passesigmapositif}). This means
that the transformation to Bessel functions with only positive indices
has effected the summation over resonances for these favourably phased terms.
Resonances only remain explicitely
present in the unfavourably phased ones, that originate in the term  $\cos \sigma \pi \, J_\sigma(x)$ 
in equation~(\ref{passesigmapositif}).  
It is shown in Section~\ref{sectresonancesnegl} 
that these residual resonant terms eventually turn out to be negligible when $\omega/\!\mid \!\!\Omega\!\!\mid \gg 1$.

\subsection{Polarization transfer coefficients from the conductivity}
\label{sectcoefftransfert1}

\begin{figure}
\begin{center}
\resizebox{0.5 \hsize}{!}{\includegraphics{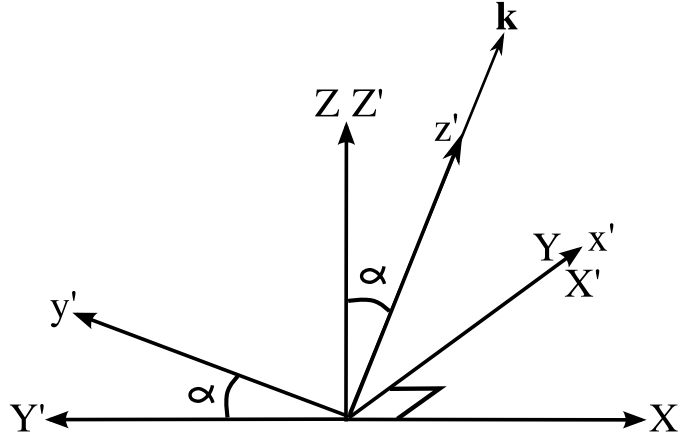}}
\caption{ The Z-axis is along the unperturbed magnetic field. The $z'$-axis is along the wavevector $\mathbf{k} =
k \cos \alpha \, {\mathbf{e}}_Z + k \sin \alpha \, {\mathbf{e}}_X$, where $0 \leq \alpha \leq \pi$. The axes Y and $x'$
are identical}
\,.\label{axesJeanJerome}
\end{center}
\end{figure}

The elements of the transfer matrix are given in equation~(\ref{coeffdesigmaHA})
in terms of the components of the conductivity tensor in the plane perpendicular 
to the direction of propagation of the radiation.
Since the components of this tensor are known in a frame in which the static magnetic field is along the $Z$ axis,
the components in equations (\ref{MzzJN}) must be transformed to the new reference frame 
$x'$, $y'$, $z'$ represented in Fig. \ref{axesJeanJerome} in which the wave vector 
$\mathbf{k}$ is along the $z'$ axis. For our purpose, it suffices to
calculate the transverse components $x'x'$, $x'y'$, $y'x'$, $y'y'$ 
which are given by \cite{theseJerome}
\begin{eqnarray}  
&&M_{x'x'} = M_{YY}
\,,\qquad \qquad \qquad \,\,\,\,\,
M_{x'y'} = \sin \alpha \, M_{YZ} - \cos \alpha \, M_{YX}
\,,\qquad \qquad \qquad \,\,
M_{y'x'} = \sin \alpha \, M_{ZY} - \cos \alpha \, M_{XY}
\,,\label{Mcomptrans}\\
&&M_{y'y'} =  \cos^2 \!\alpha \ M_{XX} + \sin^2\! \alpha M_{ZZ} - \sin\alpha \cos \alpha \, (M_{ZX} + M_{XZ} )
\,.\label{Mcomptransyprimyprim}
\end{eqnarray}
These calculations are straightforward, but it nevertheless takes some algebra to reduce
them to the simple form in equations (\ref{Mxprimxprimtext})--(\ref{Myprimyprimtext}).
Some details are given in Appendix~\ref{AppLesMiprimjprim}. 
The transverse components of the matrix {\bf{\sf{M}}} can eventually be written as
\begin{eqnarray}
&&M_{x'x'} = 
\frac{i \pi}{\gamma m} \, 
x^2 \ \frac{\partial f_0}{\partial \sigma}
\left( {\cal{D}}\left(\frac{\cos \sigma \pi}{\sin \sigma \pi}\right) {J'}^2_\sigma(x) 
- J'_\sigma(x) N'_\sigma(x) + \frac{\sigma}{\pi x^2} \right)
\,.\label{Mxprimxprimtext}\\
&&M_{x'y'} =  - \, \frac{s_q\pi}{\gamma m} \ \varpi \, x \
\frac{\partial f_0}{\partial \sigma}
\left( {\cal{D}}\left(\frac{\cos \sigma \pi}{\sin \sigma \pi}\right) J_{\sigma}(x) J'_{\sigma}(x)
- J'_{\sigma}(x) N_{\sigma}(x)  - \frac{1}{\pi x} \right)\, ,
\qquad \qquad M_{y'x'} = - M_{x'y'}
\,.\label{Mxprimyprimtext} \\
&&M_{y'y'} = \frac{i\pi}{\gamma m} \ \varpi^2 \ \frac{\partial f_0}{\partial\sigma} 
\Big( {\cal{D}}\left(\frac{\cos \sigma \pi}{\sin \sigma \pi}\right) J^2_{\sigma}(x) - J_{\sigma}(x) N_{\sigma}(x) \Big)
- \, \frac{i}{\gamma m} 
\left( \left(\sigma \cos \alpha - \varpi \right) \, \, \frac{\partial f_0}{\partial \varpi}
+ \varpi \, \cos \alpha \, \frac{\partial f_0}{\partial \sigma} \right)
\,.\label{Myprimyprimtext}
\end{eqnarray}
From equations (\ref{Mxprimxprimtext})--(\ref{Myprimyprimtext}), we may calculate the corresponding
components
of the conductivity. 
The density $n$ of the particles is factored out of the distribution function by writing it as
$f_0 = n F_0$, defining a reduced distribution function normalized to unity and a plasma frequency
$\omega_{\rm pr}$ for the relativistic species of particles considered: 
\begin{equation}
f_0 = n\, F_0 \,,\qquad \qquad \qquad \int_0^\infty \!\!\! 2\pi p_\perp dp_\perp \!\!\int_{-\infty}^{+\infty} \!\!\! dp_\parallel F_0 = 1
\,,\qquad \qquad \qquad  \omega_{\rm pr}^2 = \frac{n q^2}{\varepsilon_0 m}\,.
\label{definiGrandF0}
\end{equation}
The components of the anti-Hermitian and Hermitian parts of the conductivity can then be written as
\begin{eqnarray}
\frac{\sigma^A_{x'x'}}{\varepsilon_0}\!\!\! &=&  -2\pi^2 i \frac{\omega_{\rm pr}^2 \Omega^2}{\omega^3}
\int\!\!\!\!\int\!\! \frac{m^3c^3}{\sin^2 \alpha} \, d\varpi d\sigma \ x^2 \frac{\partial F_0}{\partial \sigma} 
\left({\cal{P}}\left(\frac{\cos \sigma \pi}{\sin \sigma \pi}\right) J^{'2}_\sigma(x) 
- J'_\sigma(x) N'_\sigma(x) +\frac{\sigma}{\pi x^2} \right)
\, .\label{sigmaAxprimxprim}\\
\frac{\sigma^A_{x'y'}}{\varepsilon_0}\!\!\! &=&  + \, 2\pi^2 s_q \frac{\omega_{\rm pr}^2 \Omega^2}{\omega^3}
\int\!\!\!\!\int\!\! \frac{m^3c^3}{\sin^2 \alpha} \, d\varpi d\sigma \ \varpi x \frac{\partial F_0}{\partial \sigma} 
\left({\cal{P}}\left(\frac{\cos \sigma \pi}{\sin \sigma \pi}\right) J_\sigma(x) J'_\sigma(x) 
- J'_\sigma(x) N_\sigma(x) - \frac{1}{\pi x}\right)\,,
\qquad \sigma^A_{y'x'} = - \sigma^A_{x'y'}
\,.\label{sigmaAxprimyprim}\\
\frac{\sigma^A_{y'y'}}{\varepsilon_0}\!\!\! &=& - 2\pi^2 i \frac{\omega_{\rm pr}^2 \Omega^2}{\omega^3} \!
\int\!\!\!\!\int \!\! \frac{m^3c^3}{\sin^2 \alpha} \, d\varpi d\sigma \!\left[
\varpi^2 \frac{\partial F_0}{\partial \sigma} 
\left(\!{\cal{P}}\left(\frac{\cos \sigma \pi}{\sin \sigma \pi}\right)\!\! J_\sigma^2(x) - J_\sigma N_\sigma(x)\right) 
\! - \! \frac{1}{\pi}\! \left(\! (\sigma \cos \alpha -\varpi) \frac{\partial F_0}{\partial \varpi}
+ \varpi \cos \alpha \frac{\partial F_0}{\partial \sigma}\right)\! \right]\! .
\label{sigmaAyprimyprim} \\
\frac{\sigma^H_{x'x'}}{\varepsilon_0}\!\!\! &=& - \,  2\pi^2 \frac{\omega_{\rm pr}^2 \Omega^2}{\omega^3} \sum_{n \in {\mathbb{N}}}
\int\!\!\!\int \frac{m^3c^3}{\sin^2 \alpha} \, d\varpi d\sigma  \ x^2 \frac{\partial F_0}{\partial \sigma} 
J^{'2}_\sigma(x) \, \delta_{\rm D}(\sigma -n)
\,.\label{sigmaHxprimxprim}\\
\frac{\sigma^H_{x'y'}}{\varepsilon_0}\!\!\! &=& - 2\pi^2 i s_q \frac{\omega_{\rm pr}^2 \Omega^2}{\omega^3}
\sum_{n \in {\mathbb{N}}}
\int\!\!\!\!\int\!\! \frac{m^3c^3}{\sin^2 \alpha} \, d\varpi d\sigma  \ \varpi x  \frac{\partial F_0}{\partial \sigma} 
J_\sigma(x) J'_\sigma(x) \, \delta_{\rm D}(\sigma -n)  \, ,  \qquad \qquad \qquad
\sigma^H_{y'x'} = - \sigma^H_{x'y'}
\,.\label{sigmaHxprimyprim}\\
\frac{\sigma^H_{y'y'}}{\varepsilon_0}\!\!\! &=&  - 2\pi^2 \frac{\omega_{\rm pr}^2 \Omega^2}{\omega^3}
\sum_{n \in {\mathbb{N}}}
\int\!\!\!\!\int\!\! \frac{m^3c^3}{\sin^2 \alpha} \, d\varpi d\sigma  \ \varpi^2 \frac{\partial F_0}{\partial \sigma} 
J^{2}_\sigma(x) \, \delta_{\rm D}(\sigma -n)
\,.\label{sigmaHyprimyprim}
\end{eqnarray}
The elements of the transfer matrix in equation~(\ref{Lequationdetransfert}) can be found
from these results by using equation~(\ref{coeffdesigmaHA}). Our choice of reference axes in the plane perpendicular to $\mathbf{k}$
results in the vanishing of the coefficients $K_{QV}$ and $K_{IU}$, since $\sigma^A_{x'y'}$ is real and
$\sigma^H_{x'y'}$ is imaginary. In equation~(\ref{sigmaAyprimyprim}) the integration over $\varpi$ and $\sigma$
of the terms
$\cos \alpha (\sigma \partial_\varpi F_0 + \varpi \partial_\sigma F_0)$ can be reduced to an integral on the
boundary of the physical domain which vanishes. 

\subsection{Formal expression of the polarization transfer coefficients}
\label{sectcoeffpremierjus}

The components of the Hermitian part of the conductivity 
in equations (\ref{sigmaHxprimxprim})--(\ref{sigmaHyprimyprim}) are
associated with dissipative radiative effects. Equation (\ref{coeffdesigmaHA})
relates them to the absorption coefficients per unit time 
which appear in equation (\ref{Lequationdetransfert}). The transfer 
coefficients per unit length in a medium of negligible dispersion are obtained from them 
by dividing by the velocity of light.
The discrete sum over the large positive integers $n$ in equations (\ref{sigmaHxprimxprim})--(\ref{sigmaHyprimyprim})
can be approximated by an integral, so that
\begin{eqnarray}
&&\frac{K_{II}}{c} 
= - \,  \pi^2 \, \frac{\omega_{\rm pr}^2 \Omega^2}{c\omega^3}
\int\!\!\!\int \frac{m^3c^3}{\sin^2 \alpha} \, d\varpi d\sigma  \ \frac{\partial F_0}{\partial \sigma} \, 
\left[x^2 \, J^{'2}_\sigma(x) + \varpi^2 \, J^{2}_\sigma(x)\right]
\,.\label{kIIcont}\\
&&\frac{K_{IQ}}{c}  
= - \,  \pi^2 \, \frac{\omega_{\rm pr}^2 \Omega^2}{c \omega^3} 
\int\!\!\!\int \frac{m^3c^3}{\sin^2 \alpha} \, d\varpi d\sigma  \ \frac{\partial F_0}{\partial \sigma} \, 
\left[x^2 \, J^{'2}_\sigma(x) - \varpi^2 \, J^{2}_\sigma(x)\right]
\,.\label{kIQcont}\\
&&\frac{K_{IV}}{c} 
= - 2\pi^2 s_q \frac{\omega_{\rm pr}^2 \Omega^2}{c \omega^3}  
\int\!\!\!\int \frac{m^3c^3}{\sin^2 \alpha} \, d\varpi d\sigma \ \frac{\partial F_0}{\partial \sigma} \,
\varpi x J_\sigma(x) J'_\sigma(x)
\,.\label{kIVcont}
\end{eqnarray}
We return to the dissipative coefficients in Section~\ref{sectcoefftransfertdissip}.
There are two non-dissipative coefficients, which in a stationary
medium are usually defined per unit length.
Assuming again no dispersion, they are
$f = K_{QU}/c$ and $h= K_{UV}/c$. When acting alone on radiation propagating in the direction
of the unit vecor $\mathbf{n}$, they cause the Stokes parameters $Q$, $U$, $V$ to vary according to the equation:
\begin{equation} 
{\mathbf{n}} \cdot {\boldsymbol{\nabla}}
\left( \begin{array}{c}
I\\
Q\\
U\\
V
\end{array} \right)
= 
-
\left( \begin{array}{cccc}
0&0&0&0\\
0&0&f&0\\
0&-f&0&h\\
0&0&-h&0
\end{array} \right)  
\left( \begin{array}{c}
I\\
Q\\
U\\
V
\end{array} \right)
\,.\label{Lequationdetransfertnondiss}
\end{equation}
This simplified transfer equation leaves the intensity $I$ invariant as well as the 
degree of polarisation $\sqrt{Q^2 +U^2+V^2}/I$.
The argument of all Bessel functions implicitely being $x$, 
the coefficients $f$ and $h$ can be written as
\begin{eqnarray}
f\!\!\! &=& 2\pi^2 s_q \ \frac{\omega_{\rm pr}^2 \Omega^2}{c\, \omega^3}
\int\!\!\!\!\int \!\! \frac{m^3c^3}{\sin^2\! \alpha} \, d\varpi \, d\sigma
\ \ \varpi \, x \ \frac{\partial F_0}{\partial \sigma} \
\left[{\cal{P}}\left(\frac{\cos \sigma \pi}{\sin \sigma \pi}\right)
J_\sigma J'_\sigma - J'_\sigma N_\sigma - \frac{1}{\pi x}\right]
\,,\label{fpremierjus}
\\
h\!\!\! &=& \pi^2 \frac{\omega_{\rm pr}^2 \Omega^2}{c\, \omega^3}
\int\!\!\!\!\int \!\! \frac{m^3c^3}{\sin^2 \alpha} \, d\varpi d\sigma
\left[\frac{\partial F_0}{\partial \sigma} \Big({\cal{P}}\left(\frac{\cos \sigma \pi}{\sin \sigma \pi}\right)\!\!
\left(\varpi^2 J_\sigma^2 - x^2 J^{'2}_\sigma\right) 
+ \left(x^2 J'_\sigma N'_\sigma - \varpi^2 J_\sigma N_\sigma\right)\Big)
+ \frac{1}{\pi} \left(\varpi \frac{\partial F_0}{\partial \varpi} - \sigma \frac{\partial F_0}{\partial \sigma}\right)
\right]
\!.\label{hpremierjus}
\end{eqnarray}

\subsection{The residual contribution of resonances to non-dissipative coefficients is negligible}
\label{sectresonancesnegl}

The principal value terms  in equations (\ref{fpremierjus}) and (\ref{hpremierjus}) are negligible in the
limit $\omega \gg \mid \!\Omega\! \mid$ in which $\sigma$ and $x$ are large. 
To prove this, we use the fact that the interval between successive zeroes of $\sin \sigma \pi$ 
is unity, whereas the Bessel 
functions vary on a much longer scale because both their index and argument are large, the
argument remaining smaller than their index, though. This can be seen from 
the definition of $\sigma$ and $x$ in equation (\ref{defxsigma}).
The contribution to the integral over $\sigma$ of 
each unit interval $[n -  \frac{1}{2} , n + \frac{1}{2}]$ can then be calculated at 
any accuracy by Taylor-expanding about $n$ the function which factors $\cot \sigma \pi$ in equations (\ref{fpremierjus})
and (\ref{hpremierjus}). This provides the  result of the integration on $[n - \frac{1}{2} , n + \frac{1}{2}]$
in the form of a series. 
The summation of these functions of $n$ over all unit intervals centred on integer values  
can then be approximately replaced, provided $\sigma$ is large,
by an integral over $\sigma$ since they slowly vary with $n$.
This integral turns out to be extremely small, owing to particular properties of the
functions $J_\sigma(x)$ and $J'_\sigma(x)$ for large $\sigma$ and small $x$. 
Details are to be found in Appendix~\ref{AppPPnegligeables}.
With this further simplification, the polarization transfer coefficients in equations (\ref{fpremierjus}) and
(\ref{hpremierjus}) lose their principal value terms. Terms in which no Bessel functions are involved can be 
reduced to a line integral on the boundary $\cal{B}$ of the 
physical domain in the $\varpi$--$\sigma$ plane (Appendix~\ref{Apprhosigma}), which is particularly 
useful to transform equation~(\ref{hpremierjus}) in which the term devoid of Bessel functions
is the divergence of the vector with components $v_\varpi = \varpi F_0$ and $v_\sigma = - \sigma F_0$.
All calculations done, we get
\begin{eqnarray}
&&f = -\, \frac{2\pi^2 s_q}{c} \frac{\omega_{\rm pr}^2 \Omega^2}{\omega^3} \!\!
\int\!\!\!\!\int \!\! \frac{m^3\!c^3}{\sin^2\!\! \alpha}  d\varpi \, d\sigma
\ \ \varpi \, x \ \left(J'_\sigma(x) N_\sigma(x) + \frac{1}{\pi x} \right) \frac{\partial F_0}{\partial \sigma} 
\,,\label{f2djus}
\\
&&h = \!\frac{\pi^2\! \omega_{\rm pr}^2 \Omega^2}{c \, \omega^3} \!\!\!
\int\!\!\!\!\int \!\! \frac{m^3\!c^3}{\sin^2\!\! \alpha} d\varpi d\sigma \!
\left(x^2 J'_\sigma(x) N'_\sigma(x) \!- \varpi^2 J_\sigma(x) N_\sigma(x)\right)\! \frac{\partial F_0}{\partial \sigma} 
\!+\! \frac{\pi \omega_{\rm pr}^2 \Omega^2}{c \, \omega^3} \!\!\!
\int_{-\infty}^{+\infty} \!\!\frac{m^3\!c^3}{\sin^2\!\! \alpha} \, d\varpi  \ \frac{2 \varpi^2 +\sigma_0^2}{\sqrt{\varpi^2 + \sigma_0^2}} 
F_0(\varpi, \sigma_b(\varpi))
,\label{h2djus}
\end{eqnarray}
The notation $\sigma_b(\varpi) = \sqrt{\sigma_0^2 + \varpi^2}$ denotes the value of $\sigma$ at a point 
of abcissa  $\varpi$  on the boundary $\cal{B}$ of the physical domain in the $\varpi$--$\sigma$ plane. 
It will soon be shown that, when expanding  the Bessel functions 
in $\mid \!\!\Omega\!\!\mid\!/\omega$, the non-Bessel terms in equations (\ref{f2djus}) and (\ref{h2djus})
almost exactly cancel the zeroth-order terms of the development. 
Equations (\ref{f2djus}) and (\ref{h2djus}) are applicable to any distribution function.
The integrals over $\sigma$ and/or $\varpi$ extend to the full range of physically 
relevant values and no other approximation
has been made than neglecting the series of residual principal value terms, which is well justified. 
Equations (\ref{f2djus}) and
(\ref{h2djus}) are then close to being exact. They however feature complicated kernels,
which makes it desirable to find simpler and suitable approximations to them.

\section{NR and QR parts of the transfer coefficients}
\label{sec:NR-QR}

\subsection{QR contribution to transfer coefficients from Nicholson's approximation }
\label{contribQR}

The phase integrals in equation~(\ref{defintphase}) involve the integration of trigonometric functions which may vary
more or less rapidly with the delay time $\tau$, depending on the 
values of the particle's parameters and the gyration angle $\phi$. 
Let us denote by $\vartheta$ the pich angle of a particle, that is, the angle of 
the particle's velocity with the magnetic field.
According to equation~(\ref{defintphase}), the characteristic variation frequency of the phase
most often is of the order of $\omega$.
When however the modulus of the particle's velocity is close to the speed of light, the characteristic frequency of 
phase variations may occasionally be much less, when the angle between 
the particle's velocity and the wave vector becomes small enough. For example, 
equation~(\ref{defintphase}) 
indicates that this frequency  would be of order $\mid \! \Omega\!\mid$ or less when
\begin{equation}
\omega - k_\parallel v_\parallel - k_\perp v_\perp \cos \phi \ \leq \ \mid\!\Omega\!\mid 
 \,,\qquad 
\mbox{that is}
 \qquad  
\left(1 - \frac{v}{c}\right) + \frac{v}{c} \left(1 - \cos (\vartheta - \alpha)\right) + 
\frac{v}{c}  \sin\alpha \sin \vartheta  \, (1 - \cos \phi) \ \leq \  \frac{\mid\!\Omega\!\mid}{c k}
\,.\label{ecartquasiresonancev1}
\end{equation}
Since $u = ck/\!\! \mid\!\Omega\!\mid$ is very large, this requires that each term on the left
of the second inequality in equation~(\ref{ecartquasiresonancev1})
be less than the term on the right, i.e. that 
\begin{equation}
\mid \!\! \alpha - \vartheta \!\!\mid \, \leq \, \left(\frac{\mid\!\!\Omega\!\!\mid}{\omega}\right)^{1/2}
\,,\qquad \qquad \qquad
\mid \!\! \phi \!\! \mid \, \leq \, \left(\frac{\mid\!\!\Omega\!\!\mid}{\omega}\right)^{1/2}   \,,\qquad \qquad \qquad
(1/\gamma) \, \leq \, \left(\frac{\mid\!\!\Omega\!\!\mid}{\omega}\right)^{1/2}
\,.\label{conditionsQR}
\end{equation}
For quasi-resonance, the third inequality in equation~(\ref{conditionsQR}) requires that
$\omega$ should be less 
than $\gamma^2 \! \mid\!\!\Omega\!\!\mid$ which is of order 
of the characteristic frequency of the synchrotron emission spectrum by 
a particle of Lorentz factor $\gamma$ (equation (\ref{defomegacritsynch})).
The other inequalities in equation~(\ref{conditionsQR}) require that the angle between the wave vector and the 
particle's velocity  be, for a frequency at the peak of synchrotron emission, less than $1/\gamma$.
These inequalities are only satisfied by a restricted class of particles, which
we refer to as quasi-resonant (QR) particles. For QR particles,
the characteristic frequency of the variations of the phase (the integrand in equation~(\ref{defintphase}))
is, during a small fraction of the synchrotron gyration period, 
much less than $\omega$ and, because of this slow variation with $\tau$,
the phase integral is exceptionnally large.  

This induced some authors 
\citep{Sazonov69, theseJean, MelroseJplPhys, MelrosePhysRev} 
to consider that 
QR particles entirely determine the non-dissipative transfer coefficients, just as
they determine the dissipative ones and the emission coefficients \citep{Westfold}. 
This shows up when phase integrals are evaluated by the method of the
(quasi-) stationary phase, resulting in the presence in the results of Airy-type functions.
The dominance of the contribution of QR particles to non-dissipative transfer coefficients
is however not granted. The functions which determine the dissipative coefficients
in equations (\ref{sigmaHxprimxprim})--(\ref{sigmaHyprimyprim}) happen to be
vanishingly small for NR particles, 
but those determining the non-dissipative ones have a much wider
support and are associated with a large number of NR particles. 
At this point, we do not know whether the QR contribution to 
the non-dissipative coefficients is negliglible compared to
the NR one, or comparable to it. This will be discussed in Section~\ref{comparQRNR}.

The first inequality in equation~(\ref{ecartquasiresonancev1}), which defines QR
particles, can be written for $\phi = 0$ in terms of the variables $\sigma$ and $x$ in equation~(\ref{defxsigma}) as
$(\sigma - x) \, < \, \gamma$.
The pitch angle $\vartheta$ of the particle and the propagation angle $\alpha$ of the wave 
being almost equal at quasi-resonance,
$\sigma \approx \gamma u \sin^2 \alpha$. Considering the general case when 
$\alpha$ is not very small,  this yields $\sigma \sim \gamma u$, a very large value.
Imposing the condition $(\sigma - x) \, < \, \gamma$ to a wave with 
frequency in the peak of the the synchrotron emission, such that 
$\omega \sim \gamma^2 \mid \! \Omega \! \mid$,
we get $ \sigma \approx \gamma^3$, which then implies that for quasi-resonance
\begin{equation}
\frac{\sigma - x}{\sigma} < \frac{1}{\sigma^{2/3}}
\,.\label{bornesigmamoinxsurx}
\end{equation}
Thus, for QR particles, $\sigma$ and $x$ are very close to each other, $x$ being 
however slightly smaller, since, from equation~(\ref{defxsigma}), 
$(\sigma -x)$ must be positive. The inequality 
in equation~(\ref{bornesigmamoinxsurx})  places the argument $x$
in the intermediate region where the so-called Nicholson's approximation to  Bessel functions of large index and argument
is appropriate (\citet{Watson} chapter 8, \citet{Olver52}).
Nicholson's approximation to 
$J_\sigma(x)$ is well known but the corresponding approximation to $N_\sigma(x)$ is not. 
Its derivation is outlined in Appendix~\ref{AppNicholson}.  The functions 
$J_\sigma$, $N_\sigma$ and their derivatives 
may be represented in this range by modified Bessel functions of the second kind $K_\nu(g)$, with index 
$\nu = 1/3$ or $2/3$, and by a similar combination $L_\nu(g)$ of modified Bessel functions
of the first kind, $I_{\pm \nu}(g)$.
$K_\nu$ and $L_\nu$ are defined by 
\begin{equation}
K_\nu(g) = \frac{\pi}{2} \, \frac{I_{-\nu}(g) - I_\nu(g)}{\sin \nu \pi}
\,,\qquad \qquad \qquad \qquad \qquad
L_\nu(g) = \frac{\pi}{2} \, \frac{I_{-\nu}(g) + I_\nu(g)}{\sin \nu \pi}
\,.\label{KLnu}
\end{equation}
The definition of the functions $K_\nu$ is standard \citep{AbramStegun}
and applies to integer values of the index in a limit sense.
Our definition of the functions $L_\nu$ in equation~(\ref{KLnu})
does not make sense for integer $\nu$. However, we only deal here with 
indices $ \nu = 1/3$ or $2/3$.
The argument $g$ on which these functions depend is 
\begin{equation}
g = \frac{2^{3/2} }{3} \ \, \frac{\left(\sigma -x\right)^{3/2} }{x^{1/2} }
\,.\label{gtexte}
\end{equation}
From equation~(\ref{bornesigmamoinxsurx}), it can be seen that the variable $g$ is ${\cal{O}}(1)$ or smaller
wherever the  Nicholson's approximation applies. The latter yields the following approximate representations:
\begin{eqnarray}
&&J_\sigma(x)  \approx    \frac{1}{\pi} \, \sqrt{{2}\over{3}} \ \sqrt{\frac{\sigma - x}{x}} \, K_{\frac{1}{3}}(g)
\,,\qquad \qquad \qquad 
N_\sigma(x) \approx  - \, \frac{\sqrt{2}}{\pi}  \ \ \sqrt{\frac{\sigma - x}{x}}  \, L_{\frac{1}{3}}(g)
\,.\label{NLtiers}
\\
&&J'_\sigma(x) \approx  \frac{2}{\pi \sqrt{3}}  \ \left(\frac{\sigma -x}{x}\right) \, K_{\frac{2}{3}}(g) 
\,,\qquad \qquad \qquad 
N'_\sigma(x) \approx  \ \ \,  \frac{2}{\pi} \ \  \left(\frac{\sigma -x}{x}\right) \, L_{\frac{2}{3}}(g)
\,.\label{NprimL2tiers}
\end{eqnarray}
The QR contributions  to the Faraday coefficients $f$ and $h$ are
obtained by substituting equations (\ref{NLtiers}) and (\ref{NprimL2tiers}) into
equations (\ref{f2djus}) and (\ref{h2djus}), the domain of integration in the
$\varpi$--$\sigma$ plane then being restricted to the QR one. 
\begin{figure}
\begin{center}
\resizebox{0.5 \hsize}{!}{\includegraphics{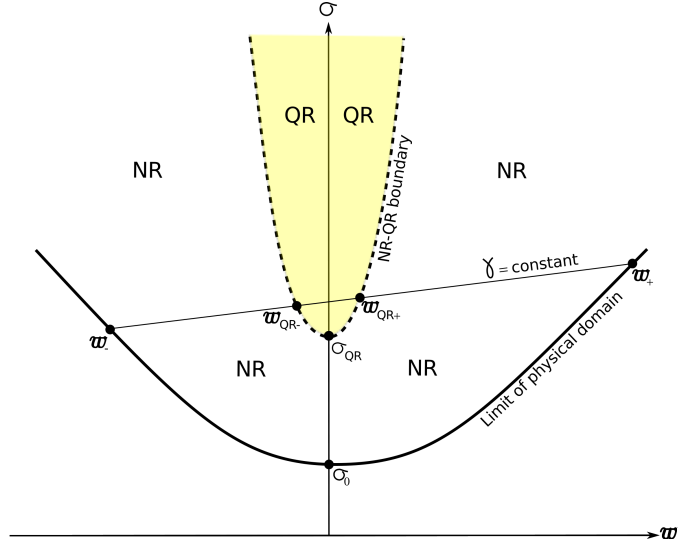}}
\caption{The physical domain of the $\varpi$-$\sigma$ plane is above the
hyperbolic line ${\cal{B}}$, of equation $\sigma^2 = \varpi^2 + \sigma_0^2$.
The smallest possible value of $\sigma$ is $\sigma_0 = \omega \sin \alpha/\! \mid \!\! \Omega\!\!\mid$. 
The QR domain is the yellow shaded area defined by $g \leq 1$ (equation \ref{gtexte}). 
It is bounded by the dashed line ${\cal{B}}_{\rm QR}$, represented by equation 
(\ref{condiQRplanpisigma}).
The smallest value, $\sigma_{{\rm QR}}$, of $\sigma$
on ${\cal{B}}_{\rm QR}$ is given by equation (\ref{condiQRplanpisigma}) for $\varpi = 0$.
The oblique line is the locus of a constant value of the Lorentz factor $\gamma$. 
It intersects ${\cal{B}}$ at $\varpi = \varpi_-(\gamma)$ and $\varpi_+(\gamma)$ and ${\cal{B}}_{\rm QR}$
at $\varpi = \varpi_{QR-}(\gamma)$ and $\varpi_{QR+}(\gamma)$.
}
\label{figBQR}
\end{center}
\end{figure}
This domain is characterized by the inequality in equation (\ref{bornesigmamoinxsurx}),
or equivalently, since at quasi-resonance $x$ and $\sigma$ are nearly equal, by the requirement that the variable $g$
in equation (\ref{gtexte}) be less than or equal to unity.
From equations (\ref{gtexte}) and (\ref{pperpderhoq}), the condition that $g = 1$ translates 
into 
\begin{equation}
\varpi^2 + \sigma_0^2 =  3^{2/3}   \sigma^{4/3}\,.
\label{condiQRplanpisigma}
\end{equation}
The variable $g$ can then be less than unity only when $\sigma$ exceeds a threshold $\sigma_{{\rm QR}}$ 
and the Lorentz factor exceeds a related one, $\gamma_{{\rm QR}}$, such that the line $\gamma = \gamma_{{\rm QR}}$
in Fig. \ref{figBQR} be tangent to the boundary ${\cal{B}}_{\rm QR}$. The Lorentz factor $\gamma_{{\rm QR}}$
is of order of the $\gamma$ variable associated with $\sigma_{{\rm QR}}$ and $\varpi = 0$. To sum up, 
\begin{equation}
\sigma_{{\rm QR}} = \frac{\sigma_0^{3/2}}{\sqrt{3}}\,, \qquad \qquad \qquad \gamma_{{\rm QR}} \approx  \sqrt{\frac{u}{3 \sin \alpha}}\,.
\label{sigmagammaqr}
\end{equation}
Thus, a particle can only quasi-resonantly interact with
the radiation when $\sigma$ exceeds $\sigma_{{\rm QR}}$.
Since $\sigma_0$ defined in equation (\ref{defxsigma})
is usually large, $\sigma_{{\rm QR}} \gg \sigma_{0}$.
The characteristic frequency $\omega_c(\gamma, \alpha)$
of synchrotron emission 
by particles of Lorentz factor $\gamma_{{\rm QR}}$, defined in equation (\ref{defomegacritsynch}),
is almost $\omega$.  
Fig. \ref{figBQR} represents the QR domain in the $\varpi$--$\sigma$ plane. When 
$u = \omega/\!\mid \!\!\Omega\!\!\mid$ becomes very large, $\sigma_{{\rm QR}}$ diverges as $u^{3/2}$ while the 
$\sigma$ value corresponding to a given Lorentz factor $\gamma$ diverges as $u$. This means that at high enough frequency
there will be a negligible number of particles in quasi-resonance
and the transfer coefficients will then
be essentially given by the NR contribution.
When on the contrary $\sigma > \sigma_{{\rm QR}}$ for relevant values of $\gamma$,
the QR region partly contributes to $f$ and $h$. These contributions are
\begin{eqnarray}
&&f_{\rm QR} = 2 s_q\, \frac{\omega_{\rm pr}^2 \Omega^2}{c\, \omega^3} \!\!
\int\!\!\!\!\int_{\rm QR}   \frac{m^3c^3}{\sin^2 \alpha}  d\varpi \, d\sigma \ \, 
\varpi \, x \, \left(
\frac{2\sqrt{2} }{\sqrt{3}} \,
\left(\frac{\sigma -x}{x}\right)^{\frac{3}{2}} \, K_{\frac{2}{3}}(g) L_{\frac{1}{3}}(g) - \frac{\pi}{x}\right) \ 
\frac{\partial F_0}{\partial \sigma}
\,,\label{fQR}
\\
&&h_{\rm QR} = \!\frac{\omega_{\rm pr}^2 \Omega^2}{c \, \omega^3} \!\!\!
\int\!\!\!\!\int_{\rm QR}   \frac{m^3c^3}{\sin^2 \alpha} d\varpi d\sigma \!
\left( \frac{4 x^2}{\sqrt{3}} \,
\left(\frac{\sigma -x}{x}\right)^2  K_{\frac{2}{3}}(g) L_{\frac{2}{3}}(g) 
+ \frac{2 \varpi^2}{\sqrt{3}} \, \left(\frac{\sigma -x}{x}\right) K_{\frac{1}{3}}(g) L_{\frac{1}{3}}(g)  
\right) \frac{\partial F_0}{\partial \sigma}
\,.\label{hQR}
\end{eqnarray}
where the subscript $\scriptstyle{\rm QR}$ indicates that the integration should be carried over the QR domain only.

\subsection{Polarization-dependent absorption coefficients}
\label{sectcoefftransfertdissip}

The dissipative coefficients per unit length are given by equations (\ref{kIIcont})--(\ref{kIVcont}).
Since the product of Bessel functions declines exponentially out of the QR domain, 
it is appropriate to use the Nicholson's approximation to represent them 
(equations (\ref{gtexte})--(\ref{NprimL2tiers})).
Equations (\ref{kIIcont})--(\ref{kIVcont}) are then expressed in terms of 
the Lorentz factor $\gamma$ and the pitch angle $\vartheta$ of the particles, or the difference $\psi$
of the latter and the propagation angle $\alpha$, which remains small in the QR domain. 
This results in
$\partial_\sigma F_0 \approx (\partial_\gamma F_0- \psi \, \partial_\vartheta F_0/\gamma)/u \sin^2 \alpha$.
Owing to the smallness of $\psi$, the angular derivative term may sometimes be neglected.
Similarly, the argument $g$ of the $K$ Bessel functions in 
equations (\ref{NLtiers}) and (\ref{NprimL2tiers}) may sometimes be approximated by 
$g_0\approx (\omega/3\gamma^2 \Omega \sin \alpha) \,
(1 + \gamma^2\psi^2)^{3/2}$. The critical frequency $\omega_c$ of synchrotron emission by a particle of 
Lorentz factor $\gamma$ in the direction $\alpha$ is defined by \citep{Westfold}
\begin{equation}
\omega_c = \frac{3}{2} \, \gamma^2 \mid \!\Omega\! \mid \sin \alpha\,.
\label{defomegacritsynch}
\end{equation}
The integration over the particles' directions of 
the right-hand sides of equations (\ref{kIIcont})--(\ref{kIVcont}) can be performed as in 
\citet{Westfold}, which gives for the total and linear polarization absorption coefficients:
\begin{eqnarray}
&&\frac{K_{II}}{c} 
= - \,  \frac{\omega_{\rm pr}^2\mid\!\Omega\!\mid }{c \, \omega^2} \frac{\sqrt{3}\, \pi}{2}  \, \sin \alpha 
\int m^3\!c^3 \, \gamma^2  d\gamma \,  \frac{\partial F_0}{\partial \gamma} \ 
\ \frac{\omega}{\omega_c} \! \int_{\frac{\omega}{\omega_c}}^\infty K_{5/3}(u)\, du\,,
\label{kIInous}\\
&&\frac{K_{IQ}}{c}  
= - \,  \frac{\omega_{\rm pr}^2 \mid\!\Omega\!\mid}{c \, \omega^2} \frac{\sqrt{3}\, \pi}{2}  \, \sin \alpha
\int m^3\!c^3\, \gamma^2  d\gamma \,  \ \frac{\partial F_0}{\partial \gamma} \ 
\ \frac{\omega}{\omega_c} K_{2/3}\! \left(\frac{\omega}{\omega_c}\right)\,.
\label{kIQnous}
\end{eqnarray}
Here, $\partial_\gamma F_0$ is meant to be taken at $\vartheta = \alpha$, the propagation angle.
The results (\ref{kIInous}) and (\ref{kIQnous}) coincide with those of \citet{SazonovSA}, 
considering his use of the CGS system of units,
the definition of his distribution function, and the presence of an unfortunate typo in
the first of his equations (2.2), where $\int_{\nu/\nu_c}^\infty K_{5/3}(u) \, du$ should be replaced by 
$\frac{\nu}{\nu_c} \int_{\nu/\nu_c}^\infty K_{5/3}(u) \, du$.

\medskip

The calculation of the absorption coefficient for circular polarization in equation (\ref{kIVcont})
is less straightforward because its 
dominant order contribution involves the integral of an odd function of $\varpi$, or $\psi$, which vanishes. 
Thus $K_{IV}/c$ generically is much smaller than the other two absorption coefficients.
Using Nicholson's approximation for $J_\sigma(x)$ and $J'_\sigma(x)$, equation (\ref{kIVcont}) becomes
\begin{equation}
\frac{K_{IV}}{c}
= - \frac{4 \sqrt{2}}{3}  \, s_q \, \frac{\omega_{pr}^2 \Omega^2}{c\, \omega^3} \int\!\!\!\int u^2 \, m^3c^3
\sin \vartheta \sqrt{\gamma^2 -1} \ d \gamma \, d\psi \ \frac{\partial F_0}{\partial \sigma} 
\ \varpi \, \frac{(\sigma - x)^{3/2} }{x^{1/2}} \ K_{1/3}(g)K_{2/3}(g)\,.
\label{KIVNichmixte}
\end{equation}
The integrand on the right-hand side of equation (\ref{KIVNichmixte})
should be expressed in terms of $\gamma$ and $\vartheta$
from equations (\ref{relapisigmagammappar}) and 
expanded to the first non-vanishing even order in $\psi$. 
This implies that the
angle derivative term in $\partial_\sigma F_0$ (equation (\ref{dersigmadegammatheta})), 
which is of order $\psi$, should be accounted for, that the derivative $\partial_\gamma F_0$
be Taylor expanded about $\vartheta =\alpha$ and that all other factors involved in equation (\ref{KIVNichmixte})
be similarly expanded. 
This applies in particular to 
the product of Bessel functions, owing to the fact
that the actual value of their argument $g$, given
by equation (\ref{gtexte}),  slightly differs
from its lowest order approximation in $\psi$.  With two terms, this argument $g$
may actually be expanded as
\begin{equation}
g = g_0 + g_1 = \frac{\gamma u}{3\, \sin \alpha} \left(\psi^2 + \frac{1}{\gamma^2}\right)^{3/2} 
- \, \frac{\gamma u \cos \alpha}{6\sin^2 \!\alpha}  \ \psi \left(\psi^2 + \frac{1}{\gamma^2}\right)^{3/2}\,.
\label{corrimpairetog}
\end{equation}
Integrating the correction to the product of Bessel functions
over $\psi$ by parts, these expansions lead to
\begin{eqnarray}
\frac{K_{IV}}{c}
&=& - \frac{2}{3}  s_q \frac{\omega_{\rm pr}^2 }{c  \mid\!\Omega\!\mid}  
\int\!\!\!\int \frac{m^3\!c^3}{\sin \alpha} \, \gamma^3 d\gamma \, d\psi \ 
\frac{\partial F_0}{\partial \gamma} \, \
 \frac{\cos \alpha}{\sin \alpha} \left(2 \psi^2 + \frac{2}{3 \gamma^2} \right) 
\left(\psi^2 + \frac{1}{\gamma^2} \right)^{3/2} \!\! K_{1/3}(g_0) K_{2/3}(g_0)\,,
\nonumber \\ 
&& - \frac{2}{3}  s_q \frac{\omega_{\rm pr}^2}{c \mid \!\Omega\! \mid}  
\int\!\!\!\int \frac{m^3\! c^3}{\sin \alpha}  \, \gamma^3 d\gamma \, d\psi \ 
\left(\frac{\partial^2 F_0}{\partial \gamma \partial \vartheta}
- \frac{1}{\gamma} \frac{\partial F_0}{\partial \vartheta} \right)
\, \psi^2  \left(\psi^2 + \frac{1}{\gamma^2}\right)^{3/2}\!\!   K_{1/3}(g_0) K_{2/3}(g_0)\,,
\label{KIV2}
\end{eqnarray}
where, again, all derivatives of $F_0$ are to be taken at $\vartheta = \alpha$. 
The integration over $\psi$ is carried out 
following a procedure similar to that described by \citet{Westfold}, which gives
\begin{eqnarray}
\frac{K_{IV}}{c} &=& 
- \frac{2 \pi}{\sqrt{3}}  s_q \frac{\omega_{\rm pr}^2 \mid \!\Omega \! \mid}{c \, \omega^2 } \, \cos \alpha \,
\int m^3\!c^3 \gamma d\gamma \ 
\frac{\partial F_0}{\partial \gamma} \, \left(\int_{\frac{\omega}{\omega_c}}^\infty \! \! 
K_{1/3}(u) \, du + 
\frac{\omega}{\omega_c}\! K_{1/3}\!\left(\frac{\omega}{\omega_c}\right) \right)  
\nonumber \\
&&\quad - \frac{\pi}{\sqrt{3}}  s_q \frac{\omega_{\rm pr}^2 \mid \! \Omega\! \mid}{c\, \omega^2 } \, \sin \alpha \,\int m^3\!c^3 \gamma d\gamma \ 
\left(\frac{\partial^2 F_0}{\partial \gamma \partial \vartheta}
- \frac{1}{\gamma} \frac{\partial F_0}{\partial \vartheta} \right)  
\int_{\frac{\omega}{\omega_c}}^\infty \! \! K_{1/3}(u) \, du\,.
\label{KIVnousfinal}
\end{eqnarray}
The result in equation (\ref{KIVnousfinal}) coincides 
with that given by \citet{SazonovSA}
in his equation (2.2).

\subsection{NR contribution to transfer coefficients from Debye's expansion}
\label{contribNR}

In the NR regime, the variable $g$ of equation (\ref{gtexte}) is larger than unity. The variables
$x$ and $\sigma$ both remain large but need not be almost equal. 
These conditions are suitable for using the Debye expansion 
of Bessel functions of large indices and argument \citep{Watson, Matviyenko}, 
subject to the condition that $(\sigma -x)\gg \sigma^{1/3}$ and,
in our case, that $x < \sigma$. 
The Debye expansion represents the Bessel functions of large index $\sigma$ for a given value of the 
argument-to-index ratio $x/\sigma$, represented by a parameter $\xi$ such that
\begin{equation}
\frac{x}{\sigma} = \frac{1}{\cosh \xi}\,.
\end{equation}
The Debye expansions of $J_\sigma(x)$ and $N_\sigma(x)$ can be written as \citep{Watson}
\begin{eqnarray}
&& J_\sigma(x) \equiv J_\sigma\left(\frac{\sigma}{\cosh \xi}\right) =
+ \ \  \frac{e^{\sigma (\tanh \xi - \xi)}  }{\sqrt{2\pi \sigma \tanh \xi}} \  \, 
\sum_{m=0}^{\infty} \frac{\Gamma(m+ 1/2) }{\Gamma(1/2)}  \ \ \frac{2^m  \ A_m(\xi) }{(\sigma \tanh \xi)^m}
\,,\label{DebyeJ} \\
&&N_\sigma(x) \equiv N_\sigma\left(\frac{\sigma}{\cosh \xi}\right) =
- \, \frac{\sqrt{2} \, e^{\sigma (\xi - \tanh \xi)} }{ \sqrt{2\pi \sigma \tanh \xi}}  \, 
\sum_{m=0}^{\infty} \frac{\Gamma(m+ 1/2) }{\Gamma(1/2)}  \ \ \frac{ (-1)^m \, 2^m  \ A_m(\xi) }{(\sigma \tanh \xi)^m}
\,.\label{DebyeN} \end{eqnarray}
where the coefficients $A_m$ depend on $\xi$, but for $A_0$ that equals unity. $\Gamma(y)$ denotes the 
gamma function of argument $y$.
The $n$-th term in the sum is of order $\sigma^{-n}$, that is of order $(\mid \!\!\Omega\!\!\mid\!/\omega)^n$.
Here we only need to proceed to second order.
It can be checked that at this order the Debye approximation continuously merges into Nicholson's approximation 
at their common limit of validity.
While $\sigma$ is regarded as a 
large parameter in the expansions in equations (\ref{DebyeJ}) and (\ref{DebyeN}), the ratio $x/\sigma$ 
should be considered of order unity.
Approximations to $J'_\sigma(x)$ and $N'_\sigma(x)$ 
are obtained by differentiating equations (\ref{DebyeJ}) and (\ref{DebyeN}) with respect to $x$ at fixed $\sigma$, taking care of the fact that 
the auxiliary variable $\xi$ depends on $x$, and thus also the coefficients $A_1$ and $A_2$. Some terms resulting from this derivation
contribute at this order. All calculations done, we obtain
\begin{eqnarray}
&& J_\sigma(x)N_\sigma(x) = \frac{(-1) }{\pi \sqrt{\sigma^2 - x^2} }
\, \left(1 + \frac{6A_2 - A_1^2}{\sigma^2 -x^2} \right)
\,,\label{JNDebye} \\
&&J'_\sigma(x)  N_\sigma(x) =  \frac{ (-1)}{\pi x} \left( 1 +  \frac{1}{2}  \, \frac{x^2}{(\sigma^2 -x^2)^{3/2} }
+ \frac{6 A_2 + x A'_1 - A_1^2}{\sigma^2 -x^2} + \frac{3}{2}  \, \frac{A_1 x^2}{(\sigma^2 -x^2)^2} \right)
\,,\label{JprimNDebye}
\\
&&J'_\sigma(x) N'_\sigma(x) = \frac{\sqrt{\sigma^2 - x^2} }{ \pi x^2} 
\, \left(1 + \frac{6A_2 + 2 xA'_1}{\sigma^2 -x^2} 
+ \frac{2 A_1 \, x^2}{(\sigma^2 -x^2)^2} - \frac{A_1^2}{\sigma^2 -x^2} - \frac{1}{4} \, \frac{x^4}{(\sigma^2 -x^2)^3}
\right)
\,.\label{JprimNprimDebye} 
\end{eqnarray}
where 
\begin{equation}
A_1 = \frac{1}{8} - \frac{5}{24} \, \frac{\sigma^2}{\sigma^2 - x^2}  \,,\qquad \qquad 
A_2 = \frac{3}{128} - \frac{77}{576} \, \frac{\sigma^2}{\sigma^2 - x^2}
+ \frac{385}{3456} \, \frac{\sigma^4}{(\sigma^2 - x^2)^2}\,,
\qquad \qquad 
x A'_1 = - \frac{5}{12} \, \frac{\sigma^2 x^2 }{(\sigma^2 - x^2)^2}
\,.\label{lesfonctionsA}
\end{equation}
The first terms 
in the parentheses of equations (\ref{JNDebye})--(\ref{JprimNprimDebye}), that are equal to unity,
remain in the unmagnetized limit, $\Omega \rightarrow 0$. The non-Bessel terms in equations (\ref{fpremierjus})--(\ref{hpremierjus})
are comparable to them.  
The NR contributions to equations (\ref{f2djus}) and (\ref{h2djus}) are obtained by 
integrating these expansions of the Bessel functions over the NR domain, 
denoted by the suffix $\scriptstyle{\rm NR}$:
\begin{eqnarray}
f_{\rm NR} &=& \frac{2\pi^2 s_q}{c} \frac{\omega_{\rm pr}^2 \Omega^2}{\omega^3} \!\!
\int\!\!\!\!\int_{\rm NR}  \frac{m^3\!c^3}{\sin^2\!\! \alpha}  d\varpi d\sigma
\ \, \frac{\varpi}{\pi} \,  \frac{\partial F_0}{\partial \sigma} \
\left( \frac{1}{2}  \, \frac{x^2}{(\sigma^2 -x^2)^{3/2} }
+ \frac{6 A_2 + x A'_1 - A_1^2}{\sigma^2 -x^2} + \frac{3}{2}  \, \frac{A_1 x^2}{(\sigma^2 -x^2)^2} \right)
\,.\label{fNR}
\\
h_{\rm NR} &=& \!\frac{\pi^2 \omega_{\rm pr}^2 \Omega^2}{c \, \omega^3} \!\!\!
\int\!\!\!\!\int_{\rm NR} \frac{m^3\! c^3}{\sin^2\!\! \alpha} d\varpi d\sigma \!
\ \, \frac{\partial F_0}{\partial \sigma} 
\ \frac{ 2 \sqrt{\sigma^2 - x^2} }{\pi} \, \left( 1 + \frac{6 A_2 - A_1^2 + xA'_1}{\sigma^2 -x^2}
+ \frac{A_1 x^2}{(\sigma^2 -x^2)^2} - \frac{1}{8}  \, \frac{x^4}{(\sigma^2 -x^2)^3} \right)
\nonumber \\
&-&  \!\!\!\! \frac{\pi^2 \omega_{\rm pr}^2 \Omega^2}{c \, \omega^3} \!\!\!
\int\!\!\!\!\int_{\rm NR} \frac{m^3\!c^3}{\sin^2\!\! \alpha} d\varpi d\sigma \ \,
\frac{\partial F_0}{\partial \sigma} \
\frac{\sigma_0^2}{\pi \sqrt{\sigma^2 - x^2} } \left(\!1 + \frac{6A_2 - A_1^2}{\sigma^2 -x^2}\! \right)
\!\! + \!\! \ \frac{\pi \omega_{\rm pr}^2 \Omega^2}{c \, \omega^3} \!\!\!
\int_{-\infty}^{+\infty} \!\!\frac{m^3\!c^3}{\sin^2\!\! \alpha} \, d\varpi \frac{2 \varpi^2 +\sigma_0^2}{\sqrt{\varpi^2 + \sigma_0^2}} 
\, F_0(\varpi, \sigma_b(\varpi)).
\label{hNR}
\end{eqnarray}
It is remarkable that the terms independent of magnetization eventually disappeared from equation~(\ref{fNR}). They also cancel out
in equation~(\ref{hNR}) as we now show. Gathering and arranging them, these terms can be written as
\begin{equation}
h_{\rm NR}^{(0)} = \!\frac{\pi \omega_{\rm pr}^2 \Omega^2}{c \, \omega^3} \!\!\!
\int\!\!\!\!\int_{\rm NR} \frac{m^3\!c^3}{\sin^2\!\! \alpha} d\varpi d\sigma \!
\ \, \frac{\partial F_0}{\partial \sigma} 
\frac{2 \varpi^2 + \sigma_0^2}{\sqrt{\varpi^2 +\sigma_0^2}} 
+ \ \frac{\pi \omega_{\rm pr}^2 \Omega^2}{c \, \omega^3} \!\!\!
\int_{-\infty}^{+\infty} \!\!\frac{m^3\!c^3}{\sin^2\!\! \alpha} \, d\varpi  \ \frac{2 \varpi^2 +\sigma_0^2}{\sqrt{\varpi^2 + \sigma_0^2}} 
\, F_0(\varpi, \sigma_b(\varpi))\,.
\end{equation}
The double integral term may be integrated explicitly over $\sigma$ at given $\varpi$ 
since only $\partial_{\sigma}F_0$ depends on this variable. The integration is on the values of $\sigma$ between its
value $\sigma_b(\varpi)$ on the boundary $\cal{B}$ of the physical domain and 
its value $\sigma_{\rm QR}(\varpi) = (\varpi^2 + \sigma_0^2)^{3/4}/\sqrt{3}$ 
on the boundary ${\cal{B}}_{\rm QR}$ between the NR and QR domains. 
The contribution from the lower boundary, $\sigma_b(\varpi)$, cancels the term that was already in the form
of a single integral over $\varpi$. The contribution from the upper boundary at $\sigma_{\rm QR}(\varpi)$ remains and
could as well be considered to be  part of the QR contribution since it can be written as 
an integral over the QR domain. Naming this contribution $h_{\rm BQR}$:
\begin{equation}
h_{\rm BQR} = \frac{\pi \omega_{\rm pr}^2 \Omega^2}{c \, \omega^3} \!\!\!
\int_{-\infty}^{+\infty} \!\!\frac{m^3\!c^3}{\sin^2\!\! \alpha} \, d\varpi  \ \frac{2 \varpi^2 +\sigma_0^2}{\sqrt{\varpi^2 + \sigma_0^2}} 
\, F_0(\varpi, \sigma_{bqr}(\varpi)) = -\ \frac{\pi \omega_{\rm pr}^2 \Omega^2}{c \, \omega^3}\!\! 
\int\!\!\!\! \int_{\rm QR} \frac{m^3\!c^3}{\sin^2\!\! \alpha} \, d\varpi d\sigma \ \frac{2 \varpi^2 +\sigma_0^2}{\sqrt{\varpi^2 + \sigma_0^2}} 
\, \frac{\partial F_0}{\partial \sigma} \,.
\label{hBQR}
\end{equation}
We refer to the remainder of the NR contribution to $h$ as the reduced NR contribution
${\hat{h}}_{\rm NR}$:
\begin{equation}
{\hat{h}}_{\rm NR}  = \pi \, \frac{\omega_{\rm pr}^2 \! \Omega^2}{c \, \omega^3} \!
\int\!\!\!\!\int_{\rm NR}  \frac{m^3\!c^3}{\sin^2\!\! \alpha} \, d\varpi  \, d\sigma 
\left(  2 \left(\frac{6 A_2 - A_1^2 + xA'_1}{(\sigma^2 -x^2)^{1/2}}
+ \frac{A_1 x^2}{(\sigma^2 -x^2)^{3/2}} - \frac{1}{8}  \, \frac{x^4}{(\sigma^2 -x^2)^{5/2}} \right)
- \sigma_0^2 \left(\frac{6A_2 - A_1^2}{(\sigma^2 -x^2)^{3/2}} \right)\! \right) \! \frac{\partial F_0}{\partial \sigma}
.\label{hNRreduit}
\end{equation}
The ${\cal{B}}_{\rm QR}$ contribution to $h$ in equation (\ref{hBQR}) is then associated with the QR contribution 
$h_{\rm QR}$ in equation (\ref{hQR}) to form a reduced QR contribution ${\hat{h}}_{\rm QR}$:
\begin{equation}
{\hat{h}}_{\rm QR} = \!\frac{\omega_{\rm pr}^2 \Omega^2}{c \, \omega^3} \!\!\!
\int\!\!\!\!\int_{\rm QR}   \frac{m^3\!c^3}{\sin^2\!\! \alpha} d\varpi d\sigma \!
\left( \frac{4 x^2}{\sqrt{3}} \,
\left(\frac{\sigma -x}{x}\right)^2  K_{\frac{2}{3}}(g) L_{\frac{2}{3}}(g) 
+ \frac{2 \varpi^2}{\sqrt{3}} \, \left(\frac{\sigma -x}{x}\right) K_{\frac{1}{3}}(g) L_{\frac{1}{3}}(g) 
- \pi \, \frac{2 \varpi^2 +\sigma_0^2}{\sqrt{\varpi^2 + \sigma_0^2}}
\right) \frac{\partial F_0}{\partial \sigma}
\,.\label{hQRreduit}
\end{equation}
The combinations of the functions $A_1$, $A_2$ and $xA'_1$ that 
appear in equations (\ref{fNR}) and (\ref{hNRreduit}) may be 
expressed in terms of $\sigma$ and $x$ by using equation~(\ref{lesfonctionsA}). 
When the integration over the $\rm NR$ domain can be extended to the full domain, 
this results in equations (\ref{ffinalordr1pisigma})--(\ref{hfinalrecappisigma}).

\subsection{NR versus QR contribution to the Faraday coefficients}
\label{comparQRNR}

The QR contributions to the Faraday coefficients 
are given by equations (\ref{fQR}) and (\ref{hQRreduit}) and the NR ones
by equations (\ref{fNR}) and (\ref{hNRreduit}). We now compare them,
making a simple ansatz concerning the distribution function, namely
that $F_0$ is a linear function of $\varpi$, so that
$\partial_\sigma F_0 = F_{0\sigma}(\sigma) + \varpi F_{0\sigma\varpi}(\sigma)$.
The $F_{0\sigma}$ term produces a nil contribution to $f$ if, as assumed here, it depends on $\sigma$ alone.
The $\varpi F_{0\sigma\varpi}$ term produces
a vanishing contribution to $h$ since its kernel is an even function of $\varpi$.
This ansatz 
is sufficient to give a hint on the dependence of the transfer coefficients on the distribution function.
The contributions $f_{\rm NR}$ and $f_{\rm QR}$ to $f$, or $h_{\rm NR}$ and $h_{\rm QR}$ to $h$, 
may then be written in a form
involving a kernel depending on $\sigma$, such that
\begin{equation}
f_{\rm R} = 2 s_q \frac{\omega_{\rm pr}^2 \Omega^2}{c\, \omega^3} \!\!
\int_{\rm R}   \frac{m^3\!c^3}{\sin^2\!\! \alpha}   d\sigma \ 
K^{(f)}_{\rm R}\! (\sigma) F_{0\sigma\varpi}(\sigma) \, ,
\qquad \qquad \qquad \qquad
h_{\rm R} = \frac{\omega_{\rm pr}^2 \Omega^2}{c\, \omega^3} \!\!
\int_{\rm R}   \frac{m^3\!c^3}{\sin^2\!\! \alpha}  \, d\sigma \ 
K^{(h)}_{\rm R}\! (\sigma)  F_{0\sigma}(\sigma) \, .
\label{formK}
\end{equation}
where 
${\rm R}$ stands either for the ${\rm QR}$ or ${\rm NR}$  
domain and $K^{(f)}_{\rm R}(\sigma)$ and $K^{(h)}_{\rm R}(\sigma)$ are corresponding one-variable kernels,
integrated over $\varpi$ at given $\sigma$. For Faraday rotation:
\begin{equation}
K^{(f)}_{\rm NR}(\sigma) = \int_{\rm NR} \!\! d\varpi \
\frac{\pi}{2} \, \frac{\varpi^2 x^2}{(\sigma^2 - x^2)^{3/2} } \, ,
\qquad \qquad \qquad 
K^{(f)}_{\rm QR}(\sigma) = \int_{\rm QR} \!\! d\varpi \
\frac{2^{3/2} }{3^{1/2}} \ \varpi^2 x \, \left(\frac{\sigma -x}{x} \right)^{3/2}  \,
\left(K_{2/3}(g) L_{1/3}(g) - \frac{\pi}{\sqrt{3} \, g}\right)\, .
\label{kernelKf}
\end{equation}
The QR kernel for $f$, on the right of equation (\ref{kernelKf}),
can be expressed in terms of the variable $g$  from
equation (\ref{gmin}) and then
be integrated over the QR domain $g_{\rm m}(\sigma) < g <1$. The
lower bound $g_{\rm m}$, defined in equation (\ref{gmin}), approaches zero
when $\sigma$ diverges. 
The integral on the right of equation (\ref{kernelKf}) 
converges as $g_{\rm m}$ approaches zero, which it does when $\sigma$ grows much larger than $\sigma_{{\rm QR}}$.
The asymptotic NR part of the kernel of $f$ can be calculated as shown in Appendix~\ref{AppcomparQRNR}. The 
results, valid for $\sigma \gg \sigma_{\rm QR}$, are:
\begin{equation}
K^{(f)}_{\rm NR}(\sigma) \approx \frac{\pi}{3} \ \sigma^2 \,  \ln\left(\frac{\sigma}{3}\right)\,,
\qquad \qquad \qquad \qquad 
K^{(f)}_{\rm QR}(\sigma) \approx 0.6\, \sigma^2 \,.
\label{kernelsf}
\end{equation}
The NR contribution to $f$ slightly dominates when  $\sigma \gg \sigma_{{\rm QR}}$.
These results partly depend on our assumption that
the distribution function depends linearly on $\varpi$. Otherwise, $f_{\rm NR}$ 
would also have a term proportional to $F_{0\sigma}$ 
that is absent from equation (\ref{formK}) by imparity. We return to this in Section~\ref{isotropicfLF}.

The $\varpi$-integrated kernel for $h$ being defined 
as in equation (\ref{formK}), 
its NR part results from equation (\ref{hNRreduit}) and is
calculated in Appendix~\ref{AppcomparQRNR}.
The QR part ${\hat{h}}_{\rm QR}$ is given by equation (\ref{hQRreduit}) and
its calculation is also outlined in this appendix.
The kernels $K^{(h)}_{\rm NR}(\sigma)$ and $K^{(h)}_{\rm QR}(\sigma)$ 
turn out to be approximately given, for $\sigma \gg \sigma_{{\rm QR}}$ by 
\begin{equation}
K^{(h)}_{\rm NR}(\sigma) \approx - \frac{\pi}{8} \, \left(\frac{\sigma}{3}\right)^{4/3}\, ,
\qquad \qquad \qquad 
K^{(h)}_{\rm QR} (\sigma) \approx + \frac{\pi}{2} \ \sigma^{4/3}\,.
\label{kernelsh}
\end{equation}
A glance at equation (\ref{kernelsh}) shows that both contributions to the kernel $K^{(h)}$ 
vary when $\sigma \gg \sigma_{{\rm QR}}$ as $\sigma^{4/3}$, 
the QR one being numerically larger.

The scaling in $\sigma^{4/3}$ anticipated for the $h$ kernel from equations 
(\ref{kernelsh}) does not conflict with the fact
that the Faraday conversion coefficient $h$ 
decreases to zero at high temperature for thermal distributions \citep{HuangCerbakov}. 
We have confirmed this  decline of $h^{\rm th}$
with temperature 
by numerically integrating equation (\ref{h2djus}) 
for a thermal distribution function and checked its compatibility wit a kernel 
in $\gamma^{4/3}$. 
We return to this is Section~\ref{subsubthermallowfreq}.
The Faraday conversion coefficient $h$ is largely affected by the QR contribution to it, which is of
an opposite sign to the NR one. 
Equations (\ref{fcoupure}) and (\ref{hcoupure}) propose a simple, though crude, way to account for 
the increasing importance of the QR contribution which
should be good enough 
wherever the balance passes from NR to QR domination.
For a given $\omega$, the QR domain becomes narrower in angle as $\gamma$ increases and
the QR kernel has an effectively compact support in this domain.
Therefore our calculation in equation (\ref{kernelsh})
of the QR contribution to $h$ should be close to being exact at large $\gamma$'s,
since $\partial F_0/\partial \sigma$ certainly is almost constant with $\varpi$ 
over the narrow QR domain, as was assumed in this section.

The condition for the QR domain not to contribute to the transfer coefficients 
is that the distribution function be negligible in it. 
From  equation (\ref{bornesigmamoinxsurx}), this happens when for most particles
$\sigma - \sqrt{\sigma^2 - \sigma_0^2 } >  \sigma^{1/3}$, or equivalently when the variable $g$
in equation (\ref{gtexte}) is larger than unity.
Expanding the expression of the latter condition in $\sigma_0/\sigma$ and noting that
$\sigma = \gamma u \sin^2\!\alpha$
at $\varpi = 0$, this inequality 
reduces again to the condition that equation~(\ref{regimechange-maintext}) be satisfied.
Up to a somewhat arbitrary factor, the parameter on the left of equation (\ref{regimechange-maintext})
is similar to the square of the regime-change parameters $X$ and $\gamma_0 X_A$, respectively, *
defined by \citet{Cerbakov2008}  and  \citet{HuangCerbakov}.
When the inequality in equation (\ref{regimechange-maintext})
is not satisfied, the coefficients $f$ and $h$ differ from their HF approximations 
in equations (\ref{ffinalpisigma}) and (\ref{hfinalpisigma}) below, where the integration is meant
to extend over the full $\varpi$--$\sigma$ domain. The intrusion of QR contributions 
is at the origin of the change in the trend of the variations of $f$ and $h$ with temperature
described by \citet{Cerbakov2008} for thermal distributions. 
This author attributes this change to a failure of the HF expansion, which is correct in the sense 
that the QR contribution cannot be expressed in the form of a series expansion,
no partial sum of the Debye series being able to represent the Bessel function in 
the domain of validity of the Nicholson's approximation.

\section{LF kernels of Faraday coefficients of isotropic distributions}
\label{seclowfrequencylimit}

\subsection{Remarks on the LF  limit of isotropic kernels for the Faraday coefficients}

In the limit of large $\gamma$'s, the QR domain contributes little to 
the Faraday rotation coefficient $f$, given by equation (\ref{f2djus}),
because it involves the integral of an odd function 
of the difference $\psi =\vartheta - \alpha$ on the small QR domain. This remark holds also for
the integral over this same QR domain of the NR contribution.
For an isotropic distribution, there is however a non-vanishing NR contribution to $f$
that is proportional to $\partial_\gamma F_0$.
Such a contribution is absent for distributions, as considered in Section~\ref{comparQRNR}, that are linear in $\varpi$.

By contrast, the QR domain, if substantially populated with particles, 
largely contributes the Faraday conversion coefficient $h$.
Particles in the QR domain satisfy the inequality inverse to that in equation 
(\ref{regimechange-maintext}) in
a strong sense.
We refer to the case when this strong inequality is satisfied 
as the LF limit.
Since any isotropic distribution function depends on $\gamma$ only,
the Faraday coefficients for such distributions can be reduced to a single
quadrature over the Lorentz factor involving a $\gamma$-dependent kernel as defined in equations
(\ref{defnoyauxisotropes}).
The exact expression of the isotropic kernel for $f$ results from equation (\ref{f2djus}) 
in which the variables $\varpi$--$\sigma$ should be changed to
$\varpi$--$\gamma$, giving on integrating over $\varpi$ at a given $\gamma$
\begin{equation} 
F^{\rm iso}(\gamma) = - \frac{2\pi^2s_q}{\sin^2\!\!\alpha} \int_{\varpi_-}^{\varpi_+} \!\! d\varpi
\ \varpi \, x \left(J'_\sigma(x) N_\sigma(x) + \frac{1}{\pi x} \right)\,.
\label{Fisogeneral}
\end{equation}
The exact isotropic kernel for $h$ is similarly obtained from equation (\ref{hpremierjus}), 
in which the principal value term should be neglected.
Using equations (\ref{relapisigmagammappar})--(\ref{dervarpidegammatheta}),
the isotropic kernel for $h$ can be written as:
\begin{equation} 
H^{\rm iso}(\gamma) = \frac{\pi^2}{\sin^2\!\!\alpha} \int_{\varpi_-}^{\varpi_+} \!\! d\varpi
\left(x^2 J'_\sigma(x) N'_\sigma(x) \!- \varpi^2 J_\sigma(x) N_\sigma(x) - \frac{1}{\pi}
\left(\gamma u \sin^2\!\!\alpha + 2\varpi \cos \alpha\right)\right)\,.
\label{Hisogeneral}
\end{equation}
In equations (\ref{Fisogeneral}) and (\ref{Hisogeneral}), the bounds $\varpi_\pm(\gamma)$ 
are defined by equation (\ref{varpiplusoumoins}). 
The kernel $H^{\rm iso}(\gamma)$ could have been deduced as well from
equation (\ref{h2djus}), reformulating the line-integral term in it as a surface integral.
These different expressions are equivalent because the non-Bessel integrals in them,
though apparently different, are actually equal, given the particular values of the bounds.

\subsection{Thermal Faraday conversion coefficient in the LF limit}
\label{subsubthermallowfreq}

\begin{figure}
\begin{center}
\resizebox{0.5 \hsize}{!}{\includegraphics{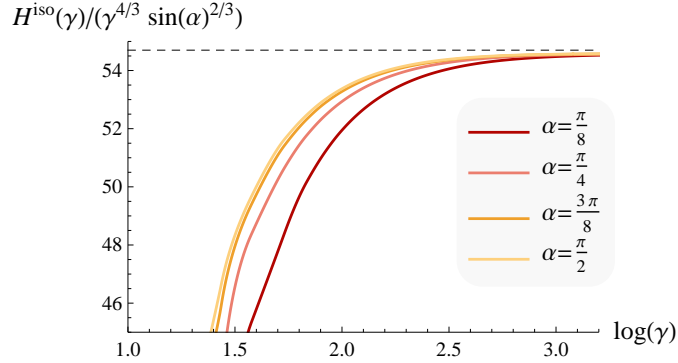}}
\caption{
The evolution of $H^{\rm iso}(\gamma)/(\gamma^{4/3} \sin^{2/3}\! \alpha)$ [computed via
equation~(\ref{Hisogeneral}) using the Olver expansion] as a function of $\gamma$ for different values of $\alpha$.
The radiation's parameter $u = 15$.
For all values of $\alpha$, this ratio asymptotes to a constant for large $\gamma$. The dashed line corresponds to the
asymptotic prediction of equation~(\ref{kernelisotroplowfreq}).
 }
\label{figKgammaOverGamma4third}
\end{center}
\end{figure}

To check the behaviour of the Faraday conversion coefficient $h$ in the LF limit,
we have numerically integrated its quasi-exact expression in equation (\ref{Hisogeneral})
with {\tt Mathematica}  for a thermal distribution function. At a given frequency, the LF limit would then correspond
to the limit of high temperatures. \citet{HuangCerbakov} have found that in this limit
the Faraday conversion coefficient decreases with temperature. We confirm this and find that this decrease
scales as $-T^{-5/3}$
by extending our numerical calculations to high values
of the temperature $T$ (see Fig.~\ref{figisothermalhasymptotes}).
This has been possible by using the uniform expansion for high order Bessel functions derived
by \citet{Olver54}  which
allows a fast and reliable calculation of high order Bessel functions (see Appendix~\ref{app:olver}).
Returning the result of the integration over the directions of the particles before performing
that over $\gamma$
reveals the high energy properties of the isotropic kernels derived in Sections
\ref{isotropicfLF} and \ref{kernelHlowfreqanalytiq}.
This numerically supports the analytical result, reached in Section~\ref{kernelHlowfreqanalytiq}
and shown in Fig.~\ref{figKgammaOverGamma4third}, that the kernel of $h$ asymptotically grows with the Lorentz factor
as $\gamma^{4/3}$
and shows that this increase with energy of
the kernel is consistent with the fast decline with temperature of the thermal Faraday conversion coefficient.
Note finally that, considering equations~(\ref{defFthermal}) and~(\ref{defnoyauxisotropes}),
equation~(\ref{kernelisotroplowfreq}) predicts at large $\gamma$
the result of the numerical integration and the $T^{-5/3}$ asymptote for the thermal distribution.

\subsection{Isotropic kernel of the Faraday rotation coefficient in the LF limit}
\label{isotropicfLF}

The integral in equation (\ref{Fisogeneral}) may be separated into NR  and  QR parts as follows.
The Faraday rotation kernel $F^{\rm iso}(\gamma)$ is the sum of the expression in 
equation (\ref{fNR}), integrated over the NR domain and limited to its lowest order term in $\Omega/\omega$, and of 
the QR contribution from equation (\ref{fQR}).
The integral over the NR domain in equation (\ref{fNR}) can be extended to the full domain at the expense of adding, 
if needed, to the contribution from equation (\ref{fQR})
a correction to account
for the undue integration over the QR domain so introduced.
After the necessary change of variables, and considering the
result in equation (\ref{fisotfinal}), the kernel $F^{\rm iso}(\gamma)$ assumes the, still exact, form
\begin{eqnarray} 
F^{\rm iso}(\gamma) &=&   4 \pi s_q \frac{\omega \cos \alpha}{\mid \!\!\Omega\!\!\mid} 
\left(\gamma\, {\cal{L}}(\gamma)
-  \sqrt{\gamma^2 -1} \right)
- \pi s_q \frac{\Theta_H(\gamma -\gamma_{{\rm QR}})}{\sin^2 \alpha}  
 \int_{\rm QR} \!\! d\varpi \ \frac{\varpi \, x^2}{(\sigma^2 -x^2)^{3/2}}
\nonumber \\
&&+ 2 s_q \frac{\Theta_H(\gamma -\gamma_{{\rm QR}})}{\sin^2 \alpha}  
\int_{\rm QR} \!\! d\varpi \  \varpi\left(\sqrt{3} g K_{2/3}(g) L_{1/3}(g) - \pi\right)\,.
\label{Fisoarragement}
\end{eqnarray} 
The first term in the first line of equation (\ref{Fisoarragement})
is the NR isotropic kernel, calculated by integration over the full physical domain.
The second term in the first line is an integral at given $\gamma$ 
over the QR domain correcting for the fact that the
NR domain really does not extend over the full physical one.
The second line is the QR contribution proper.
The variables $\sigma$, $x$ and $g$ are functions of $\gamma$ and $\varpi$ 
that may be found from  Appendix~\ref{Apprhosigma} and equation (\ref{gtexte}). The QR integration over $\varpi$
is between $\varpi_{QR-}$ and $\varpi_{QR+}$ defined in the caption of Fig.~\ref{figBQR}.

As $\gamma$ approaches infinity, the interval $[\varpi_{QR-},\, \varpi_{QR+}]$ becomes more
and more symmetrical about zero while $g(+\varpi, \gamma)$ and $g(-\varpi, \gamma)$ converge to each other.
The second and third terms in equation (\ref{Fisoarragement}) then approach the integral of an odd function
over an interval symmetrical with respect to $\varpi = 0$, which then vanishes. 
As for the circular polarization absorption coefficient $K_{IV}$ calculated in         
Section~\ref{sectcoefftransfertdissip}, higher order terms determine the QR contribution.
Since only isotropic distributions are considered
in this section, corrections caused by the anisotropy of the distribution function are absent. 
However, the interval $[\varpi_{QR-},\, \varpi_{QR+}]$ 
is slightly asymmetrical with respect to $\varpi = 0$, which gives rise to a first order 'offset' correction. 
Moreover, $\sigma(+\varpi, \gamma)$ and $\sigma(-\varpi, \gamma)$ slightly differ when $\varpi \ll \gamma u \sin^2\!\alpha$,
as can be seen from equation (\ref{relapisigmagammappar}),
and then $g(+\varpi, \gamma)$ and $g(-\varpi, \gamma)$ differ
by the quantity $g_1$ in equation (\ref{corrimpairetog}). Thus 
the integrands in the QR integrals in equation (\ref{Fisoarragement}) have a small even part
that gives rise to a 'parity' correction to these integrals. 
At large $\gamma$'s, the offset $\Delta\varpi_{\rm QR}$ is given by
the expansions in equations
(\ref{varpiqrplusexp}) and (\ref{varpiqrmoinsexp}), eventually giving for the offset correction (Appendix~\ref{appfenLF}):
\begin{equation}
F^{\rm iso}_{\rm off}(\gamma) = - \frac{4 \pi\, s_q}{3}  \, \gamma u \cos\alpha + 8 s_q \gamma u \cos\alpha \,
\left(\sqrt{3} K_{2/3}(1) L_{1/3}(1) - \pi\right)\,.
\label{Fisooff}
\end{equation}
The parity corrections are evaluated by integrating
the even part of the integrands in equation (\ref{Fisoarragement}) over the symmetric interval
$[- 3^{1/3} (\gamma u \sin^2\alpha)^{2/3}, \, +3^{1/3} (\gamma u \sin^2\alpha)^{2/3}]$. 
The even part of the 
integrand in the second line of equation (\ref{Fisoarragement}) is obtained by Taylor-expanding in $g$ the function 
in the parentheses about $g_0$ as in equation (\ref{corrimpairetog}), then performing the integration by
using $g$ as the integration variable. To sufficient accuracy, $g$ is related to $\varpi$ 
by the lowest-order relation in equation (\ref{gvarpiapproximatif}).  
The minimum value $g_m$ of $g$, 
reached at $\varpi = 0$, is given by equation (\ref{gmin}) and
approaches zero as $\gamma$ grows larger. 
The parity correction from the second line
of equation (\ref{Fisoarragement}) is thus evaluated taking $g_m = 0$. The calculation of the parity correction
from the first line is more straightforward. All calculations done (Appendix~\ref{appfenLF}), it is found that 
\begin{equation}
F^{\rm iso}_{\rm par}(\gamma) =  - \pi s_q \gamma u \cos \alpha \left[\frac{8}{3} \ln(\gamma)
+ \frac{4}{3}\ln\left(\frac{\sin \alpha}{u}\right) + 2 \ln(4) + \frac{4}{3} \ln(3) -4
+ \frac{8\sqrt{3}}{\pi} \,\int_0^1\! dg \, g\frac{d}{dg}\!\left(\sqrt{3} g K_{2/3}(g)L_{1/3}(g)\right) \right]\,.
\label{Fisoimparite}
\end{equation}
An asymptotic expansion of $F^{\rm iso}$ for large $\gamma$'s is finally obtained by expanding the first term of equation
(\ref{Fisoarragement}) up to order $\gamma$, integrating by parts the last term of equation (\ref{Fisoimparite}), which partly 
simplifies with the second term of equation (\ref{Fisooff}), and numerically calculating 
the coefficients of the terms of the expansion.
This gives
\begin{equation}
F^{\rm iso}_{\rm LF}(\gamma) = \pi s_q \, \gamma \, u \cos \alpha 
\left(\frac{4}{3}\ln\left(\frac{\gamma u}{\sin\alpha}\right) - 1.260\, 724\, 39\right)\,.
\label{FLFdetail}
\end{equation}
\begin{figure}
\begin{center}
\resizebox{0.45 \hsize}{!}{\includegraphics{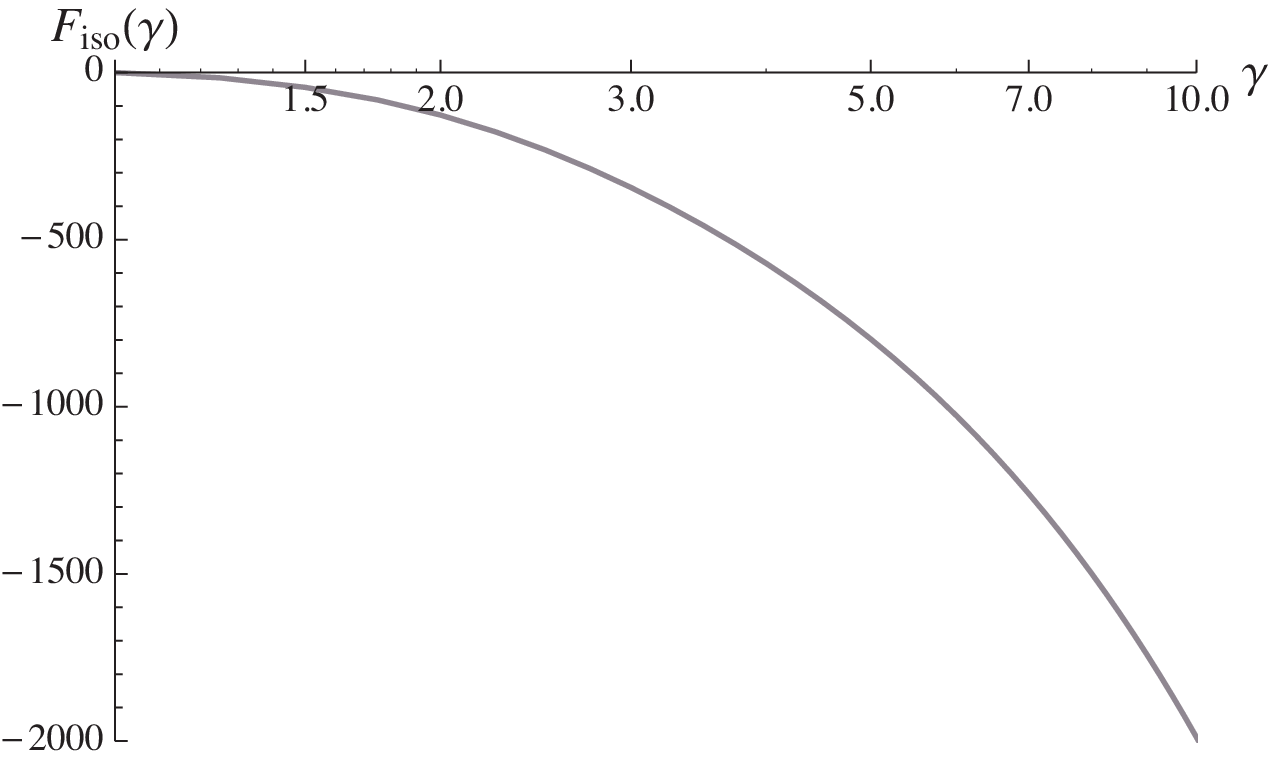}}
\resizebox{0.45 \hsize}{!}{\includegraphics{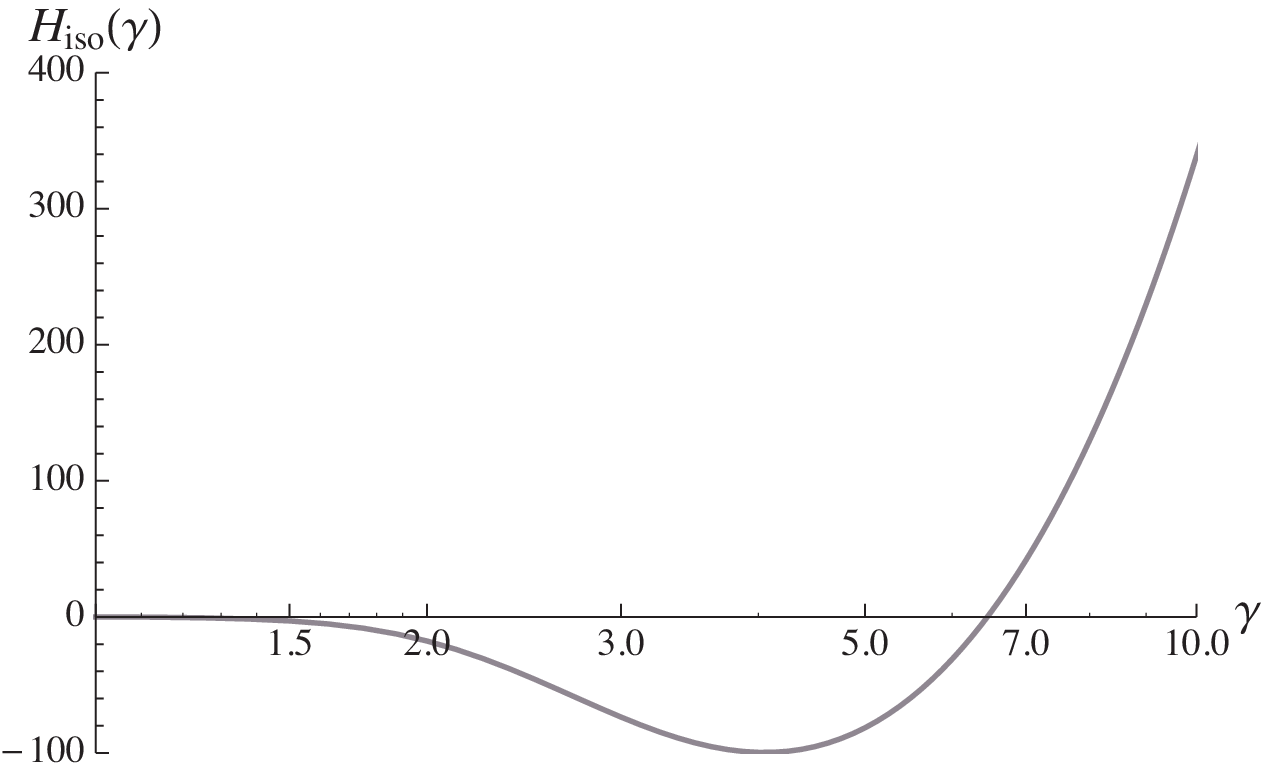}}
\caption{The isotropic kernels $F^{\rm iso}(\gamma)$ (left-hand panel) and $H^{\rm iso}(\gamma)$ (right-hand panel)
as given by equations (\ref{Fisogeneral}) and (\ref{Hisogeneral}).
The  parameters of the radiation are $\omega = 15 \mid \!\Omega\!\mid$ and
$\alpha =\pi/4$
}
\label{fig:kernel-iso}
\end{center}
\end{figure}

\subsection{Isotropic kernel of the Faraday conversion coefficient in the LF limit}
\label{kernelHlowfreqanalytiq}

The integral in equation (\ref{Hisogeneral})
can be separated into an NR and a QR contribution.
This is best achieved by writing the kernel $H^{\rm iso}(\gamma)$ as the sum of the contributions to it from 
equations (\ref{hNRreduit}) and (\ref{hQRreduit}), which may be modified by
extending the former integral to the full $\sigma$--$\varpi$ domain and
adding to equation (\ref{hQRreduit}) a correction to account 
for the undue integration over the QR domain so introduced. 
The variables are then changed to $\gamma$ and $\varpi$.
The extended NR contribution then provides the result in equation (\ref{hisotfinal})
while the expression for the correction to the QR contribution is similar to that in
equation (\ref{hfinalpigamma}), the boundaries of the integral over
$\varpi$ being however placed at $\varpi_{QR+}(\gamma)$ and
$\varpi_{QR-}(\gamma)$ on the edge of the QR domain
shown in Fig.~\ref{figBQR}. 
The kernel $H^{\rm iso}(\gamma)$ then takes the form
\begin{eqnarray}
&&H^{\rm iso}(\gamma)= -\frac{\pi}{2} \, \sin^2\!\!\alpha 
\left(\gamma \, (2 \gamma^2\! -\! 3)\, \sqrt{\gamma^2 -1} + {\cal{L}}(\gamma)\right)
\!+\! \frac{\Theta_H(\gamma -\gamma_{{\rm QR}})}{\sin^2 \alpha} \! 
\int_{\varpi_{QR-}}^{\varpi_{QR+}} \!\! d\varpi \,
\frac{\pi}{8} \!\left(\frac{2 x^4}{(\sigma^2 - x^2)^{5/2}} + \sigma_0^2 \, 
\frac{x^2 (4\sigma^2 + x^2)}{(\sigma^2 - x^2)^{7/2}}\right) 
\nonumber \\
&&\,\,\,\, + \frac{\Theta_H(\gamma -\gamma_{{\rm QR}})}{\sin^2 \alpha}  \!
\int_{\varpi_{QR-}}^{\varpi_{QR+}} \!\! d\varpi \
\left(
\frac{4x^2}{\sqrt{3}} \left(\frac{\sigma -x}{x}\right)^2K_{2/3}(g)L_{2/3}(g) 
+ \frac{2\varpi^2}{\sqrt{3}}  \left(\frac{\sigma -x}{x}\right) K_{1/3}(g) L_{1/3}(g) - \pi \, 
\frac{2\varpi^2 + \sigma_0^2}{\sqrt{\varpi^2 + \sigma_0^2}}   \right)\,.
\label{Kisogeneral}
\end{eqnarray}
The first term in the first line of equation (\ref{Kisogeneral})
is the NR isotropic kernel, calculated by integration over the full physical domain.
The second term in the first line is an integral over the QR domain correcting for the fact that the
NR domain really does not extend over the full physical one. 
The second line is the QR contribution proper.
The LF limit of this kernel
is to be obtained from an asymptotic expansion in $\gamma$ of
the right-hand side of equation (\ref{Kisogeneral}).
This LF development is 
valid when $\gamma \gg \gamma_{{\rm QR}}\approx (u/(3 \sin\alpha))^{1/2}$.
The domain of validity of the LF limit is where
$\omega  \ll 3 \gamma^2 \sin \alpha\mid\!\Omega\!\mid$ (equation \ref{regimechange-maintext}), which
is equivalent to $\omega \ll\omega_c(\gamma, \alpha)$, where $\omega_c$
is the characteristic frequency of synchrotron emission defined in equation (\ref{defomegacritsynch}). 

As discussed in Section~\ref{contribQR}, 
the variable $g$ in equation (\ref{gtexte}) is less than unity in the QR domain.
At the order required by the calculations in this subsection, the lowest order approximation
to the values of $\varpi_{QR\pm}(\gamma)$ is sufficient, namely
\begin{equation}
\varpi_{QR\pm}(\gamma) \approx  \pm \left(3 \gamma^2 u^2 \sin^4\!\! \alpha\right)^{1/3}\, .
\label{varpiedgeordre0}
\end{equation}
From equations (\ref{relapisigmagammappar}) and (\ref{pperpderhoq}),
the integrand in the first line of equation (\ref{Kisogeneral}) can be written in terms of $\varpi$ and $\gamma$, 
or $G=\gamma u \sin^2\!\alpha/3$, as
\begin{eqnarray}
&&\frac{2 x^4}{(\sigma^2 - x^2)^{5/2}} + \sigma_0^2 \, 
\frac{x^2 (4\sigma^2 + x^2)}{(\sigma^2 - x^2)^{7/2}} \quad = \quad \frac{162 G^4}{(\varpi^2 + \sigma_0^2)^{5/2}}
\left(1 + \frac{2\varpi \cos\! \alpha}{3G} - \frac{\varpi^2 \sin^2\! \alpha +\sigma_0^2}{9 G^2} \right)^2
\nonumber \\
&&\qquad \qquad \qquad +\frac{405 \sigma_0^2 G^4 }{(\varpi^2 + \sigma_0^2)^{7/2}}
\left(1 + \frac{2\varpi \cos \alpha}{3G} - \frac{\varpi^2 \sin^2 \!\alpha +\sigma_0^2}{9 G^2} \right)
\left(1 + \frac{2\varpi \cos\! \alpha}{3G} + \frac{5 \varpi^2 \cos^2\! \alpha -\varpi^2 - \sigma_0^2}{45 G^2} \right) \, .
\label{lafonctionaintegrer}
\end{eqnarray}
The primitive of this function is 
shown in Appendix~\ref{Applowfrequencyapprox}.  The integral over $\varpi$ 
that appears in the second term of the first line of equation (\ref{Kisogeneral}) can be obtained from it
and expanded to the required order in $\gamma$. It is inferred  from Section~\ref{comparQRNR} 
that this order should be ${\cal{O}}(\gamma^{4/3})$. 
The expansion procedure, outlined in Appendix~\ref{Applowfrequencyapprox}, leads to
\begin{equation}
\frac{\pi}{8\sin^2\alpha} \int_{\varpi_{QR-}}^{\varpi_{QR+}} \! d\varpi \, 
\left(\frac{2 x^4}{(\sigma^2 - x^2)^{5/2}} + \sigma_0^2 \, 
\frac{x^2 (4\sigma^2 + x^2)}{(\sigma^2 - x^2)^{7/2}} \right) 
\quad \approx \quad \pi \sin^2\!\alpha \, \gamma^4 - 2 \pi \sin^2\! \alpha \, \gamma^2 
- \frac{\pi}{8} \, \sin^{2/3}\!\!\alpha
\, \left(\frac{\gamma u}{3}\right)^{4/3} \, .
\label{correctionNRQR}
\end{equation}
The proper QR contribution to $H^{\rm iso}(\gamma)$ originates from the integration
over $\varpi$ of the second line in equation (\ref{Kisogeneral}).
This contribution is calculated to order $\gamma^{4/3}$ 
by changing the variable $\varpi$
for the variable $g$ on which the $K$ and $L$ functions depend, much as was done 
in Section~\ref{comparQRNR}, but for the fact
that the integration over $\varpi$ is now performed at constant $\gamma$ instead of at constant $\sigma$. 
Some details are presented in Appendix~\ref{Applowfrequencyapprox}.
From these calculations it is found that the LF limit to the QR contribution to 
the kernel $H^{\rm iso}(\gamma)$ is
\begin{equation}
H^{\rm iso}_{\rm QR}(\gamma)=  
2  \left(\gamma u\right)^{4/3} \left(\sin\!\alpha\right)^{2/3} \ 3^{1/6} 
\int_0^1 dg \ g^{2/3} \!\left(K_{2/3}(g)L_{2/3}(g) + K_{1/3}(g) L_{1/3}(g) - \frac{2\pi}{\sqrt{3} \, g} \right)\, .
\label{Khisointegre}
\end{equation}
The integrand in equation
(\ref{Khisointegre}) declines rapidly to zero out of the QR domain  (see Section~\ref{comparQRNR})
and the value of the definite integral over $g$ is, with a relative error of only $10^{-4}$, 
equal to $\pi/(4 \ \cdot  3^{1/6})$. Adopting this value, we eventually get
\begin{equation}
H^{\rm iso}_{\rm QR}(\gamma) \approx \frac{\pi}{2} \ \left(\sin\!\alpha\right)^{2/3} 
\left(\gamma u\right)^{4/3} \, .
\label{HisoQRfinally}
\end{equation}
Accounting for the slightly different definitions of $H^{\rm iso}(\gamma)$ and $K^{(h)}_{\rm QR}(\sigma)$ in equations
(\ref{defnoyauxisotropes}) and (\ref{formK}), the results in equations (\ref{HisoQRfinally}) and (\ref{kernelsh})
are entirely equivalent since 
at fixed $\gamma$ the QR domain is concentrated near $\varpi = 0$  at
$\sigma = \gamma u \sin^2\!\!\alpha$. This was
not unexpected because the QR domain is of a very small angular extent, 
so that the QR contribution is insensitive to the fact that the distribution function is
isotropic or otherwise.
Finally, the asymptotic expansion in $\gamma$ of the first term in equation (\ref{Kisogeneral}), 
limited to the order $\gamma^{4/3}$, is
\begin{equation}
-\frac{\pi}{2} \, \sin^2\!\!\alpha 
\left(\gamma \, (2 \gamma^2 -3)\, \sqrt{\gamma^2 -1} + {\cal{L}}(\gamma)\right)
\approx -\frac{\pi}{2} \, \sin^2\!\!²\alpha \, (2\gamma^4 - 4 \gamma^2)\,.
\label{devasymptNRtotal}
\end{equation}
The complete $H^{\rm iso}(\gamma)$ kernel in the LF limit is the sum of the three contributions 
in equations (\ref{correctionNRQR}), (\ref{HisoQRfinally}) and (\ref{devasymptNRtotal}). The large terms
proportional to $\gamma^4$ and $\gamma^2$ cancel out and the net result, valid in the LF limit, is:
\begin{equation}
H^{\rm iso}_{\rm LF}(\gamma) \approx \frac{\pi}{8} \left(\gamma u\right)^{4/3}  \left(\sin\!\alpha\right)^{2/3} 
\left(4  - \frac{1}{3^{4/3}}\right)\,.
\label{kernelisotroplowfreq}
\end{equation}
This result is in complete agreement with those in equation (\ref{kernelsh}) and
confirms that in the LF limit neither the QR nor the NR 
contribution to $h$ predominates.

\subsection{Approximate expressions for the isotropic kernels over the full energy range}
\label{kernelsinterpol}

Equations (\ref{fisotfinal}) and (\ref{hisotfinal})
provide expressions for the isotropic kernels in the HF regimes that are valid
in the interval $[1, \,\gamma_{\rm QR}]$, although when $\gamma$ approaches $\gamma_{\rm QR}$ from
below more terms should be retained in the Debye expansion. Equations 
(\ref{FLFdetail}) and (\ref{kernelisotroplowfreq})
provide expressions of the kernels in 
the asymptotic LF regime at $\gamma \gg \gamma_{\rm QR}$. The transition between the HF regime and the 
asymptotic LF regime should take place over an interval $\gamma_{\rm QR1} < \gamma < \gamma_{\rm QR2}$,
where $\gamma_{\rm QR1} =A_1 \gamma_{\rm QR}$ and $\gamma_{\rm QR2} = A_2 \gamma_{\rm QR}$  and
$A_1<1$ and $A_2>1$ are coefficients of order unity. 
For the kernel $F^{\rm iso}(\gamma)$ these coefficients should be close to unity
because the change from the HF regime to the LF regime for this coefficient is dim.
For $H^{\rm iso}(\gamma)$, or its derivative ${H'}^{\rm iso}(\gamma)$, $A_2$ 
is more likely to be of order of several units since Fig.~\ref{figKgammaOverGamma4third} 
shows that the asymptotic regime is in this case slowly reached.
Fig.~\ref{fig:kernel-iso} shows that exact kernels are continuous and at least once differentiable.
Our proposed approximation adopts the HF expressions (\ref{fisotfinal}) and (\ref{hisotfinal})
at $\gamma < \gamma_{\rm QR1}$ and extends them into the LF domain by a function 
that connects at $\gamma_{\rm QR1}$ to the HF expression and to its first order derivative and 
asymptotically merges into the LF limit in equations (\ref{FLFdetail}) or (\ref{kernelisotroplowfreq}).
The kernels themselves or their derivatives could be interpolated that way, according to whether 
one whishes to use equations (\ref{defnoyauxisotropes}) as they stand or to calculate them
by integrating by parts. We opted for the second method for the coefficient $h$,
which minimizes the inaccuracies due to the existence of singularities in the derivative
of the distribution functions of truncated power-laws.
We interpolate from $\gamma_{\rm QR1}$ to infinity by connection functions $C_{c}(\gamma-\gamma_{\rm QR1})$ 
that we define for coefficient $c$ ($= f$ or $h$) and element $i$ in the connection
($i=1$ for HF and $i=2$ for LF) by
\begin{equation}
C_{ci} (\gamma-\gamma_{\rm QR1}) = \left(1 - \exp\left(-\frac{(\gamma -\gamma_{\rm QR1})}{\lambda_{ci} \, \gamma_{\rm QR1}}\right)
\right)^2\,.
\label{connectionfunction}
\end{equation}
The fact that this function is a square  warrants that the interpolated
function and its first derivative are continuous at $\gamma_{\rm QR1}$
when the functions are themselves continuously differentiable.
The extrapolations of the HF expressions to the LF domain that appear on the second lines of 
equations (\ref{Fisointerpol}) and (\ref{Hisointerpol})
must then be continuous and derivable at $\gamma_{\rm QR1}$
but are otherwise unconstrained. A linear extrapolation proved satisfactory for $F^{\rm iso}$, taking in this case
$\gamma_{\rm QR1} = \gamma_{\rm QR2} = \gamma_{\rm QR}$. 
Due to the non-negligible gap that the interpolation should bridge between the HF
and the LF regime, the interpolation formula for the derivative ${H'}^{\rm iso}(\gamma)$ of the kernel
of the $h$ coefficient is more sophisticated. A value of $\gamma_{\rm QR1}$ smaller than $\gamma_{\rm QR}$ has been chosen,
and the extrapolation of the HF behaviour in the LF domain at $\gamma > \gamma_{\rm QR1}$ has been endowed
with a local bump that is meant to represent the cumulative effect of higher order terms in the Debye expansion when approaching
its limit of validity. 

The decrement parameter $\lambda_{ci}$ is a constant, chosen
to optimize the fit to the exact kernels in Fig.~\ref{fig:kernel-iso}. For  the kernel $F^{\rm iso}$ we have adopted 
$\lambda_{f1} = \lambda_{f2}= 1/(1.7)$.
For the derivative ${H'}^{\rm iso}(\gamma)$ of the kernel of $h$
we have adopted $\gamma_{\rm QR1} = 0.7 \gamma_{\rm QR}$ and different decrement parameters 
$\lambda_{ci}$ in equation (\ref{connectionfunction}), 
namely $\lambda_{h1} = 0.4$ and $\lambda_{h2}=1.8$. Other relevant parameters are described below.
The resulting interpolation formulae for $F^{ \rm iso}(\gamma)$ and ${H'}^{\rm iso}(\gamma)$  at $\gamma > \gamma_{\rm QR1}$ are
\begin{eqnarray}
F^{\rm iso}_{\rm int}(\gamma) &=& \pi s_q u \cos \alpha\, \Big[ 
C_{f}(\gamma-\gamma_{\rm QR})\
\gamma
\left(\frac{4}{3} \ln\left(\frac{\gamma u}{\sin \alpha}\right)
- 1.260\, 724\, 39\right) + \left(1 - C_{f}(\gamma-\gamma_{\rm QR})\right)\times 
\nonumber \\
&& \,\,\,\,\,\,\,\,\,
\left(4 \gamma_{\rm QR} \ln\left(\gamma_{\rm QR}\! +\! \sqrt{\gamma_{\rm QR}^2 -1}\right) - 4 \sqrt{\gamma_{\rm QR}^2 -1}
+ 4 \ln\left(\gamma_{\rm QR}\! +\! \sqrt{\gamma_{\rm QR}^2 -1}\right) (\gamma - \gamma_{\rm QR}) \right)\Big]\,, 
\label{Fisointerpol}\\
{H'}^{\rm iso}_{\rm int}(\gamma) &=& C_{h2}(\gamma -\gamma_{\rm QR1}) \left( - \frac{\pi}{6} \left(u^2 \sin \alpha\right)^{2/3} 
\left(4 - \frac{1}{3^{4/3}}\right) \, \gamma^{1/3}\right)
+ \left(1 - C_{h1}(\gamma-\gamma_{\rm QR1})\right)\times
\nonumber \\
&&\frac{\pi \, \sin^2 \alpha}{2} \Big[
4 (2 \gamma_{\rm QR1}^2 -1)\sqrt{\gamma_{\rm QR1}^2 -1} + 
\frac{4 (6\gamma_{\rm QR1}^2 -5)\gamma_{\rm QR1}}{\sqrt{\gamma_{\rm QR1}^2 -1}} \, \left(\gamma - \gamma_{\rm QR1}\right)
+ \eta \, \frac{\left(\frac{\gamma - \gamma_{\rm QR1}}{\mu \gamma_{\rm QR1}}\right)^2 
}{
1 + \left(\frac{\gamma - \gamma_{\rm QR1}}{\mu \gamma_{\rm QR1}}\right)^3} \Big]\,.
\label{Hisointerpol}
\end{eqnarray}
The coefficient $\eta = 103$ and $\mu = \lambda_{h1}/1.5$.
The transition value $\gamma_{\rm QR}$ rigorously is the value of $\gamma$ that causes the  line of constant $\gamma$
in Fig.~\ref{figBQR} to tangent the NR-QR boundary. It slightly differs from the 
$\gamma$ value associated to $\sigma_{\rm QR}$ at $\varpi = 0$ that 
is approximately representative of it and to which it reduces for $\cos \alpha = 0$. 
In our numerical evaluations we used the exact value, given by
\begin{equation}
\gamma_{\rm QR} \, u \sin ^2 \alpha = \sqrt{\frac{u \sin \alpha}{3}} \ \left(u \sin \alpha -\cos^2 \alpha\right)
\end{equation}
In Appendix~\ref{AppinterpolKernels} we give some examples of Faraday coefficients derived from 
the interpolating formulae (\ref{Fisointerpol}) and (\ref{Hisointerpol}) 
and contrast them with the quasi-exact ones obtained 
by double integration over $\varpi$ and $\sigma$ of the expressions in equations (\ref{fexacteanticip}) and (\ref{hexactanticip}). 
The accuracy of the fit provided by the interpolating formulae
is of order or better than about 10\% in this frequency range.
%%%%%%%%%%%%%%%
\section{Conclusion}
\label{SecDiscussion}
%%%%%%%%%%%%%%%

\subsection{Summary of the results in the different regimes}
\label{secsynthese}

We have derived an almost exact expression for the Faraday transfer coefficients
in the form of equations (\ref{fexacteanticip})--(\ref{hexactanticip}), which we repeat here, 
as expressed in terms of the $\varpi$--$\sigma$ variables defined in equation (\ref{defxsigma}):
\begin{eqnarray}
f &=& - \, \frac{2\pi^2 s_q}{c} \ \frac{\omega_{\rm pr}^2 \Omega^2}{\omega^3}
\int\!\!\!\!\int \!\! \frac{m^3\!c^3}{\sin^2\!\! \alpha} \, d\varpi \, d\sigma
\ \ \varpi \, x \ \frac{\partial F_0}{\partial \sigma} \
\left[
J'_\sigma(x) N_\sigma(x) + \frac{1}{\pi x}\right]
\,.\label{fexactsummary}
\\
h &=& \frac{\pi^2}{c} \frac{\omega_{\rm pr}^2 \Omega^2}{\omega^3}
\int\!\!\!\!\int \!\! \frac{m^3\!c^3}{\sin^2\!\! \alpha} \, d\varpi d\sigma
\left[
\frac{\partial F_0}{\partial \sigma} 
\left(x^2 J'_\sigma(x) N'_\sigma(x) - \varpi^2 J_\sigma(x) N_\sigma(x)\right)
+ \frac{1}{\pi} \left(\varpi \frac{\partial F_0}{\partial \varpi} - \sigma \frac{\partial F_0}{\partial \sigma}\right)
\right]
\,.\label{hexactsummary}
\end{eqnarray}
These expressions, which involve a two-real-variables integration, are 
completely general, applying
to isotropic as well as to non-isotropic distributions and encompassing all regimes of particle-wave interaction.
The Olver uniform expansions of high order Bessel functions
presented in Appendix~\ref{app:olver} can be used
to speed up the integration in equations (\ref{fexactsummary}) and (\ref{hexactsummary}) when 
very high values of $\sigma$ are considered. 
In the HF limit, when QR contributions to these integrals are negligible,
the transfer coefficients 
can be more simply written as
\begin{eqnarray}
&& f_{\rm HF} = 2\pi s_q \, \frac{\omega_{\rm pr}^2 \Omega^2}{c \, \omega^3} \!\!
\int\!\!\!\int \, \frac{m^3c^3}{\sin^2 \alpha}  d\varpi \, d\sigma
\ \, 
\left( \frac{1}{2}  \, \frac{\varpi x^2}{(\sigma^2 -x^2)^{3/2} }\right)\frac{\partial F_0}{\partial \sigma}
\, ,\label{ffinalpisigma}
\\
&&h_{\rm HF}=-\, \pi \, \frac{\omega_{\rm pr}^2 \! \Omega^2}{c \, \omega^3} \!
\int\!\!\!\int \, \frac{m^3c^3}{\sin^2\! \alpha} \, d\varpi  \, d\sigma \ \,  
\left(\frac{1}{8} \, \frac{2 x^4 (\sigma^2 - x^2) + \sigma_0^2 x^2 (4 \sigma^2 + x^2) }{\left(\sigma^2 - x^2 \right)^{7/2} } \right)
\frac{\partial F_0}{\partial \sigma} .
\label{hfinalpisigma}
\end{eqnarray}
Within the HF limit, these expressions 
are also completely general. The integrands in equations (\ref{ffinalpisigma}) and (\ref{hfinalpisigma})
are more regular than those in equations (\ref{fexactsummary})--(\ref{hfinalpisigma}).
The integration over $\varpi$ in equations (\ref{ffinalpisigma}) and (\ref{hfinalpisigma}),
although it is meant to cover the NR domain only, extends in
the HF limit to the full physical domain by lack of a QR contribution,
the $\rm QR$ domain then being ill-populated.

We have particularized equations (\ref{ffinalpisigma}) and (\ref{hfinalpisigma}) to a number of different physical
situations. For isotropic distribution functions they reduce to equations (\ref{fisotfinal}) and (\ref{hisotfinal}),
which coincide with known results for thermal distribution functions, as shown in Section~\ref{sec:thermal}. 
A quadrupolar anisotropy, represented by a distribution functions as in equation (\ref{eq:ansaztFaniso}), produces
the transfer coefficients shown in equations (\ref{fanisotfinal}) and (\ref{hanisotfinal}) while anisotropies
of higher multipolar orders generate the coefficients compiled in Appendix~\ref{app:Anisotrotropic}.  
In all these cases, the expression of the transfer coefficients is reduced to a one-variable quadrature
over the energy of the particles.
For anisotropies that cannot be expanded in a sum of a few multipolar terms, the HF coefficients
are obtained by performing the double integrals in equations (\ref{ffinalpisigma}) and (\ref{hfinalpisigma}), either
in these variables or in others, as shown in the case of a beam in Section~\ref{sec:beam}.

The HF approximation progressively loses validity as 
the fraction of particles with a Lorentz factor $\gamma$ large enough to interact with the wave in
the LF mode increases. A regime
change occurs when the QR contribution ceases to be negligible, which, for a given frequency and
a given direction of propagation,
happens for Lorentz factors such that their characteristic
synchrotron frequency, defined in equation (\ref{defomegacritsynch}), is of the order of or larger than the
frequency of the radiation considered.
We offer clear physical and mathematical explanations for this behaviour, that has been
previously well observed in the results of \citet{Cerbakov2008} and \citet{HuangCerbakov}: 
the regime change occurs when particles in quasi-resonance make a 
contribution comparable to 
the NR one.
From a mathematical standpoint, the two-variable kernels 
in equations (\ref{fexactsummary}) and (\ref{hexactsummary}) must be represented
differently in these two regimes, in terms 
of Nicholson's approximations for quasi-resonance or in terms of the 
Debye expansion of high order Bessel functions for non-resonance.

In full generality, the QR contributions to the transfer coefficients
are given by equations (\ref{fQR}) and (\ref{hQR}). The QR contribution to $h$
is however better expressed as in equation (\ref{hQRreduit}).
When QR contributions are important, 
the NR and QR contributions should eventually be added and 
care should be taken to integrate the QR contributions in equations (\ref{fQR}) and (\ref{hQRreduit})
over the QR domain only and the NR contributions 
in equations (\ref{ffinalpisigma}) and (\ref{hfinalpisigma}) over the NR domain  only.
We have compared the QR and NR contributions to the Faraday coefficients
in Section~\ref{comparQRNR}, and shown that in the LF limit
none of them predominates.
For a given frequency, the QR contribution to the
Faraday conversion coefficient $h$ grows when $\sigma \gg \sigma_{{\rm QR}}$ to a value comparable
to, and in fact numerically larger than, the NR one.
In the limit of large energies, the angular integration over the QR domain
covers a very small interval in pitch angle space,
as can be seen from equation (\ref{conditionsQR}). 
Across this interval, the partial derivative $\partial_\sigma F_0$ usually varies by 
only a little amount.
Our estimation in Section~\ref{comparQRNR} of ${\hat{h}}_{\rm QR}$ is for this reason 
expected to be a reliable one, that can be written in the LF limit as
\begin{equation}
{\hat{h}}^{\rm QR}_{\rm LF}  = \frac{\omega_{\rm pr}^2 \Omega^2}{c\, \omega^3}  
\int_{×\sigma \gg \sigma_{{\rm QR}}}^\infty \ \, \frac{m^3\!c^3}{\sin^2\!\!\alpha}\,  d\sigma \ \
\frac{\pi}{2} \, \sigma^{4/3} \, \partial_\sigma F_0(\varpi = 0, \sigma) \,.
\label{KhQRsidominant}
\end{equation}
Equation (\ref{KhQRsidominant}) is one among different contributions to $h$. It is
is meant to apply only to values of $\sigma$ that are relevant to the asymptotic LF limit.
It is otherwise
general and valid for any distribution function whatever the
pitch angle distribution, the latter
being anyway irrelevant to the QR contribution.

The complication of accounting for both
NR and QR contributions has been overcome for isotropic distribution functions by actually calculating their sum 
in the LF limit. This has lead to
the results in equations (\ref{defnoyauxisotropes}), (\ref{FLFdetail}) and (\ref{kernelisotroplowfreq}). 
The asymptotic LF contribution to the Faraday coefficients is in this case
\begin{eqnarray} 
&&f^{\rm iso}_{\rm LF} = \pi s_q \frac{\omega_{\rm pr}^2 \mid\!\Omega\!\mid}{c \, \omega^2} \, \cos \alpha 
\int_{\gamma \gg \gamma_{{\rm QR}}}^\infty \!\! m^3\!c^3 \, d\gamma \, \gamma 
\left(\frac{4}{3} \ln\left(\frac{\gamma u}{\sin \alpha}\right)- 1.260\, 724\, 39\right)
\  \frac{dF_0}{d\gamma}\ \, ,
\label{flowfreqsummary} \\
&&h^{\rm iso}_{\rm LF} = \frac{\omega_{\rm pr}^2 \mid\!\Omega\!\mid^{2/3}}{c \, \omega^{5/3}} \sin\!\alpha\!^{2/3}
\, \frac{\pi}{8} \left(4  - \frac{1}{3^{4/3}}\right)
\int_{\gamma \gg \gamma_{{\rm QR}}}^\infty \!\! m^3\!c^3 \, d\gamma \, \gamma^{4/3} \ \frac{dF_0}{d\gamma}\ \,.
\label{hlowfreqsummary}
\end{eqnarray}
These simple analytical results, which apply to the asymptotic LF regime $\gamma \gg \gamma_{{\rm QR}}$,
do not seem to have
appeared so far in the literature. For example, \citet{Cerbakov2008} and \citet{HuangCerbakov}
provide intermediate and LF regime fits to numerical results. 
For application to specific distributions functions, equations
(\ref{fexacteanticip}) and (\ref{hexactanticip}), 
(\ref{fcoupure}) and (\ref{hcoupure}),  (\ref{ffinalpisigma}) and (\ref{hfinalpisigma}) or (\ref{KhQRsidominant})
may of course be expressed in terms of any set of physical variables, such as the Lorentz factor $\gamma$ 
and pitch angle $\vartheta$ of the particles, as was partly done in Sections \ref{SecIsotrope} and \ref{SecAnisotrope}.

We proposed in equations (\ref{Fisointerpol})--(\ref{Hisointerpol})
approximate expressions of the kernels of the Faraday coefficients for isotropic distribution functions
that interpolate between the HF regime and the asymptotic LF  one.
These approximations to the kernels generate reasonably accurate expressions of
the coefficients $f$ and $h$ that can be written
in the form of a simple one-variable quadrature over the entire energy range.
In the HF limit,
we have similarly reduced to a one-variable quadrature on energy the expression of the
Faraday coefficients 
for a large class of anisotropic distribution functions proportional to Legendre polynomials 
of low degree depending on the cosine of the pitch angle. 
These functions may be used as a basis for expanding distribution functions of any simple kind of anisotropy.
For distribution functions
with sharp anisotropies our results can be made completely explicit by performing a 
double integration over energy and angle.

%%%%%%%%%%%%%%%%%%%%%%%%%%%%%%%%%%%%%%%%%
\subsection{Discussion and prospects}

Our analytical results have been deduced from a new, but straightforward, calculation of
the anti-Hermitian part of the conductivity tensor expressed in a form in which the
usual sum over discrete synchrotron resonances is exactly replaced by an integration over a continuous variable.
This particular representation of the conductivity tensor has not been hitherto used in this context. 
The integral over this continuous variable 
has principal value singularities that can be exactly reduced to the sum of a regular part
and of a residual part that has been shown to be safely negligible. As a result, simple, general, quasi-exact 
and non-singular expressions for the transfer coefficients have been obtained in a form that parallels
that of the familiar expressions of the synchrotron emission and absorption coefficients, 
but differs from the form in which Faraday coefficients are usually expressed in the literature.
Not-withstanding this difference
our results exactly coincide with known exact ones for thermal distributions.

Unlike the Hermitian part of the conductivity, which describes synchrotron absorption,
the anti-Hermitian part is never dominated by the contribution of a small 
QR population travelling close to the direction of propagation of the considered wave.
On the contrary, NR particles make  most of the contribution to the non-dissipative coefficients
in the HF limit and cannot be ignored otherwise.

The transfer properties of relativistic plasmas
may sensibly differ from those of cold plasmas, which may be a useful diagnostic. 
For example, the Faraday rotation coefficient $f$ of a symmetric pair plasma, with equal densities
of electrons and positrons having identical distribution functions, 
exactly vanishes but its Faraday conversion coefficient $h$ does not. The
study of a field-aligned beam in Section~\ref{sec:beam} also revealed 
that the Faraday rotation coefficient is enhanced in this case for radiation travelling in the beam. 
The knowledge of transport coefficients in all physical regimes, HF or LF, should allow 
optimal use of inversion algorithms of multiwavelength polarization data 
to yield the density and magnetic structure of the source and
of the intervening medium. 

The reconstruction of the magnetic field distribution in volume, from multiwavelength polarization observations of 
synchrotron emitting astrophysical media, has been attempted either partially \citep['Faraday synthesis';][]{Brentjens} 
or fully \citep{Thiebautetal}, but in the latter case assuming a known distribution of the electronic population(s) 
responsible for the synchrotron emission and/or the Faraday rotation. In the case of cold plasmas 
(e.g. the interstellar medium of our Galaxy or nearby galaxies) where the circular polarization is negligible (both from emission and transfer), 
these assumptions about the underlying electronic spatial distribution(s) are a real limitation to reconstructing the magnetic field structure. 
However, it was also shown \citep{Thiebautetal} that in the case of relativistic sources where a single electronic distribution would be responsible 
for both synchrotron emission and radiative transfer effects (through their contribution to the dielectric properties of the plasma), 
the multiwavelength observation of circular polarization (in addition to intensity and linear polarization) should 
in principle allow the simultaneous reconstruction of the magnetic field {\it and} electronic spatial distributions. In this context, 
the Faraday rotation and conversion coefficients derived in this work are of particular interest.  The diagnostic of relativistic jets 
from galactic nuclei and of pulsar winds by these methods would be of particular interest. It should be noted
however that our present results only yield the transfer coefficients in the plasma rest frame and should be transformed
to the observer's frame for use in the inversion algorithm for example along the lines described by \citet{GammieLeung}. 

%===============================================
\section*{Acknowledgments}
We thank M. Lemoine and G. Pelletier for early discussions and the 'Programme National de Cosmologie et Galaxies'  for funding.
CP thanks the community of 
 {\tt http://mathematica.stackexchange.com} for technical advice.

{}

\bigskip

\noindent
The following appendices are provided as online Additional Supporting Information to
the printed version of this paper and can be found at URL\\
{\tt http://mnras.oxfordjournals.org/lookup/suppl/doi:10.1093/mnras/stt135/-/DC1}

\bigskip

\appendix

\section{Variables suited to the relativistic particle-wave interaction}
\label{Apprhosigma}

The variables $\sigma$ and $\varpi$ 
are defined in equation~(\ref{defxsigma}). They depend on the propagation 
angle $\alpha$ of the radiation (Fig.~\ref{axesJeanJerome}) and
may be substituted to other dynamical variables of a particle,  such as  
$p_\perp$ and $p_\parallel$, or the modulus $p$ 
of its momentum (or its Lorentz factor $\gamma$) and the pitch angle $\vartheta$ of the particle's velocity.
The variables $\sigma$, $\varpi$ are convenient since some often-met operators 
acting on the distribution function take a simple form when expressed in terms of them,
as equation~(\ref{Df0etlopdsurdsigma}) shows.
Assuming the vacuum dispersion relation to be valid, equation~(\ref{defxsigma}) relates $\varpi$ and $\sigma$ to $\gamma$ and 
$p_\parallel/mc$ by 
\begin{equation}
\displaystyle \varpi =  \displaystyle u \ \left( \gamma \cos \alpha - \frac{p_\parallel}{mc} \right) \,,
\qquad 
\displaystyle \sigma \,   = \displaystyle u \ \left( \gamma - \frac{p_\parallel}{mc} \, \cos \alpha\right) \,,
\qquad \qquad \qquad
\displaystyle \gamma = \ \frac{\sigma - \varpi \cos \alpha}{u \, \sin^2 \alpha}\,,
\qquad 
\displaystyle \frac{p_\parallel}{mc} = \ \frac{\sigma \cos \alpha - \varpi}{u \, \sin^2 \alpha}
\,.\label{relapisigmagammappar}
\end{equation}
From this, $p_\perp$ and the variable $x$ defined in equation~(\ref{defxsigma}) may be deduced.
Using the notation $\sigma_0 = u \sin \alpha$, defined in equation~(\ref{defxsigma}), we have
\begin{equation}
\frac{p_\perp^2}{m^2 c^2} = \frac{\sigma^2 - \varpi^2 - \sigma_0^2}{\sigma_0^2}
\,,\qquad \qquad \qquad \qquad 
x = \sqrt{\sigma^2 -\varpi^2 - \sigma_0^2} \,,\qquad \qquad \qquad \qquad {\mathrm{where}} \qquad \sigma_0 = u \sin \alpha 
\,.\label{pperpderhoq}
\end{equation}
The physical region of the $\varpi$-$\sigma$ plane is where $\sigma$ is positive and  $p_\perp^2$ positive, 
that is, $\sigma^2 \geq \varpi^2 + \sigma_0^2$. Thus $\sigma$ assumes  a minimum value,
$\sigma_{0}$ that is usually large and can only be reached when $\varpi = 0$. 
The dynamical state of particles that are represented by points on the border of the physical domain
are those for which $p_\perp = 0$. 
For particles traveling at a velocity close to the speed of light,
the sign of $\varpi$ on the boundary of the domain is opposite to that of the field-aligned component of the
particle's velocity, as can be seen from equations (\ref{relapisigmagammappar}). 
Particles with vanishing $\varpi$ are those for which
$\gamma \cos \alpha = \sqrt{\gamma^2 -1} \cos \vartheta$, which means that the field-aligned
particle velocity equals the field-aligned wave velocity. For large $\gamma$'s this relation approximately reduces to 
$\vartheta = \alpha$.
When $\varpi$ does not vanish, its sign 
and value are indicative of the direction of the velocity of the particle. This
makes the variable $\varpi$ akin to an angular one.
By calculating the jacobian of the transformation from $p_\perp$, $p_\parallel$ to $\varpi$, $\sigma$, it is found that
\begin{equation}
p_\perp dp_\perp dp_\parallel = \frac{m^3 c^3 }{u^2 \sin^2 \alpha} \, \gamma \ d\varpi \, d\sigma\,,
\end{equation}
where $\gamma$ is the function of $\varpi$ and $\sigma$ given by equation~(\ref{relapisigmagammappar}). 
The operators acting on the distribution function $f_0$ that are most often met in these calculations
translate in $\varpi$-$\sigma$ variables as
\begin{eqnarray}
&&\frac{\partial f_0}{\partial p_\perp} = \frac{u}{m^2 c^2} \, \frac{p_\perp}{\gamma}  
\left( \frac{\partial f_0}{\partial \sigma } + \cos \alpha \frac{\partial f_0}{\partial \varpi}\right) 
\,,\qquad \qquad \qquad 
\frac{\partial f_0}{\partial p_\parallel} = - \frac{1}{\gamma mc} \, \left(\varpi \,  \frac{\partial f_0}{\partial \sigma} +
\sigma  \, \frac{\partial f_0}{\partial  \varpi}  \right) 
\,.\label{derf0simples}\\
&&D(f_0)  =  -\frac{u \, v_\perp}{mc} \ 
\left(\frac{\partial f_0}{\partial  \varpi} + \cos \alpha \, \frac{\partial f_0}{\partial \sigma}\right)
\,,\qquad \qquad \qquad 
\left(\omega \, \frac{\partial f_0}{\partial p_\perp} + k_\parallel \, D(f_0) \right)
= \frac{\omega \, u\, v_\perp \sin^2\alpha }{mc^2} \ \frac{\partial f_0}{\partial \sigma}
\,.\label{Df0etlopdsurdsigma}
\end{eqnarray}
\begin{figure}
\begin{center}
\resizebox{0.5 \hsize}{!}{\includegraphics{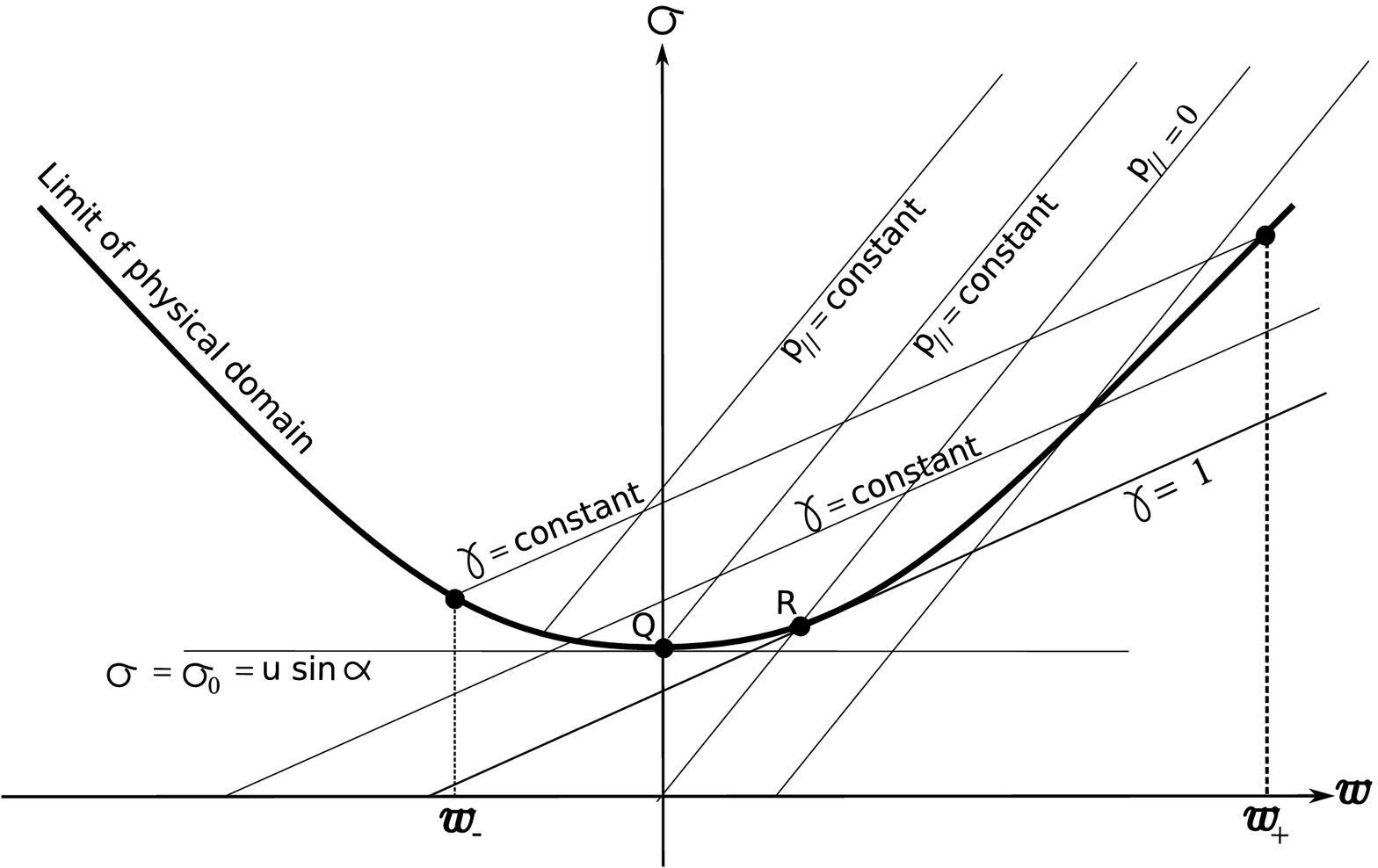}}
\caption{In the $\varpi$-$\sigma$ plane, the physical domain is above the
thick hyperbolic line that represents states with zero perpendicular momentum.
Lines of constant Lorentz factor $\gamma$, and of constant parallel momentum $p_\parallel$,
are shown. The state of rest is R. The state of lowest value of $\sigma$, $\sigma_{0} = u \sin \alpha$, at Q,
differs from it. }
\label{figpisigma}
\end{center}
\end{figure}
Conversely, in terms of the Lorentz factor $\gamma$ and pitch angle $\vartheta$:
\begin{eqnarray}
\frac{\partial f_0}{\partial \sigma} &=& \frac{1}{u \sin^2\!\!\alpha} \left(\frac{\partial f_0}{\partial \gamma} - \frac{1}{\sin \vartheta}
\left(\frac{\cos \alpha}{\sqrt{\gamma^2 -1}} - \frac{\gamma \cos \vartheta}{\gamma^2 -1} \right) \frac{\partial f_0}{\partial \vartheta}\right)\,.
\label{dersigmadegammatheta}
\\
\frac{\partial f_0}{\partial \varpi} &=& \frac{1}{u \sin^2\!\!\alpha}  \left(- \cos \alpha \, \frac{\partial f_0}{\partial \gamma}+
\frac{1}{\sin \vartheta} \left(\frac{1}{\sqrt{\gamma^2 -1}} 
- \frac{\gamma \cos \vartheta \cos \alpha}{\gamma^2 -1} \right)  \frac{\partial f_0}{\partial \vartheta}\right)\,.
\label{dervarpidegammatheta}
\end{eqnarray}

\section{The continuous-spectrum representation of the conductivity}
\label{LesGQin}

This appendix gives a few hints to calculate the matrix elements $G_{ab}$ 
in equation~(\ref{Qabforme2}), the functions $T_a$ being defined by equation (\ref{lesfonctionsT}).
Similar calculations being
described in some detail in the paper by \citet{Qin}, we 
only sketch here the main methods that should be used. 
The definitions of some quantities in \citet{Qin} 
may however differ from ours
by a sign and the gyrofrequencies are defined in their paper 
with a sign different from ours. Consider first the case in which $a= b=1$, the matrix element then  being
\begin{equation}
G_{11}(\sigma, x)=  \int_0^{2\pi}\! \frac{d\phi}{2\pi}  \int_0^{2\pi} \! \frac{dw}{2\pi} \ \, 
\exp\Big({i \big(\sigma w + x (\sin (\phi - w)-\sin \phi)\big)}\Big) 
\,.\label{Q11avantmodif}
\end{equation}
The identity $\sin (\phi - w)-\sin \phi
= -2 \sin(w/2) \cos(\phi - w/2)$ allows to integrate over $\phi$ by using the identity
\begin{equation}
\int_0^{2\pi} \frac{d\theta}{2\pi} \ \,  e^{iy \sin \theta} = J_0(y)
\,.\label{reprintegJ0}
\end{equation}
$J_0$ is the Bessel function of index $0$. For some other coefficients, the integration over $\phi$ yields 
a result proportional to the first or second derivative of $J_0$. Changing the angle $w$ into $2 \beta$,
$G_{11}$ is reduced to the quadrature (\citet{Watson} chapter 5)
\begin{equation}
G_{11}(\sigma, x)= \! \int_0^{\pi}\!\!  \frac{d\beta}{\pi} \ \,  e^{2 i \sigma \beta} \, J_0(2 x \sin \beta)
= e^{i\sigma \pi} \, J_\sigma(x) J_{-\sigma}(x)
\,.\label{intJ0enJJsigma}
\end{equation}
Some other elements, like $G_{31}$, involve after integrating over $\phi$
the derivative of the Bessel functions $J_0$. The integration over $w$ 
is then completed by integrating by parts.
The calculation of elements $G_{23}$, $G_{33}$ and $G_{22}$ is less straightforward
because the integrals over $w$ cannot be directly obtained from the identity in equation~(\ref{intJ0enJJsigma}). 
This difficulty is overcome by reducing the matrix element to a form that can be integrated by parts, 
by using Bessel's equation for $J_0(t)$
\begin{equation}
\frac{d^2 J_0(t)}{dt^2} + \frac{1}{t} \, \frac{dJ_0(t)}{dt} + J_0(t) = 0
\,.\label{eqBesselJ0}
\end{equation}
Consider for example the matrix element 
\begin{equation}
G_{33} (\sigma, x)= \int_0^{2\pi} \frac{d\phi}{2\pi} \ \cos \phi \int_0^{2\pi} \frac{dw}{2\pi} 
\, e^{i\sigma w} \ \cos (\phi - w) \ \, e^{i x (\sin(\phi - w) - \sin\phi)}
\,.\label{G33app}
\end{equation}
Changing $\phi$ to $\theta$ by $\phi = \theta + w/2 + \pi/2$ and 
then $w/2$ to $\beta$, 
$G_{33}$ in equation (\ref{G33app}) is changed to
\begin{equation}
G_{33}(\sigma, x) = \int_0^{2\pi} \frac{dw}{2\pi}  \, e^{i\sigma w}  \int_0^{2\pi} \frac{d\theta}{2\pi} 
\left(\sin^2 \theta - \sin^2 \frac{w}{2}\right) 
e^{2ix \sin \frac{w}{2} \sin \theta} =  
- \int_0^{\pi} \frac{d\beta}{\pi}  \, e^{2i\sigma \beta} 
\left(J^{"}_0\left(2x \sin \beta\right) + \sin^2 \beta \ J_0\left(2x \sin \beta \right)\right)
\,.\label{G33avecJsecond0}
\end{equation}
On integrating over $\theta$, use has been made of equation~(\ref{reprintegJ0}) and of its second derivative with respect to $y$.
The Bessel equation (\ref{eqBesselJ0}) being written for an argument $t= t_\beta \equiv 2x \sin \beta$, the second derivative term 
with respect to $t$ in equation~(\ref{eqBesselJ0}) is expressed in terms of derivatives with respect to $\beta$. The first order derivative term
proportional to $dJ_0(t_\beta) /d\beta$ happens to be $\tan^2\! \beta \,  (J'_0(t_\beta)/t_\beta)$.
All terms proportional to $J'_0(t_\beta)$
are then eliminated by using equation~(\ref{eqBesselJ0}) again, which yields
\begin{equation}
4 x^2 \left( J^{"}_0\!\left(t_\beta\right) + \sin^2 \beta \ J_0\!\left(t_\beta\right)\right)
= \frac{d^2 J_0\!\left(t_\beta\right)}{d \beta^2} \,,\qquad \qquad {\mathrm{with}} \qquad \qquad t_\beta \equiv  2 x \sin \beta
\,.\label{eqBesseltransfo}
\end{equation}
Using this, the expression of $G_{33}$ in equation~(\ref{G33avecJsecond0}) becomes
\begin{equation}
G_{33} =  -\int_0^{\pi} \frac{d\beta\, e^{2i\sigma \beta}}{4\pi x^2} 
\frac{d^2\! J_0\! \left(t_\beta\right)}{d \beta^2}  = 
- \left[ \frac{ e^{2i\sigma \beta}}{4\pi x^2} 
\frac{d J_0\!\left(2x \sin \beta \right)}{d\beta} \right]_0^\pi 
+ \frac{2 i\sigma}{4 x^2} 
\int_0^\pi \frac{d \beta}{\pi} \, e^{2i\sigma \beta} \frac{d\!J_0\!\left(2x \sin \beta \right)}{d\beta}
\,.\label{G33intpart1}
\end{equation}
which can be integrated by parts again, noting that the square bracket term in equation~(\ref{G33intpart1})
vanishes because $J'_0(t)$ vanishes at $t = 0$, to eventually give, using equation~(\ref{intJ0enJJsigma})
\begin{equation}
G_{33}(\sigma, x) \, =\,   
+ \frac{2 i\sigma}{4\pi x^2} \left[e^{2i\sigma \beta} J_0\left(2x \sin \beta \right)\right]_0^\pi 
+\frac{\sigma^2}{x^2}  \int_0^\pi \frac{d\beta}{\pi}  \, e^{2i\sigma \beta} J_0\left(2x \sin \beta \right) 
\, =  \, \frac{i \sigma}{2\pi x^2}  \left(e^{2i\sigma \pi} -1\right) 
+\frac{\sigma^2}{x^2}  e^{i\sigma \pi} J_\sigma(x) J_{-\sigma}(x)\,.
\end{equation}

\section{The distribution associated with multiple resonances}
\label{Appdistrib}

Let us consider the limit for $N \rightarrow \infty$ of the following function of $\sigma$:
\begin{equation}
\left(1 - e^{2i\pi N\sigma}\right) \frac{\cos \sigma \pi}{\sin \sigma \pi}\,.
\label{ladistribresonapp}
\end{equation}
This function 
enters expressions involving the integration of its product with a regular function of $\sigma$.
When $N$ diverges, $\exp(2i\pi N\sigma)$ oscillates infinitely rapidly, leaving
in the limit a vanishing integral when multiplied by any regular function, which is the case in any interval 
containing no integer value.
At integer values of $\sigma$, the denominator $\sin \sigma \pi$ vanishes, as does also 
the factor $(1 - \exp(2i\pi N\sigma))$. To find the limit value, we expand by setting 
$\sigma = n +\eta$, where $\eta$ is ${\cal{O}}(1/N)$. This gives 
\begin{equation}
g(\sigma) \ \left(1 - e^{2i\pi N\sigma}\right)  \frac{\cos \sigma \pi}{\sin \sigma  \pi} \approx 
g(n + \eta) \left(2 \pi N^2 \eta  - i \, 2N\,  \frac{\sin 2\pi N \eta}{2\pi N \eta}\right)\,.
\end{equation}
Near the singularity, the real term, which is proportional to $2 \pi N^2 \eta$, appears 
as a symmetrical cut replacing the diverging part of the function $\cos \sigma \pi/\sin \sigma  \pi$
by an odd function of $\eta$ that vanishes at $\eta =0$
in an interval that decreases with $N$ as $1/N$.
This means that as
$N \rightarrow \infty$, the real part of the diverging integral of $g(\sigma) (\cos \sigma \pi/\sin \sigma  \pi)$
should be understood as a principal value. The imaginary part is
proportional near the resonance to $2N (\sin 2\pi N \eta/2\pi N \eta)$ which is large at $\eta = 0$, and then 
quickly decreases in the vicinity of $\sigma = n$, oscillating rapidly on a scale ${\cal{O}}(1/N)$. 
The integral of the product with a regular function $g(\sigma)$ in the vicinity
of $\sigma =n$  can be evaluated by changing from $\eta$ to $y = 2\pi N \eta$, which yields, in this vicinity
\begin{equation}
\int d\sigma \,  g(\sigma) \Big(- i \, 2N \frac{\sin 2\pi N \eta}{2\pi N \eta}\Big) \approx
- i g(n) \, \frac{1}{\pi} \int_{-\infty}^{+\infty}  \frac{\sin y \, dy}{y} = - i \, g(n) \,.
\end{equation}
In the limit $N \rightarrow \infty$, the function in equation (\ref{ladistribresonapp}) then converges to the distribution
\begin{equation}
\lim_{N \rightarrow \infty} \left(1 - e^{2i\pi N\sigma}\right)  \frac{\cos \sigma \pi}{\sin \sigma  \pi}  \equiv
{\cal{D}}\left(\frac{\cos \sigma \pi}{\sin \sigma  \pi}\right) =
{\cal{P}} \left(\frac{\cos \sigma \pi}{\sin \sigma  \pi}\right) - i \pi \sum_n \delta_{\rm D}(\sigma -n)\,.
\end{equation}
where $\delta_{\rm D}$ is a Dirac distribution.
This leads to the result in equation~(\ref{lafameusedistrib}). 
The distribution $ {\cal{D}}(\cos \sigma \pi/\sin \sigma  \pi)$
is causal, which was obvious from the beginning
since the integration has been carried over positive delay times only. This explicitly shows up in the fact that when
$\sigma$ is close to a singularity, at
$\sigma =n$ say, the distribution ${\cal{D}}$ reduces, according to Plemelj's formula, to
\begin{equation}
\frac{1}{\pi} \left({\cal{P}} \frac{1}{\sigma -n} - i\pi \delta_{\rm D}(\sigma -n)\right) = \frac{1}{\pi} \frac{1}{(\sigma -n) + i0} \,.
\end{equation}
The presence of the infinitesimal positive imaginary part at the denominator of the right-hand side member reveals
the causal character of the distribution.

\section{The transverse components of the conductivity}
\label{AppLesMiprimjprim}

The components $XYZ$ of the tensor ${\mathbf{M}}$ in the X,Y,Z frame being given by equations 
(\ref{MzzJN}), its transverse components 
in the $x'$, $y'$, $z'$ frame are obtained from them by 
equations (\ref{Mcomptrans}) and (\ref{Mcomptransyprimyprim}). This first gives
\begin{eqnarray}
M_{x'x'} &=& 
\frac{i \pi\, v_\perp}{\mid \Omega_*\!\mid} \,
\Big(\omega \frac{\partial f_0}{\partial p_\perp} + k_\parallel D(f_0) \Big)
\left( {\cal{D}}\!\left(\frac{\cos \sigma \pi}{\sin \sigma \pi}\right) {J'}^2_\sigma(x) 
- J'_\sigma(x) N'_\sigma(x) + \frac{\sigma}{\pi x^2} \right)\,,
\nonumber \\
M_{x'y'} &=& \frac{s_q \pi v_\perp}{\mid \Omega_*\!\mid}  
\! \left(
\sin \!\alpha \!\left(\! \omega \frac{\partial f_0}{\partial p_\parallel} 
- k_\perp D(f_0) \frac{ \sigma}{x}\!\right) 
- \cos \alpha \frac{\sigma}{x} \Big(\!\omega \frac{\partial f_0}{\partial p_\perp} + k_\parallel D(f_0)\! \Big)
\!\right)\!
\!\left(\! {\cal{D}}\!\left(\frac{\cos \sigma \pi}{\sin \sigma \pi}\right)\!  J_{\sigma}(x) J'_{\sigma}(x)\!
- J'_{\sigma}(x) N_{\sigma}(x)\!  - \! \frac{1}{\pi x}\! \right)\!,
\nonumber \\
M_{y'x'}  &=&  (-s_q\pi) \, \left(
\sin \alpha \frac{v_\parallel}{\mid \Omega_*\!\mid} \, 
- \cos \alpha \frac{v_\perp}{\mid \Omega_*\!\mid} \, \frac{\sigma}{x} \right) 
\Big(\omega \frac{\partial f_0}{\partial p_\perp} + k_\parallel D(f_0) \Big)
\left( {\cal{D}}\!\left(\frac{\cos \sigma \pi}{\sin \sigma \pi}\right) J_{\sigma}(x) J'_{\sigma}(x)
- J'_{\sigma}(x) N_{\sigma}(x)  - \frac{1}{\pi x} \right)\,,
\nonumber \\
M_{y'y'} &=& \cos^2\alpha 
\frac{i\pi v_\perp}{\mid \Omega_*\!\mid} \,
\Big(\omega \frac{\partial f_0}{\partial p_\perp} + k_\parallel D(f_0) \Big)
\frac{\sigma^2}{x^2} \, 
\left( \, {\cal{D}}\!\left(\frac{\cos \sigma \pi}{\sin \sigma \pi}\right) 
J^2_{\sigma}(x) - J_{\sigma}(x) N_{\sigma}(x) - \frac{1}{\sigma \pi} \right)
\nonumber\\
&+& \sin^2 \alpha  \
\frac{i \pi v_\parallel}{\mid \Omega_* \! \mid} 
\left(\!
\left(\omega \frac{\partial f_0}{\partial p_\parallel}  - k_\perp D(f_0)  \frac{\sigma}{x}\right)
\Big({\cal{D}}\!\left(\frac{\cos \sigma \pi}{\sin \sigma \pi}\right)  J^2_{\sigma}(x) - J_{\sigma}(x) N_{\sigma}(x)\Big)
\ + \ \frac{ k_\perp D(f_0) }{\pi x} \, 
\right) 
\nonumber\\
&-& \sin \alpha \cos \alpha \
\frac{i\pi v_\perp}{\mid \Omega_*\!\mid} \, \frac{\sigma}{x}\, 
\left( \ \omega \frac{\partial f_0}{\partial p_\parallel} - k_\perp D(f_0)  \frac{\sigma}{x} \right)
\left({\cal{D}}\!\left(\frac{\cos \sigma \pi}{\sin \sigma \pi}\right)  J^2_\sigma(x) 
- J_\sigma(x) N_\sigma(x) - \frac{1}{\sigma \pi} \right)
\nonumber \\
&-& \sin \alpha \cos \alpha  \ 
\frac{i \pi v_\parallel}{\mid \Omega_*\!\mid} \, \frac{\sigma}{x} \,
\left(\omega \frac{\partial f_0}{\partial p_\perp} + k_\parallel D(f_0) \right) 
\left( {\cal{D}}\!\left(\frac{\cos \sigma \pi}{\sin \sigma \pi}\right)
 J^2_{\sigma}(x) - J_{\sigma}(x) N_{\sigma}(x) -\frac{1}{\sigma \pi} \right)\,.
\nonumber
\end{eqnarray}
Although this is not obvious, the elements $M_{x'y'}$ and $M_{y'x'}$ differ by only a sign,
since their factors of $\sin \alpha$ are equal. This can be seen by calculating the difference  of these factors of $\sin \alpha$,
accounting for the definition of $D(f_0)$ in equation~(\ref{defDanisotrop}). 
The expression of $M_{y'x'}$ is improved by arranging 
the first parenthesis using the definitions in equation~(\ref{defxsigma}). The writing of $M_{y'y'}$ 
can be similarly improved by gathering the terms proportional to Bessel functions on the one hand,
and those which are not on the other hand, then
changing some factors accompanying 
$\sin^2\!\! \alpha$ and $\sin \alpha \, \cos \alpha$  by using 
equations (\ref{defDanisotrop}) and (\ref{defxsigma}). 
Denoting $D(f_0)$ by $D$ for brevity, this eventually gives, for example
\begin{eqnarray}
&&v_\parallel \, \left( \omega \frac{\partial f_0}{\partial p_\parallel}  - k_\perp D \frac{\sigma}{x} \right)
= \frac{v^2_\parallel}{v_\perp} \ \, \left(\omega \frac{\partial f_0}{\partial p_\perp} + k_\parallel D \right)\,,
\\
&&v_\perp \,  \left( \ \omega \frac{\partial f_0}{\partial p_\parallel} - k_\perp D \, \frac{\sigma}{x} \right)
+ v_\parallel \left(\omega \frac{\partial f_0}{\partial p_\perp} + k_\parallel D \right) 
= 2 v_\parallel \left(\omega \frac{\partial f_0}{\partial p_\perp} + k_\parallel D \right)\,,
\\
&&
- v_\perp \cos^2\!\alpha \, \frac{\sigma}{x^2}\left(\!\omega \frac{\partial f_0}{\partial p_\perp} + k_\parallel D\! \right)
\!+\!  v_\parallel \sin^2\! \alpha \frac{ k_\perp D}{x} 
+\, \frac{\sin \alpha \cos \alpha}{x} 
\left(\! v_\perp\! \left(\!\omega \frac{\partial f_0}{\partial p_\parallel} - k_\perp D \frac{\sigma}{x} \right)
+ \! v_\parallel \! \left(\!\omega \frac{\partial f_0}{\partial p_\perp} + k_\parallel D\! \right)\! \right) 
\nonumber\\
&& \qquad \qquad = \mid\!\Omega_*\!\mid \ \left( 
\frac{\omega \sin \alpha \cos \alpha}{k_\perp} \ \, \frac{\partial f_0}{\partial p_\parallel}
- \left( \omega \cos \alpha - k v_\parallel\right) 
\Big(\frac{ D \sin \alpha}{k_\perp v_\perp} + \frac{\cos \alpha}{k_\perp^2 v_\perp }
\, \Big(\omega \frac{\partial f_0}{\partial p_\perp} + k_\parallel  D \Big) 
\Big) \right)\,.
\end{eqnarray}
The last equation has been obtained by using the vacuum dispersion relation. 
The matrix elements can be given a simpler expression by using, instead of $p_\perp$ and $p_\parallel$,
the variables $\sigma$ and $\varpi$ defined in equation~(\ref{defxsigma}) and presented in Appendix~\ref{Apprhosigma}. 
This results in the equations (\ref{Mxprimxprimtext})--(\ref{Myprimyprimtext}).
\section{Residual principle value terms are negligible}
\label{AppPPnegligeables}
The integrations which appear in equations (\ref{fpremierjus}) and (\ref{hpremierjus})
are over the physical domain in the $\varpi$--$\sigma$ plane. 
This domain is bounded from below by the curve $\cal{B}$
of equation $\sigma^2 = \varpi^2 + u^2 \sin^2 \alpha$. This constrains
$\sigma$ to be larger than a minimum value, $\sigma_0 = u \sin \alpha$, that is usually large
(unless $\sin \alpha$ is very small) because $u =\omega/\!\!\mid\! \Omega\! \mid$ is large.
For a given value of $\sigma$, $\varpi$ spans the interval $[-\varpi_b(\sigma),  + \varpi_b(\sigma)]$, where
$\varpi_b(\sigma) = \sqrt{\sigma^2 - \sigma_0^2}$. Each principal value term in equations (\ref{fpremierjus})
and (\ref{hpremierjus}) can be written as a sum of a few contributions of the form
\begin{equation}
T = \int\!\!\!\!\int \!\! d\varpi \, d\sigma \ \, {\cal{P}}\!\left(\frac{\cos \sigma \pi}{\sin \sigma \pi}\right)
A(\varpi, \sigma) \, B_\sigma(x) \, \frac{\partial F_0}{\partial \sigma} 
= \int_{\sigma_0}^\infty {\cal{P}}\!\left(\frac{\cos \sigma \pi}{\sin \sigma \pi}\right)
C(\sigma)
\,.\label{structurePP} 
\end{equation}
where $A(\varpi, \sigma)$ is a function of $\sigma$, $\varpi$ or $x$
and $B_\sigma(x)$ is the product of two functions that may be either
$J_\sigma(x)$ or its derivative $J'_\sigma(x)$. The function $C(\sigma)$ is the integral over $\varpi$ of
$A(\varpi, \sigma)\, B_\sigma(x) \, \partial_\sigma F_0$
on the interval $[-\varpi_b(\sigma),  + \varpi_b(\sigma)]$. It varies on a scale of order $\sigma$.
On $[n-1/2, n+1/2]$, $C(\sigma)$ may be series-expanded
by setting $a= (\sigma -n)$ and can be written, on this interval, as
\begin{equation}
C(\sigma) = C(n) + a \,  C^{(1)}(n) + \frac{a^2}{2} \, C^{(2)}(n) + \cdots + \frac{a^k}{k!}  \, C^{(k)}(n) + \cdots
\end{equation}
where $C^{(k)}(n)$ is the $k$-th derivative of $C(\sigma)$ at $\sigma = n$. From this we get
\begin{equation}
\int_{n -1/2}^{n + 1/2} \!\! d\sigma \, {\cal{P}}\!\left(\frac{\cos \sigma \pi}{\sin \sigma \pi}\right) C(\sigma) =
\sum_{k =0}^\infty \Lambda_{k} \, \frac{C^{(k)}(n)}{k!}\,,
\qquad \qquad {\mathrm{where}}  \qquad \Lambda_{k} = \int_{-1/2}^{+1/2} \!\! da \
{\cal{P}}\!\left(\frac{a^k \cos a \pi}{\sin a \pi} \right)
\,.\label{serieintervaln}
\end{equation}
The integral defining $\Lambda_k$ is regular when $k > 0$.
The coefficient $\Lambda_k$ vanishes when $k$ is even because the integrand then is odd in $a$.
Therefore the summation on $k$ on the left of equation~(\ref{serieintervaln})
is on odd values $k = 2m+1$ only.  This series is absolutely convergent since
each derivative $C^{(2m +1)}(n)$ is expected to be of order $C^{(2m -1)}(n)/n^2$ , thus
decreasing rapidly with $m$, so that the factorial $(2m + 1)!$ at the denominator 
warrants a fast convergence. 
Moreover, $\Lambda_{2m +1}$ also decreases with $m$.
Let $n_1$ be the smallest integer for which $n_1 -1/2 \geq \sigma_0$, that
is, the smallest integer for which the interval $[n_1 -1/2, n_1 + 1/2]$ is entirely contained
in $[\sigma_0, \infty]$. The expression $T$ in equation~(\ref{structurePP}) can be written as
the sum of a contribution from the interval $[\sigma_0, n_1 - 1/2]$ and a sum
over $n \geq n_1$ of terms similar to that in equation~(\ref{serieintervaln}):
\begin{equation}
T = \int_{\sigma_0}^{n_1 - \frac{1}{2}} \!\! d\sigma \, 
{\cal{P}}\!\left(\frac{\cos \sigma \pi}{\sin \sigma \pi}\right) C(\sigma)
+ \sum_{n= n_1}^{\infty} \left(\sum_{m =0}^\infty \Lambda_{2m +1} \, \frac{C^{(2m + 1)}(n)}{(2m + 1)!} \right)\,.
\end{equation}
It is possible to replace the summation over $n$ by an integral because the terms
of the series in $n$ do not change by much when $n$ changes by unity. Noting that 
$C^{(2m + 1)}(\sigma)$ is the derivative of $C^{(2m)}(\sigma)$,
the second term can be exactly calculated by integrating over $\sigma$, which 
performs the summation over $n$ as an integral, giving
\begin{equation}
T = \int_{\sigma_0}^{n_1 - \frac{1}{2}}\!\! d\sigma \, 
{\cal{P}}\!\left(\frac{\cos \sigma \pi}{\sin \sigma \pi}\right) C(\sigma)
+ \sum_{m =0}^\infty \Lambda_{2m +1} \, \frac{C^{(2m)}(n_1)}{(2m + 1)!} \equiv T_{\mathrm{first}} + T_{\mathrm{sum}}
\,.\label{Tapresserieintegrale}
\end{equation}
We have separated $T$ in equation~(\ref{Tapresserieintegrale}) into a first term,
$T_{\mathrm{first}}$, the second one, $T_{\mathrm{sum}}$, being the sum over all complete
unit intervals. $T_{\mathrm{sum}}$ is
similar to the integral calculated in equation~(\ref{serieintervaln}), 
for the particular value $n = n_1$, 
the derivatives $C^{(2m)}$ replacing the derivatives $C^{(2m + 1)}$. 
The ratio $C^{(2m)}(n_1)/C^{(2m + 1)}(n_1)$ is expected to be of order of $n_1$, which itself is close to
$\sigma_0$. We may assume that there is an upper bound over all values of $m$,  $K \sigma_0$ say,
to the ratio $C^{(2m)}(n_1)/C^{(2m + 1)}(n_1)$, so that we may evaluate an upper bound on $T_{\mathrm{sum}}$:
\begin{equation}
T_{\mathrm{sum}} \leq K \sigma_0 \sum_{m =0}^\infty \Lambda_{2m +1} \, \frac{C^{(2m + 1)}(n_1)}{(2m + 1)!}
=  K \sigma_0 \int_{n_1 -1/2}^{n_1 + 1/2} \!\! d\sigma \, 
{\cal{P}}\left(\frac{\cos \sigma \pi}{\sin \sigma \pi}\right) C(\sigma) 
\,.\label{estimTsum}
\end{equation}
The contibution $T_{\mathrm{first}}$ of the first interval, defined in equation~(\ref{Tapresserieintegrale}), 
is itself of the same order of magnitude as the factor of $K \sigma_0$ in the last term of equation~(\ref{estimTsum}).
It then suffices to show that  $T_{\mathrm{sum}}$, as estimated in equation~(\ref{estimTsum}), is negligibly small.

Since $n_1$ is distant from $\sigma_0$ by less than unity, the argument $x$ of the Bessel functions that
corresponds to $\sigma = n_1 \leq \sigma_0 +1$ is very small compared to its index $n_1\approx \sigma_0$, 
the ratio $x/\sigma$ being of order $\sqrt{2/\sigma_0}$.
Now, for arguments $x \ll \sqrt{\sigma}$,  as is the case here,
the Bessel functions of the first kind $J_\sigma(x)$ 
may be represented by their Carlini approximation \citep{AbramStegun}
that can be further
simplified by using Stirling's formula for $\Gamma(\sigma)$ at large $\sigma$ \citep{AbramStegun}. This gives
\begin{equation}
J_\sigma(x) \approx \frac{1}{\sqrt{2\pi \sigma} } \ \left(\frac{ex}{\sigma}\right)^\sigma  \,,
\qquad \qquad 
J'_\sigma(x) \approx  \frac{e}{\sqrt{2\pi \sigma} } \ \left(\frac{ex}{\sigma}\right)^{\sigma -1}  
\,.\label{approxJJprim}
\end{equation}
The function $A$ in equation~(\ref{structurePP}) may be either $\varpi^2$, $\varpi x$ or $x^2$ depending on 
the corresponding Bessel factor $B_\sigma$ and the variable $x$ is
$\sqrt{\sigma^2 -\sigma_0^2 -\varpi^2}$. Setting 
$\varpi = \lambda \, \sqrt{\sigma^2 -\sigma_0^2}$, where $\lambda$ is in
the interval $[-1, +1]$, we get
\begin{equation}
C(\sigma) \approx \frac{\sqrt{\sigma^2 -\sigma_0^2}}{2\pi \sigma}  
\left(\frac{e^2 (\sigma^2 -\sigma_0^2)}{\sigma^2}\right)^\sigma
\int_{-1}^{+1}\!\!\!d \lambda \ \, 
\left( \begin{array}{c}
\lambda^2 \, (\sigma^2 - \sigma_0^2)\\
\lambda \ \, \sigma \sqrt{\sigma^2 - \sigma_0^2}\\
\sigma^2
\end{array} \right)
\left(1 - \lambda^2\right)^\sigma \
\frac{\partial F_0}{\partial \sigma}\! \left(\lambda \sqrt{\sigma^2 -\sigma_0^2}, \ \sigma\right)
\,,
\end{equation}
where the upper element in the column vector corresponds to $J_\sigma^2$, the middle
one to $J_\sigma J'_\sigma$ and the lower one to
$J^{'2}_{\sigma}$.  The integral over $\lambda$ is itself a small number
owing to the presence of the factor $(1-\lambda^2)^\sigma$ that is a large power of a number smaller than unity.
But it is the  factor in front of the integral that makes the whole expression extremely small, since
$\sigma$ does not differ from $\sigma_0$ by more than two units, implying that
\begin{equation}
\left(\frac{e^2 (\sigma^2 -\sigma_0^2)}{\sigma^2}\right)^\sigma
<   \left(\frac{4 e^2 \sigma_0}{\sigma_0^2}\right)^{\sigma_0} 
= {\cal{O}}\left(\left(\frac{1}{\sigma_0}\right)^{\sigma_0}\right)
\,.\label{PetitCarlini}
\end{equation}
For frequencies at the peak of synchrotron emission, $\sigma_0$ is of order of the square 
of the Lorentz factor $\gamma$ of the emitting particles, which means that 
the term on the right of equation~(\ref{PetitCarlini}) is of order $(\gamma^2)^{-\gamma^2}$,  
entirely negligible even at relatively small values of $\gamma$, such as $\gamma = 10$, where its value is about 10$^{-200}$. 
Then, setting $\sigma =  \sigma_0 + h$ where $h$ is ${\cal{O}}(1)$,
substituting $\sigma_0$ to $\sigma$ wherever possible and otherwise
approximating  $(\sigma^2 - \sigma_0^2)$ by $2 \sigma_0 h$, the upper bound on $T_{\mathrm{sum}}$ on the right of equation~(\ref{estimTsum}) 
can be written as
\begin{equation}
T_{\mathrm{sum}} \leq K \frac{\sigma_0^{\frac{5}{2}}}{\pi \sqrt{2}}  \, \left(\frac{2e^2}{\sigma_0}\right)^{\sigma_0} 
\!\! \int_{(n_1-\sigma_0) -\frac{1}{2}}^{(n_1-\sigma_0) +\frac{1}{2}} \!\!\! dh \ 
{\cal{P}}\!\left(\frac{\cos \sigma \pi}{\sin \sigma \pi}\right) \, h^{\sigma_0 + \frac{1}{2}} 
\left( \begin{array}{c}
\!\!(2h/\sigma_0) \!\!\\
\!\!(2h/\sigma_0)^{\frac{1}{2}}\!\!\\
\!\!1\!\!
\end{array} \right)
\ \int_{-1}^{+1}\!\!\!d \lambda \, 
\left( \begin{array}{c}
\!\lambda^2\!\\
\!\lambda\!\\
\!1\!
\end{array} \right)
\left(1 - \lambda^2\right)^\sigma 
\frac{\partial F_0}{\partial \sigma}\! \left(\lambda \sqrt{2h \sigma_0}, \sigma\right)\,.
\end{equation}
Again, the prefactor proportional  to $(\sigma_0)^{-\sigma_0}$ causes the
whole expression to be negligebly small.  This results from the
extreme smallness of the Bessel functions of the first kind and large indices
when their argument is small. 

\section{Nicholson's approximation}
\label{AppNicholson}

Nicholson's approximation provides expressions for the Bessel functions of large indices 
when their argument is close to the index, in our case slightly smaller. \citet{Watson}, chapter 6, 
gives the following 
integral representations of these functions:
\begin{eqnarray}
J_\sigma(x) &=& \frac{1}{\pi} \int_0^\pi \cos\left(\sigma \theta - x\sin \theta\right) \, d\theta
- \frac{\sin \sigma \pi}{\pi} \int_0^\infty \exp\left(- \,  \left(\sigma t + x \sinh t\right) \right) \, dt
\,.\label{Jreprintegrale}
\\
- \, N_\sigma(x) &=& \frac{1}{\pi} \int_0^\pi \sin\left(\sigma \theta - x\sin \theta\right) \, d\theta
\ + \frac{1}{\pi} \int_0^\infty \exp\left(\sigma t - x \sinh t\right) \, dt
+ \frac{\cos \sigma \pi}{\pi} \int_0^\infty \exp\left(- \,  \left(\sigma t + x \sinh t\right) \right) \, dt
\,.\label{Nreprintegrale}
\end{eqnarray}
For $x$ and $\sigma$ large and almost equal, the integrand is rapidly oscillating, or decaying,
and any of these integrals may be approximated by the method of stationary phase which consists in expanding the
arguments of the trigonometric or exponential functions 
in the vicinity of the value of $\theta$ or $t$ where they are stationary. This gives, for example
\begin{equation}
\frac{1}{\pi} \int_0^\pi \cos \left(\sigma \theta - x\sin \theta\right) \, d\theta 
\approx \frac{1}{\pi} \int_0^\pi \cos \left((\sigma - x)\theta + x {\theta^3}/{6} \right) \, d\theta
= \frac{6^{1/3}}{\pi \, x^{1/3} }  \int_0^\infty 
\cos \left( 3 \ \frac{2^{1/3} (\sigma -x) }{ 3^{2/3} \, x^{1/3} }  \ v + \, v^3\right) \ d v\,.
\end{equation}
These integrals involve Airy-Hardy integrals of order 3 \citep{Petiau} defined by
\begin{equation}
{\mathrm{Ci}}_3(g) = \int_0^\infty \!\! \cos(3 g v + v^3) \, dv \,,\qquad \qquad 
{\mathrm{Si}}_3(g) = \int_0^\infty \!\! \sin (3 g v + v^3) \, dv \,,\qquad \qquad 
{\mathrm{Ei}}_3(g) = \int_0^\infty \!\! \exp(- (3 g v + v^3)) \, dv \,.
\end{equation}
The last, negative exponential, term in equations (\ref{Jreprintegrale}) and (\ref{Nreprintegrale}) being 
negligible compared to the others for large and positive $x$ and $\sigma$, we are left with
\begin{equation}
J_{\sigma}(x) \approx \frac{6^{1/3}}{\pi\, x^{1/3} } \ {\mathrm{Ci}}_3
\left(\frac{2^{1/3} (\sigma -x) }{3^{2/3} \, x^{1/3} }\right)
\,,\qquad \qquad \qquad \qquad 
-\, N_{\sigma}(x) \approx \frac{6^{1/3}}{\pi \, x^{1/3} }  \left[
{\mathrm{Si}}_3\left(\frac{2^{1/3} (\sigma -x) }{3^{2/3} \, x^{1/3} }\right)
+ {\mathrm{Ei}}_3\left(-\, \frac{2^{1/3} (\sigma -x) }{3^{2/3} \, x^{1/3} }\right) \right]
\,.\label{JNdeAiryHardy}
\end{equation}
The function ${\mathrm{Ci}}_3$ reduces to a combination of modified Bessel functions of index $\pm 1/3$, but
the functions ${\mathrm{Si}}_3$ and ${\mathrm{Ei}}_3$,  separately, do not. However,
the combination that appears in equation~(\ref{JNdeAiryHardy}) does.
Correcting a little typo in Watson's book, p.324 equation~(8) \citep{Watson}, we have
\begin{equation}
{\mathrm{Ci}}_3(g) = \frac{\pi \sqrt{g} \, \cos \frac{\pi}{6}}{3 \, \sin \frac{\pi}{3} } \, 
\ \left(I_{- \frac{1}{3}}\left(2 g^{\frac{3}{2}} \right) - I_{+ \frac{1}{3}}\left(2 g^{\frac{3}{2}} \right) \right)\,,
\quad  
{\mathrm{Si}}_3(g) + {\mathrm{Ei}}_3(-g)= \frac{\pi \sqrt{g} \, \left(1 + \sin \frac{\pi}{6}\right) }{3 \, \sin \frac{\pi}{3} } \,
\ \left(I_{-\frac{1}{3}}\left(2 g^{\frac{3}{2}} \right) + I_{+\frac{1}{3}}\left(2 g^{\frac{3}{2}} \right) \right)
\,.\label{combinAiryKL}
\end{equation}
The results in equation~(\ref{NLtiers}) are deduced from equation~(\ref{combinAiryKL}) 
and from the definitions in equation (\ref{KLnu}).
Those in equation~(\ref{NprimL2tiers}) result from the expressions in equation~(\ref{NLtiers}) by derivating with respect to $x$.
Use is made of the recurrence relations for the functions $I_\nu$, which translate into different recurrence relations
for the functions $K_\nu$ and $L_\nu$, that have different parity under a change of sign of the index
\begin{eqnarray}
&&I_{\nu -1}(y) - I_{\nu + 1}(y) = \frac{2\nu}{y} \ I_{\nu}(y) 
\,,\qquad \qquad \qquad \qquad 
I_{\nu -1}(y) + I_{\nu + 1}(y) = 2 I'_{\nu}(y)\,,
\nonumber \\
&&K'_{\frac{1}{3}}(y) = - K_{\frac{2}{3}}(y) - \frac{1}{3y} \ K_{\frac{1}{3}}(y)
\,,\qquad \qquad \qquad \qquad 
L'_{\frac{1}{3}}(y) = + L_{\frac{2}{3}}(y) - \frac{1}{3y} \ L_{\frac{1}{3}}(y)\,.
\nonumber 
\end{eqnarray}
The results in equation~(\ref{NprimL2tiers}) have been obtained by neglecting
subdominant terms that appear as a result of the derivation.

\section{Olver uniform asymptotic expansion of Bessel functions} \label{app:olver}

The Olver approximation for the Bessel functions and their derivatives at large order and
for $z<1$ is given by \cite{Olver54}
\begin{equation}
J_\sigma(\sigma z) \sim \left(\frac{4\zeta}{1-z^2}\right)^{1/4}\left( 
\frac{{\rm Ai}(\sigma^{2/3} \zeta)}{\sigma^{1/3}} \sum_{s=0}^{\infty} \frac{A_s(\zeta)}{\sigma^{2s}}
+
\frac{{\rm Ai}'(\sigma^{2/3} \zeta)}{\sigma^{5/3}} \sum_{s=0}^{\infty} \frac{B_s(\zeta)}{\sigma^{2s}}
\right) \,,\label{eq:defOlverJ}
\end{equation}
\begin{equation}
N_\sigma(\sigma z) \sim - \left(\frac{4\zeta}{1-z^2}\right)^{1/4}\left( 
\frac{{\rm Bi}(\sigma^{2/3} \zeta)}{\sigma^{1/3}} \sum_{s=0}^{\infty} \frac{A_s(\zeta)}{\sigma^{2s}}
+
\frac{{\rm Bi}'(\sigma^{2/3} \zeta)}{\sigma^{5/3}} \sum_{s=0}^{\infty} \frac{B_s(\zeta)}{\sigma^{2s}}
\right)\,,
\end{equation}
\begin{equation}
J'_\sigma(\sigma z) \sim -\frac{2}{z} \left(\frac{4\zeta}{1-z^2}\right)^{-1/4}\left( 
\frac{{\rm Ai}(\sigma^{2/3} \zeta)}{\sigma^{4/3}} \sum_{s=0}^{\infty} \frac{C_s(\zeta)}{\sigma^{2s}}
+
\frac{{\rm Ai}'(\sigma^{2/3} \zeta)}{\sigma^{2/3}} \sum_{s=0}^{\infty} \frac{D_s(\zeta)}{\sigma^{2s}}
\right)\,,
\end{equation}
\begin{equation}
N'_\sigma(\sigma z) \sim \frac{2}{z} \left(\frac{4\zeta}{1-z^2}\right)^{-1/4}\left( 
\frac{{\rm Bi}(\sigma^{2/3} \zeta)}{\sigma^{4/3}} \sum_{s=0}^{\infty} \frac{C_s(\zeta)}{\sigma^{2s}}
+
\frac{{\rm Bi}'(\sigma^{2/3} \zeta)}{\sigma^{2/3}} \sum_{s=0}^{\infty} \frac{D_s(\zeta)}{\sigma^{2s}}
\right)  \,,\label{eq:defOlverN}
\end{equation}
where ${\rm Ai}$ and ${\rm Bi}$ are Airy functions defined as in the appendix
of the paper by \cite{Olver54} and ${\rm Ai}'$ and ${\rm Bi}'$ their derivatives. The variable $\zeta$
is related to $z$ by
\[
\frac{2}{3}\zeta^{3/2}= \cosh^{-1}\frac{1}{z} -\sqrt{1-z^2}\,,
\]
and $A_s$, $B_s$, $C_s$ and $D_s$ are given by
\begin{eqnarray}
A_s(\zeta) &= \displaystyle \sum_{m=0}^{2s} b_m \zeta^{-3 m/2} U_{2s -m}(\frac{1}{\sqrt{1-z^2}})\,, \qquad
\zeta^{1/2} B_s(\zeta)&= -\sum_{m=0}^{2s+1} a_m \zeta^{-3 m/2} U_{2s+1 -m}(\frac{1}{\sqrt{1-z^2}}) \,,\\
\zeta^{1/2} C_s(\zeta)&=\displaystyle -\sum_{m=0}^{2s+1} b_m \zeta^{-3 m/2} V_{2s+1 -m}(\frac{1}{\sqrt{1-z^2}}) \,, \qquad
D_s(\zeta) &=\sum_{m=0}^{2s} a_m \zeta^{-3 m/2} V_{2s -m}(\frac{1}{\sqrt{1-z^2}})\,.
\end{eqnarray}
The Debye polynomials $U_n$ and $V_n$ are found by solving the following recurrence
\[
U_n(x)=
\frac{1}{2} \left(1-x^2\right) x^2
   {U'}_{n-1}(x)+\frac{1}{8} \int_0^x \left(1-5
   t^2\right) {U}_{n-1}(t)
   \, dt\,, \quad
V_n(x)=
x \left(x^2-1\right) \left(x
   {U}'_{n-1}(x)+\frac{1}{2}
   {U}_{n-1}(x)\right)+
   {U}_n(x)\,, 
\] with $U_0=1$ and $ V_0=1\,.$
The coefficients,
\[
a_0=1\,, \quad b_0=1\,,\quad a_s=\frac{(2s+1)(2s+3)\cdots(6s-1)}{s! (144)^s}\,,\quad b_s=-\frac{6s +1}{6s-1} a_s
\]
arise from asymptotic developments of the Airy functions at infinity, such as:
\[
{\rm Ai}(\sigma^{2/3}\zeta)\sim \frac{1}{2\sigma^{1/6}\zeta^{1/4}\sqrt{\pi}} \exp\left(
-\frac{2}{3}\sigma \zeta^{3/2} \right) \sum_{s=0}^\infty (-1)^s \frac{ a_s}{\sigma^5 \zeta^{3 s/2}}\,,
\quad
{\rm Ai}'(\sigma^{2/3}\zeta)\sim -\frac{\sigma^{1/6} \zeta^{1/4}}{2\sqrt{\pi}} \exp\left(
-\frac{2}{3}\sigma \zeta^{3/2} \right) \sum_{s=0}^\infty (-1)^s \frac{ b_s}{\sigma^5 \zeta^{3 s/2}}\,.
\]
The Debye expansion is recovered by plugging these asymptotic expansions in equation~(\ref{eq:defOlverJ})-(\ref{eq:defOlverN}),
e.g.
\[
J_\sigma(\sigma z)\sim \frac{1}{ \sqrt{2 \pi \sigma} (1-z^2)^{1/4}} \exp\left(
-\frac{2}{3}\sigma \zeta^{3/2} \right) \sum_{s=0}^\infty \frac{U_s(1/\sqrt{1-z^2})}{\sigma^s}\,.
\]
Making the following change of variables:
\[
z=\frac{1}{\cosh \xi}\,\quad \frac{2}{3} \zeta^{3/2}=\xi -\tanh \xi\,.
\]
we recover equations~(\ref{DebyeJ}) and (\ref{DebyeN}).
In Section~\ref{subsubthermallowfreq} we make use of
equations~(\ref{eq:defOlverJ})-(\ref{eq:defOlverN}) to first order to compute
numerically equation~(\ref{Hisogeneral}). The corresponding implementation in {\tt Mathematica} is available
at the following URL: {\tt http://www. iap.fr/users/pichon/olver/}.
 
\section{Comparison of quasi-resonant and non-resonant contributions}
\label{AppcomparQRNR}

We derive in this appendix the estimations presented in Section~\ref{comparQRNR} for the NR and QR contributions to the
$\varpi$-integrated kernels of the Faraday rotation and conversion coefficients. 
Let us first consider the 
NR contributions. Equation (\ref{fNR}) gives $f_{\rm NR}$, which we may reduce to its dominant term, the first one
in the parenthesis. Equation (\ref{hNRreduit}) gives $h_{\rm NR}$.
The double integrals in these equations are over the NR domain. The latter is represented in Fig.~\ref{figBQR}
and the equations of its lower boundary ${\cal{B}}$ and upper boundary ${\cal{B}}_{\rm QR}$ respectively are 
$\varpi^2 +\sigma_0^2 = \sigma^2$ and $\varpi^2 + \sigma_0^2 = 3^{2/3} \sigma^{4/3}$.
Taking advantage of the symmetry of the integrands, the 
$\varpi$-integrated kernels in equation (\ref{formK}) can be written as:
\begin{eqnarray}
K_{\rm NR}^{(f)} &=& 2\, \int_{\sqrt{3^{2/3} \sigma^{4/3} - \sigma_0^2}}^{\sqrt{\sigma^2 - \sigma_0^2}} 
\! d\varpi \, \varpi^2  \left(\frac{\pi}{2}\right) \, \frac{\sigma^2 -(\varpi^2 +\sigma_0^2)}{(\varpi^2 +\sigma_0^2)^{3/2}}\, ,
\\
K_{\rm NR}^{(h)} &=& 2\, \int_{\sqrt{3^{2/3} \sigma^{4/3} - \sigma_0^2}}^{\sqrt{\sigma^2 - \sigma_0^2}} 
\! d\varpi \, \left(\frac{\pi}{8}\right) \, 
\left(\frac{2 (\sigma^2 - (\varpi^2 +\sigma_0^2))^2}{(\varpi^2 +\sigma_0^2)^{5/2}} 
+ \frac{\sigma_0^2 (\sigma^2 - (\varpi^2 +\sigma_0^2)) (5 \sigma^2 - (\varpi^2 +\sigma_0^2))}{(\varpi^2 +\sigma_0^2)^{7/2}} \right)
\, .
\end{eqnarray}
To evaluate these integrals in the limit of large $\sigma$'s, 
that is when $\sigma_0/\sigma \rightarrow 0$, we substitute $y=\varpi/\sigma$
to $\varpi$. The
lower bound of the integrations becomes $(3/\sigma)^{1/3}$ and the upper bound unity. The result is
\begin{equation}
K^{(f)}_{\rm NR} = \pi \, \sigma^2  \int_{(3/\sigma)^{1/3}}^{1} \ 
dy \ \frac{1 - y^2}{y}  \approx \, \frac{\pi \sigma^2}{3}  \, \ln\left((\sigma/3)\right) \,,
\qquad \qquad \quad
K^{(h)}_{\rm NR} = \left(\frac{- \pi}{\, 2}\right) \int_{(3/\sigma)^{1/3}}^{1}
dy \, \frac{\, (1- y^2)^2}{y^5} \approx - \frac{\pi}{8} \left(\frac{\sigma}{3}\right)^{4/3}\,.
\end{equation}
We now turn to the QR contributions which we evaluate in the same limit. 
Equation (\ref{fQR})
gives $f_{\rm QR}$ and equation (\ref{hQRreduit}) gives $h_{\rm QR}$, combined with the 
boundary term $h_{\rm BQR}$ in equation (\ref{hBQR}).  
The variable $g$
is defined in equation (\ref{gtexte}) and 
can be expressed in terms of $\varpi$ and $\sigma$ from equation~(\ref{pperpderhoq}).
At a given $\sigma$, $g$ reaches a minimum $g_{\rm m}(\sigma)$ at $\varpi = 0$.
In the QR domain, $x$ may be considered
equal to $\sigma$ except where the difference $(\sigma -x)$ is involved, which makes it possible 
to invert the relation between $g$ and $\varpi$,
yielding, in the QR domain:
\begin{equation}
g \approx \, \frac{\left(\sigma_0^2 +\varpi^2\right)^{3/2} }{3 \, \sigma^2}  
\,,\qquad\quad  \varpi^2 + \sigma_0^2 \approx \left(3 g \sigma^2\right)^{2/3}\, ,
\qquad \quad x \approx \sigma \,,
\qquad \quad \frac{\sigma-x}{x} \approx \frac{3^{2/3}}{2} \, \left(\frac{g}{\sigma}\right)^{2/3} \,,
\qquad \quad g_{\rm m}(\sigma) = \frac{\sigma_0^3}{3 \sigma^2} \, .
\label{gmin}
\end{equation}
Assuming as in Section~\ref{comparQRNR} that $\partial_\sigma F_0$
depends linearly on $\varpi$, 
equation (\ref{hQRreduit}) 
can be recast in the form of equation (\ref{formK}). 
When $\varpi$ is changed for $g$ and the limit $\sigma \gg \sigma_{{\rm QR}} \sim \sigma_{0}^{3/2}$
is taken, the $\varpi$-integrated kernels assume the form
\begin{eqnarray}
&&K^{(f)}_{\rm QR} = \sqrt{3} \, \sigma^2 \int_{g_{\rm m}}^{1} dg \ 
g^{2/3} \, \sqrt{g^{2/3} - g_{\rm m}^{2/3}} \, \left(K_{\frac{2}{3}}(g) L_{\frac{1}{3}}(g) - \frac{\pi}{g \sqrt{3}} \right)\,,
\label{kernelfQReng}
\\
&&K^{(h)}_{\rm QR} = 2 \ (3^{1/6}) \, \sigma^{4/3} \int_{g_{\rm m}}^{1} dg \ 
g^{2/3} \, \left( K_{\frac{2}{3}}(g) L_{\frac{2}{3}}(g) + K_{\frac{1}{3}}(g) L_{\frac{1}{3}}(g) 
- \frac{2 \pi}{g \sqrt{3}} \right)\,.
\label{kernelhQReng}
\end{eqnarray}
The parenthesis in equation (\ref{kernelfQReng}) diverges as $g$ approaches zero as $g^{-1/3}$, but the integral 
still converges as $g_{\rm m}$ approaches zero.
Similarly, the parenthesis in equation (\ref{kernelhQReng}) diverges as $g^{-4/3}$ at small values of $g$. 
Thanks to the factor $g^{2/3}$, the integral over $g$ is however convergent when its lower bound 
approaches zero,
which it does when $\sigma$ grows very large. 
The integral in equation (\ref{kernelfQReng}) can be calculated numerically for $g_{\rm m} =0$, giving the result
in equation (\ref{kernelsf}). 
From numerical calculation, it also appears that the
right-hand side of equation (\ref{kernelhQReng}) converges when $\sigma$ approaches infinity to a value that
is close to
$(\pi/2) \, \sigma^{4/3}$, with a relative accuracy of $10^{-4}$. Adopting this convenient approximation, it is found that
for $\sigma \gg \sigma_0^{3/2}$ 
\begin{equation}
K^{(h)}_{\rm QR} (\sigma) \approx \frac{\pi}{2} \ \sigma^{4/3}\,,
\end{equation}
which is the result mentioned in equation (\ref{kernelsh}).
For large values of $g$, that is out of the QR domain, the parenthesis in equation (\ref{kernelhQReng})
decreases quickly to zero, owing to the fact that any product of a $K$ and $L$ function asymptotes to $\pi/(\sqrt{3} g)$ in
this limit. The remainder of the parenthesis decreases as $g^{-3}$ whereas 
each of its terms considered individually decreases only as $g^{-1}$.
The integrand thus declines with $g$ as $g^{-7/3}$, which would cause the integral in equation (\ref{kernelhQReng})
to converge if it were to be extended to a bound much larger than unity. 
Thus, the support of the integrand in equation (\ref{kernelhQReng})
does not extend out of the QR domain. The decrease of the integrand 
in equation (\ref{kernelfQReng}) is however
slower.

\section{Anisotropic kernels of higher multipolar orders for Faraday coefficients at high-frequency} \label{app:Anisotrotropic}
%%%%%%%%%%%%%%%%%%%%%%%%%%

Let us parametrize the distribution function  as: $ F(\gamma,\vartheta)=F_n(\gamma)P_n(\cos \vartheta) $.
After a bit of algebra, following the substitutions of Section~\ref{SecAnisotrope},
we find that the  coefficient $f$ can be written as in equation (\ref{DetDprim})
with the same $A_f$ factor that appears in equation (\ref{fanisotfinal}), while the function $D'_f$ 
can be written as $D'_{f0}(\gamma) + D'_{f1}(\gamma) {\cal L}(\gamma)$.
The factors $D'_{f0}$  ({\sl left column}) and $D'_{f1}$ ({\sl right column})
are listed below for $n=0,\cdots \, 6$, where $P_i$ denotes the value of 
a Legendre polynomials $P_i$ at angle $\alpha$,   $P_i \equiv P_i(\cos \alpha)$: 
\begin{equation}
\begin{array}{ll}
 -\frac{4 \sqrt{\gamma ^2-1} P_1}{\gamma } & \scriptstyle 4 P_1 \\
\scriptstyle -6 P_2 & \frac{2 \left(2 \gamma ^2+1\right) P_2}{\gamma  \sqrt{\gamma ^2-1}} \\
 \frac{4 \left(\left(\gamma ^2-1\right) P_1-\left(11 \gamma ^2+4\right) P_3\right)}{5 \gamma  \sqrt{\gamma
   ^2-1}} & \frac{4 \left(\left(6 \gamma ^2+9\right) P_3+P_1\left(1-\gamma ^2\right)\right)}{5
   \left(\gamma ^2-1\right)} \\
 \frac{54 \left(\gamma ^2-1\right) P_2-25 \left(10 \gamma ^2+11\right) P_4}{21 \left(\gamma ^2-1\right)} &
   \frac{6 \left(-2 \gamma ^4+\gamma ^2+1\right) P_2+5 \left(8 \gamma ^4+24 \gamma ^2+3\right) P_4}{7
   \gamma  \left(\gamma ^2-1\right)^{3/2}} \\
 \frac{8 \left(11 \gamma ^4-7 \gamma ^2-4\right) P_3-\left(274 \gamma ^4+607 \gamma ^2+64\right) P_5}{18
   \gamma  \left(\gamma ^2-1\right)^{3/2}} & -\frac{8 \left(2 \gamma ^4+\gamma ^2-3\right) P_3-5 \left(8
   \left(\gamma ^2+5\right) \gamma ^2+15\right) P_5}{6 \left(\gamma ^2-1\right)^2} \\
 \frac{500 \left(10 \gamma ^4+\gamma ^2-11\right) P_4-441 \left(28 \gamma ^4+104 \gamma ^2+33\right)
   P_6}{660 \left(\gamma ^2-1\right)^2} & -\frac{20 (\gamma -1)^2 \left(8 \left(\gamma
   ^2+3\right) \gamma ^2+3\right) P_4-21 \left(16 \gamma ^6+120 \gamma ^4+90 \gamma ^2+5\right) P_6}{44
   \gamma  \left(\gamma ^2-1\right)^{5/2}} \\
 \frac{5 (\gamma^2 -1) \left(274 \gamma ^4+607 \gamma ^2+64\right) P_5-2 \left(1452 \gamma
   ^6+8132 \gamma ^4+5175 \gamma ^2+256\right) P_7}{130 \gamma  \left(\gamma ^2-1\right)^{5/2}} & -\frac{15
   (\gamma^2 -1)  \left(8 \left(\gamma ^2+5\right) \gamma ^2+15\right) P_5-14 \left(16 \gamma
   ^6+168 \gamma ^4+210 \gamma ^2+35\right) P_7}{26 \left(\gamma ^2-1\right)^3} \\
\end{array}
\end{equation}
The first line corresponds to an isotropic distribution and reflects
the result in equation~(\ref{fisotfinal}). 
The third line
corresponds to the $D'_f$ part of equation~(\ref{fanisotfinal}).
Generically, we find that both coefficients scale like 
$Q_\alpha[\gamma] P_{n-1}(\cos \alpha )+ Q_\beta[\gamma] P_{n+1}(\cos \alpha)$ 
with $Q_\alpha$ and $Q_\beta$ polynomials  in $\gamma$. The dependence of the Faraday rotation coefficient
on angle $\alpha$ for the considered anisotropy is set by these Legendre polynomials, $P_{n-1}$ and $P_{n+1}$.
The term independent of ${\cal L}(\gamma)$ scales like $1/(\gamma^2-1)^{(n-1)/2}$ 
and the term linear in ${\cal L}(\gamma)$ scales like $1/(\gamma^2-1)^{n/2}$.
Similarly, the function $D_f$ (equation (\ref{DetDprim}))
can be written as $D_{f0} + D_{f1} {\cal L}(\gamma)$.
The factors $D_{f0}$  ({\sl left column}) and $D_{f1}$ ({\sl right column})
are gathered below for $n=1,\cdots \, 6$ (no $D_f$ term is present when $n = 0$). Again,
$P_i \equiv P_i(\cos \alpha)$:
\begin{equation}
\begin{array}{ll}
\scriptstyle \frac{2 }{3 \gamma }\big(\frac{\left(5 \gamma ^2+4\right) P_2}{\gamma ^2-1}-{\scriptstyle 2}\big)
  & -\frac{2 \left(-2
   \gamma ^2+\left(2 \gamma ^2+7\right) P_2+2\right)}{3 \left(\gamma ^2-1\right)^{3/2}} \\
 \frac{\left(34 \gamma ^2+86\right) P_3-24 \left(\gamma ^2-1\right) P_1}{5 \left(\gamma ^2-1\right)^{3/2}}
   & -\frac{6 \left(\left(2 \gamma ^4+15 \gamma ^2+3\right) P_3-2 \left(\gamma ^4-1\right) P_1\right)}{5
   \gamma  \left(\gamma ^2-1\right)^2} \\
 \frac{12 \left(-5 \gamma ^4+\gamma ^2+4\right) P_2+\left(74 \gamma ^4+387 \gamma ^2+64\right) P_4}{7
   \gamma  \left(\gamma ^2-1\right)^2} & -\frac{3 \left(\left(8 \left(\gamma ^2+13\right) \gamma
   ^2+63\right) P_4-4 \left(2 \gamma ^4+5 \gamma ^2-7\right) P_2\right)}{7 \left(\gamma ^2-1\right)^{5/2}}
%   \\
   \end{array} \nonumber
\end{equation}
\begin{equation}
\begin{array}{ll}
 \frac{\left(394 \gamma ^4+3517 \gamma ^2+1759\right) P_5-20 \left(17 \gamma ^4+26 \gamma ^2-43\right)
   P_3}{27 \left(\gamma ^2-1\right)^{5/2}} & -\frac{5 \left(\left(-8 \gamma ^6-52 \gamma ^4+48 \gamma
   ^2+12\right) P_3+\left(8 \gamma ^6+160 \gamma ^4+195 \gamma ^2+15\right) P_5\right)}{9 \gamma 
   \left(\gamma ^2-1\right)^3} \\
 \frac{6 \left(276 \gamma ^6+3764 \gamma ^4+3789 \gamma ^2+256\right) P_6-20 (\gamma -1) (\gamma +1)
   \left(74 \gamma ^4+387 \gamma ^2+64\right) P_4}{88 \gamma  \left(\gamma ^2-1\right)^3} & -\frac{15
   \left(2 \left(-8 \gamma ^6-96 \gamma ^4+41 \gamma ^2+63\right) P_4+\left(16 \gamma ^6+456 \gamma ^4+930
   \gamma ^2+215\right) P_6\right)}{44 \left(\gamma ^2-1\right)^{7/2}} \\
 \frac{5 (\gamma^2 -1) \left(274 \gamma ^4+607 \gamma ^2+64\right) P_5-2 \left(1452 \gamma
   ^6+8132 \gamma ^4+5175 \gamma ^2+256\right) P_7}{130 \gamma  \left(\gamma ^2-1\right)^{5/2}} & -\frac{15
   (\gamma^2 -1)  \left(8 \left(\gamma ^2+5\right) \gamma ^2+15\right) P_5-14 \left(16 \gamma
   ^6+168 \gamma ^4+210 \gamma ^2+35\right) P_7}{26 \left(\gamma ^2-1\right)^3} \\
\end{array}
\end{equation}
The second line corresponds to the $D_f$ part of equation~(\ref{fanisotfinal}).
The same scalings are found with respect to $P_{n-1}(\cos \alpha )$ and $  P_{n+1}(\cos \alpha)$. 
The terms independent of ${\cal L}(\gamma)$ scale like $1/(\gamma^2-1)^{(n+1)/2}$ 
and the terms proportional to ${\cal L}(\gamma)$ scale like $1/(\gamma^2-1)^{(n+2)/2}$.
The anisotropic coefficient $h$
can also be written as in equation (\ref{DetDprim}),
with a prefactor $A_h$ that is identical to that in equation (\ref{hanisotfinal}) and a function $D'_h$
that can again be written as $D'_{h0} + D'_{h1} {\cal L}(\gamma)$.
The factors $D'_{h0}$  ({\sl left column}) and $D'_{h1}$ ({\sl right column})
are listed below for $n=0,\cdots \, 3$:
\begin{equation}\begin{array}{ll}
 \scriptstyle\frac{1}{3} \sqrt{\gamma ^2-1} \left(2 \gamma ^2-3\right) \left(P_2-1\right) & \frac{P_2-1}{3 \gamma } \\
 -\frac{\left(6 \gamma ^4-17 \gamma ^2-4\right) \left(P_1-P_3\right)}{15 \gamma } &
   \frac{P_3-P_1}{\sqrt{\gamma ^2-1}} \\
 \frac{7 \left(4 \gamma ^4-8 \gamma ^2+19\right)+5 \left(-20 \gamma ^4+76 \gamma ^2+31\right) P_2+36
   \left(2 \gamma ^4-9 \gamma ^2-8\right) P_4}{210 \sqrt{\gamma ^2-1}} & \frac{14 \gamma ^2+94 \gamma ^2
   P_2-108 \gamma ^2 P_4-7 P_2+7}{42 \gamma -42 \gamma ^3} \\
 \frac{9 \left(\left(36 \left(\gamma ^2-3\right) \gamma ^2+493\right) \gamma ^2+104\right) P_1+7 \left(-132
   \gamma ^6+696 \gamma ^4+509 \gamma ^2-248\right) P_3+100 \left(6 \gamma ^6-39 \gamma ^4-80 \gamma
   ^2+8\right) P_5}{1890 \gamma  \left(\gamma ^2-1\right)} & \frac{100 \gamma ^2 P_5-9 \left(2 \gamma
   ^2+3\right) P_1+\left(27-82 \gamma ^2\right) P_3}{18 \left(\gamma ^2-1\right)^{3/2}} \\
\end{array}
\end{equation}
The first line corresponds to the bracket in equation~(\ref{hisotfinal}) and the third one to that in equation~(\ref{hanisotfinal}).
The terms independent of ${\cal{L}}(\gamma)$ scale like $1/(\gamma^2-1)^{(n-1)/2}$ 
while terms proportional to ${\cal{L}}(\gamma)$ scale like $1/(\gamma^2-1)^{n/2}$. 
The number of Legendre polynomials now increases with the anisotropy order $n$.
Finally, the similar functions
associated to $D_h$ (defined as in equation (\ref{DetDprim})) 
are, for $n=1,\cdots 3$:
\begin{equation}
\begin{array}{ll}
 -\frac{\left(2 \gamma ^2+13\right) \left(P_1-P_3\right)}{15 \left(\gamma ^2-1\right)} & \frac{\left(4
   \gamma ^2+1\right) \left(P_1-P_3\right)}{5 \gamma  \left(\gamma ^2-1\right)^{3/2}} \\
 -\frac{7 \left(-6 \gamma ^4+17 \gamma ^2+4\right)+5 \left(18 \gamma ^4+65 \gamma ^2+4\right) P_2-12
   \left(4 \gamma ^4+37 \gamma ^2+4\right) P_4}{105 \gamma  \left(\gamma ^2-1\right)^{3/2}} &
   \frac{\left(24 \gamma ^2+5\right) P_2-12 \left(2 \gamma ^2+1\right) P_4+7}{7 \left(\gamma ^2-1\right)^2}
   \\
 \frac{9 \left(64 \gamma ^4-258 \gamma ^2-331\right) P_1-7 \left(168 \gamma ^4+854 \gamma ^2-197\right)
   P_3+100 \left(6 \gamma ^4+83 \gamma ^2+16\right) P_5}{630 \left(\gamma ^2-1\right)^2} & -\frac{500
   \left(4 \gamma ^2+3\right) \gamma ^2 P_5+9 \left(8 \gamma ^2 \left(\gamma ^2-22\right)-7\right) P_1+7
   \left(-296 \gamma ^4+12 \gamma ^2+9\right) P_3}{210 \gamma  \left(\gamma ^2-1\right)^{5/2}} \\
\end{array}\end{equation}
The terms independent of ${\cal{L}}(\gamma)$ scale like $1/(\gamma^2-1)^{(n+1)/2}$ 
and the terms proportional to ${\cal{L}}(\gamma)$ scale like $1/(\gamma^2-1)^{(n+2)/2}$.\\

\section{Faraday conversion coefficient in the LF limit}
\label{Applowfrequencyapprox}

The equation $g = 1$ which defines the values $\varpi_{QR\pm}$ of $\varpi$ at the edge of the QR domain for
a given value of $\gamma$ can be exactly written as:
\begin{equation}
\frac{\left(\varpi^2 + \sigma_0^2\right)^{3/2}}{3 \gamma^2 u^2 \sin^4\alpha}  = 
\left(1 +  \frac{\varpi \cos \alpha}{\displaystyle\gamma u \sin^2 \alpha}\right)^{\!2}\!
\left(1 - 
\frac{\varpi^2 +\sigma_0^2}{\gamma^2 u^2 \sin^4\alpha\, \left(1 + \frac{\varpi \cos \alpha}{\gamma u \sin^2 \alpha}\right)^2}
\right)^{\! 1/4} \!
\left(\frac{1}{2} + 
\frac{1}{2}  \sqrt{
1 - 
\frac{\varpi^2 +\sigma_0^2}{\gamma^2 u^2 \sin^4\alpha\,
\left(1 + \frac{\varpi \cos \alpha}{\gamma u \sin^2 \alpha}\right)^2}
}
\right)^{\!3/2}\!\!.
\label{basediterationvarpiQR}
\end{equation}
Since $\mid \varpi_{QR\pm}\mid$ is much less than $\gamma u \sin^2 \alpha$, the three
parentheses on the right are close to unity and the simplest approximation to $\varpi_{QR\pm}$ in the
LF limit is given by equation (\ref{varpiedgeordre0}).
Higher accuracy approximations to $\varpi_{QR\pm}$ can be generated by successively iterating on equation
(\ref{basediterationvarpiQR}), setting
$ \gamma u \sin^2 \alpha = 3G$.
The result, valid for $\mid \varpi_{QR\pm}\mid \gg \sigma_0$, is:
\begin{eqnarray}
\varpi_{QR+}&=& + \left(3 \gamma^2 u^2 \sin^4 \alpha\right)^{1/3} \, \left(1 +\frac{2\cos \alpha}{3} \frac{1}{G^{1/3}} 
+ \left(\frac{\cos^2 \alpha}{3} - \frac{5}{24}\right) \frac{1}{G^{2/3}} 
+\left(\frac{10\, \cos^3 \alpha}{81} - \frac{5 \, \cos \alpha}{36}\right) \frac{1}{G} +\cdots \right)\,,
\label{varpiqrplusexp}
\\
\varpi_{QR-}&=&  -\, \left(3 \gamma^2 u^2 \sin^4 \alpha\right)^{1/3} \, \left(1 - \frac{2\cos \alpha}{3} \frac{1}{G^{1/3}}
+ \left(\frac{\cos^2 \alpha}{3} - \frac{5}{24}\right) \frac{1}{G^{2/3}} 
- \left(\frac{10\, \cos^3 \alpha}{81} - \frac{5 \, \cos \alpha}{36}\right) \frac{1}{G} +\cdots \right)\,.
\label{varpiqrmoinsexp}
\end{eqnarray}
The primitive of the function in equation (\ref{lafonctionaintegrer}) is
\begin{eqnarray}
&&\int \! d\varpi \, \left(\frac{2 x^4}{(\sigma^2 - x^2)^{5/2}} + \sigma_0^2 \, 
\frac{x^2 (4\sigma^2 + x^2)}{(\sigma^2 - x^2)^{7/2}} \right)  =   
\frac{81 \varpi(7 \sigma_0^4 + 10 \varpi^2 \sigma_0^2 + 4\varpi^4)}{(\varpi^2 + \sigma_0^2)^{5/2}} 
\left(\frac{G}{\sigma_0}\right)^{\!4} 
- \frac{36 \cos\alpha \,\sigma_0^3\, (5 \sigma_0^2 + 2 \varpi^2)}{(\varpi^2 + \sigma_0^2)^{5/2}} 
\left(\frac{G}{\sigma_0}\right)^{\!3} 
\nonumber \\
&&- 18\varpi\, \frac{4\sin^2\!\alpha \varpi^4 + (9 - 7\cos^2\!\alpha)\sigma_0^2\varpi^2 + 5\sigma_0^4}{(\varpi^2 + \sigma_0^2)^{5/2}} 
\left(\frac{G}{\sigma_0}\right)^{\!2}\!\!
+ 12\cos\!\alpha 
\frac{2\sin^2\! \alpha \sigma_0\varpi^4+ 5 \sin^2\!\alpha \sigma_0^3 \varpi^2 +(3-2\cos^2\!\alpha) \sigma_0^5}{(\varpi^2 + \sigma_0^2)^{5/2}} 
\left(\frac{G}{\sigma_0}\right)
\nonumber \\
&& \!- 2\cos^2\!\!\alpha(\!1\! +\! \sin^2\!\! \alpha)\! \ln\!\left(\!\sqrt{\varpi^2 +\sigma_0^2} +\varpi\!\right)\!\!
+ \!\varpi \frac{(3\! + \! 6\cos^2\!\!\alpha\!-\! 5\cos^4\!\!\alpha)\varpi^4\! 
\!+\!\! (6\! +\! 14 \sin^2\!\!\alpha \!\cos^2\!\!\alpha)\sigma_0^2 \varpi^2 \!\!+\!\!
(3\! + \! 8\cos^2\!\!\alpha \!- \!6 \cos^4\!\!\alpha)\sigma_0^4}{3\, (\varpi^2 + \sigma_0^2)^{5/2}}\,.
\label{primitiveexpand}
\end{eqnarray}
In the LF limit $\sigma_0 /\!\!\mid\!\!\varpi_{QR\pm}\!\!\mid$ is small, as can be judged from equation
(\ref{varpiedgeordre0}), this ratio  being
about $(u/(3 \gamma^2 \sin \alpha))^{1/3}$ which is meant to be small in this limit.
The right-hand side of equation (\ref{primitiveexpand}) can then be expanded
in $\sigma_0/\varpi$, and the definite integral over the interval $[\varpi_{QR-}, \varpi_{QR+}]$ calculated from it
at any desired order in $\gamma$, using
equations (\ref{varpiqrplusexp})--(\ref{varpiqrmoinsexp}).
It will however be sufficient to extend the expansion to the first non-vanishing order in $\gamma$, which from 
Section~\ref{comparQRNR} is expected
to be $\gamma^{4/3}$. The ratio $(G/\sigma_0)$
is proportional to $\gamma$ and, as mentioned above, $\sigma_0/\varpi_{QR\pm}$
is of order $1/\gamma^{2/3}$. The factors of all
powers of $(G/\sigma_0)$ in equation (\ref{primitiveexpand})
remain finite or approach zero as $\sigma_0/\varpi$
approaches zero. This implies that
terms proportionnal to $(G/\sigma_0)^1$ and $(G/\sigma_0)^0$
are negligible at the order $\gamma^{4/3}$, as are also the logarithmic terms.
Similarly, the term proportionnal to $(G/\sigma_0)^3$
does not contribute at this order because its factor
is ${\cal{O}}(1/\gamma^2)$.
Those parts of equation (\ref{primitiveexpand}) that do contribute at order $\gamma^{4/3}$ are the first term,
the factor of which should be expanded to order $(\sigma_0/\varpi)^4$ and the term proportional to $(G/\sigma_0)^2$,
the factor of which would have to be expanded to order $(\sigma_0/\varpi)$, had terms of that order been present in its
expansion. This leads to equation (\ref{primitivtronq}) below, which only involves
the lowest order approximation to $\varpi_{QR\pm}$ from equations (\ref{varpiqrplusexp})--(\ref{varpiqrmoinsexp}):
\begin{equation}
\int \! d\varpi \, \left(\frac{2 x^4}{(\sigma^2 - x^2)^{5/2}} + \sigma_0^2 \, 
\frac{x^2 (4\sigma^2 + x^2)}{(\sigma^2 - x^2)^{7/2}} \right) \approx 
\frac{81 \, \varpi}{\mid\!\varpi\!\mid}  
\left(4 - \frac{1}{2} \, \frac{\sigma_0^4}{\varpi^4}\right)  \left(\frac{G}{\sigma_0}\right)^4 
- \, \frac{72\,  \varpi}{\mid\!\varpi\!\mid} 
\sin^2\! \alpha \left(\frac{G}{\sigma_0}\right)^2\,.
\label{primitivtronq}
\end{equation}
We then obtain, using equation (\ref{primitivtronq}), the definition $G = \gamma u \sin^2\!\!\alpha/3$ and
the lowest order expressions for $\varpi_{QR\pm}$,
the integral over $\varpi$ on the first line
of equation (\ref{Kisogeneral}) in the form of equation (\ref{correctionNRQR}).

\medskip

Let us now evaluate, along the lines
described in Section~\ref{seclowfrequencylimit},  the QR contribution to $H^{\rm iso}$.
In the LF limit, the variable $g$ assumes, for
a given Lorentz factor $\gamma$, a value less than unity
in the QR domain, which reaches near $\varpi=0$
a small minimum, $g_{m}$, given by equation (\ref{gmin}).
The relation between
the variables $g$ and $\varpi$ at a given $\gamma$
results from equations (\ref{relapisigmagammappar}),
(\ref{pperpderhoq}) and (\ref{gtexte}). In the LF limit,
it becomes
\begin{equation}
g \approx \frac{(\varpi^2 + \sigma_0^2)^{3/2} }{3 \, \gamma^2 u^2 \sin^4\!\alpha} \,,
\qquad \qquad {\mathrm{or}} \qquad \qquad 
\varpi^2 = \left(3 \gamma^2 u^2 \sin^4\!\alpha\right)^{2/3} \ \left(g^{2/3} - g_m^{2/3}\right)\, .
\label{gvarpiapproximatif}
\end{equation}
From equations (\ref{gvarpiapproximatif})  and (\ref{pperpderhoq}) we also get
\begin{equation}
\left(\frac{\sigma -x}{x}\right) \approx \left(\frac{3 \, g}{2^{3/2} \gamma u \sin^2\! \alpha}\right)^{2/3}\,.
\end{equation}
These results are used to convert the second line of equation
(\ref{Kisogeneral}) into functions of $g$,
so that eventually:
\begin{eqnarray}
&&H^{\rm iso}_{\rm QR}(\gamma)=  
\frac{2}{\sin^2 \alpha} (\gamma u \sin^2\!\alpha)^{4/3} \, 3^{1/6}
\int_{g_m}^{1} \frac{dg}{g^{1/3} \sqrt{g^{2/3}- g_m^{2/3}}} \
\Big(g^{4/3} \left(K_{2/3}(g)L_{2/3}(g) + K_{1/3}(g) L_{1/3}(g) - \frac{2\pi}{\sqrt{3} \, g} \right)\Big)
\nonumber \\
&& \hskip 2cm - \frac{2}{\sin^2 \alpha} (\gamma u \sin^2\!\alpha)^{4/3} \, 3^{1/6}
\int_{g_m}^{1} \frac{dg}{g^{1/3} \sqrt{g^{2/3}- g_m^{2/3}}} \
\Big(g_m^{2/3} g^{2/3} \left(K_{1/3}(g) L_{1/3}(g) - \frac{\pi}{\sqrt{3} \, g}\right) 
\Big)\,.
\label{Khisoeng}
\end{eqnarray}
The products $K_{2/3}(g)L_{2/3}(g)$ and $K_{1/3}(g) L_{1/3}(g)$ diverge at small $g$ as $g^{-4/3}$. Therefore
both integrals in equation (\ref{Khisoeng}) converge in the limit $g_m \rightarrow 0$, since the second integrand
is less than $g^{4/3} (K_{1/3} L_{1/3} - \pi/(\sqrt{3} g))$. Evaluating these integrals in the limit of vanishing $g_m$,
we obtain equation (\ref{Khisointegre}) of the main text.

\section{Faraday rotation coefficient in the LF limit} \label{appfenLF}

We derive here the low-frequency (LF) limit of equation (\ref{Fisoarragement}), 
which corresponds to $\gamma \gg \gamma_{\rm QR}$ (equation (\ref{sigmagammaqr})).
The offset correction results from the small displacement
$\Delta\varpi_{\rm QR}$
of the center of the interval $[\varpi_{\rm QR-}(\gamma), \, ,\varpi_{\rm QR+}(\gamma)]$ 
with respect to zero.
Approximations 
to $\varpi_{\rm QR\pm}(\gamma)$ are given in
equations (\ref{varpiqrplusexp})--(\ref{varpiqrmoinsexp}). 
$\Delta\varpi_{\rm QR}$ being small, the integral of an odd function $f_{\rm odd}(\varpi)$  on 
$[\varpi_{\rm QR-}, \, ,\varpi_{\rm QR+}]$ is about twice $\Delta\varpi_{\rm QR} \, f_{\rm odd}(+\varpi_{\rm QR0})$,
$\varpi_{\rm QR0}$ being the dominant term in equation (\ref{varpiqrplusexp}).
From the definition of the QR domain, $g(\gamma, \varpi_{\rm QR\pm})$ is unity. This gives
the offset correction associated with the 
the second and third terms in equation (\ref{Fisoarragement}). 
Since in the LF limit $(\sigma -x)\ll \sigma$ and $\varpi_{QR0}\ll \sigma_0$ the raw expressions
may be simplified to
equation (\ref{Fisooff}):
\begin{equation}
F^{\rm iso}_{\rm off} = - \frac{4 \pi\, s_q}{3}  \, \gamma u \cos\alpha + 8 s_q \gamma u \cos\alpha \,
\left(\sqrt{3} K_{2/3}(1) L_{1/3}(1) - \pi\right)\,.
\label{Foffsetresult}
\end{equation}
The parity correction is  the integral over the symmetrical interval $[-\varpi_{\rm QR0}, \, +\varpi_{\rm QR0}]$ of 
the almost odd integrands in equation (\ref{Fisoarragement}). 
Any of these two integrands can be factored as
$\varpi \, \Phi(\varpi)$. Only the small odd part of $\Phi(\varpi)$, 
which is defined by $\Phi_{\rm odd}(\gamma, \varpi)= (\Phi(\gamma, +\varpi)-\Phi(\gamma, -\varpi))/2$,
contributes to the integral over the
interval $[-\varpi_{\rm QR0}, \, +\varpi_{\rm QR0}]$. 
Using the relation $\sigma^2 - x^2 = \varpi^2 + u^2\sin^2 \alpha$
the function $\Phi_{\rm odd}$ associated with the integrand on the first line of 
equation (\ref{Fisoarragement}) can be written as
\begin{equation}
\Phi_{\rm odd}^{(1)}(\gamma, \varpi) =
\frac{1}{2} \ \frac{4 \gamma u \varpi \sin^2 \alpha \cos \alpha}{\left(\varpi^2 + u^2 \sin^2 \alpha\right)^{3/2}}\,.
\end{equation}
The associated integral in
equation (\ref{Fisoarragement}) can be calculated by integraton by parts, giving 
\begin{equation}
\int_{-\varpi_{\rm QR0}}^{+\varpi_{\rm QR0}} \!\! d\varpi \frac{\varpi^2}{\left(\varpi^2 +\sigma_0^2\right)^{3/2}}
=  \ln \left(\frac{\sqrt{\varpi_{\rm QR0}^2 +\sigma_0^2} 
+ \varpi_{\rm QR0}}{\sqrt{\varpi_{\rm QR0}^2 +\sigma_0^2} -  \varpi_{\rm QR0}} \right) 
- 2\ \frac{\varpi_{\rm QR0}}{\sqrt{\varpi_{\rm QR0}^2 +\sigma_0^2}}\,.
\end{equation}
This exact relation is expanded for large $\gamma$ up to terms of order $\gamma^0$ included, leading to
an associated parity correction
\begin{equation}
F^{(1)}_{\rm par} = - \pi s_q \gamma u \cos \alpha \ \left(\frac{8}{3} \, \ln\left(\gamma\right) + 
\frac{4}{3} \ln\left(\frac{\sin \alpha}{u}\right) + 2 \ln\left(4 \, . 3^{2/3}\right) -4 \right)\,.
\label{corrimparNR}
\end{equation}
The parity correction associated with the integral  on the second line of equation (\ref{Fisoarragement}) 
involves the calculation of the odd part of the function $g K_{2/3}(g) L_{1/3}(g)$, where $g(\gamma,\varpi)$ is
defined by equation (\ref{gtexte}). 
For small $\varpi/\gamma u \sin^2\alpha$, $g$ may be expanded by
using equation (\ref{relapisigmagammappar}) and the second relation in equation (\ref{pperpderhoq})
into
\begin{equation}
g = g_0 + g_1 \qquad {\mathrm{where}} \qquad g_0(\gamma, \varpi) = \frac{1}{3} \, 
\frac{(\varpi^2 + \sigma_0^2)^{3/2}}{(\gamma u \sin^2\!\!\alpha)^2}
\qquad {\mathrm{and}} \qquad g_1 = - 2 g_0 \, \frac{\varpi \cos \alpha}{\gamma u \sin^2\!\!\alpha}\,,
\label{g0g1devarpi}
\end{equation}
where $g_1 \ll g_0$ in the LF limit.
Denoting $g(\gamma, \pm \varpi)$ by $g_\pm$,
the functions $g_\pm K_{2/3}(g_\pm) L_{1/3}(g_\pm)$ are then Taylor-expanded about $g_0$ 
to provide an approximation of the odd part of $g K_{2/3}(g) L_{1/3}(g)$, that is
\begin{equation}
\frac{1}{2} \left(g_+ K_{2/3}(g_+) L_{1/3}(g_+) - g_- K_{2/3}(g_-) L_{1/3}(g_-)\right) \approx 
-  \frac{2 \varpi \cos \alpha}{\gamma u \sin^2\!\!\alpha} 
\ \, g_0 \frac{d}{dg_0} \Big(g_0 K_{2/3}(g_0) L_{1/3}(g_0)\Big)\,.
\end{equation}
The parity correction associated with the integral on the second line of 
equation (\ref{Fisoarragement}) can then be written as
\begin{equation}
F^{(2)}_{\rm par} =  -  \frac{8 \sqrt{3} s_q \cos \alpha}{\gamma u  \sin^4\!\!\alpha} 
\int_0^{+\varpi_{QR0}}  \!\! d\varpi  
\varpi^2 \ g_0 \frac{d}{dg_0} \Big(g_0 K_{2/3}(g_0) L_{1/3}(g_0)\Big)\,.
\label{integralpariteQR}
\end{equation}
Using equation (\ref{g0g1devarpi}), the integral in equation (\ref{integralpariteQR}) 
is expressed in terms of the integration variable $g_0$, denoted by $g$ below, which
ranges from $g_m = g_0(\gamma, \varpi = 0)$ to unity.
For large $\gamma$, $g_m \rightarrow 0$. 
To the same order as in equation (\ref{Foffsetresult}),
it suffices to calculate the integral over $g$  in equation (\ref{integralpariteQR})
to order ${\cal{O}}(\gamma^0)$, which gives
\begin{equation}
F^{\rm (2)}_{\rm par} = - 8 \sqrt{3} s_q \gamma u \cos \alpha \int_{0}^1 dg \ \ g  \frac{d}{dg} \Big(gK_{2/3}(g) L_{1/3}(g)\Big)\,.
\label{corrimparQR}
\end{equation}
The corrections in equations (\ref{Foffsetresult}), (\ref{corrimparNR}) and (\ref{corrimparQR}) are then added to the 
non-integral term in equation (\ref{Fisoarragement}). For consistency, the latter is expanded
for large $\gamma$ as
$ \gamma \, \ln\left(\sqrt{\gamma^2 -1} + \gamma\right) - \sqrt{\gamma^2 -1}
\approx \gamma \ln\left(2\gamma\right)- \gamma$.
The integral over 
$g$ in equation (\ref{corrimparQR}) may be integrated by parts, resulting in
the cancellation of the Bessel term in equation (\ref{Foffsetresult}) and leaving us with
\begin{equation} 
F^{\rm iso}_{\rm LF}(\gamma) = \pi s_q \gamma u \cos \alpha\Big[ 
\frac{4}{3} \ln\left(\frac{\gamma u}{\sin \alpha}\right) -\frac{4}{3}
-\frac{4}{3} \, \ln(3) 
- 8 + \frac{8\sqrt{3}}{\pi} \int_0^1 g K_{2/3}(g) L_{1/3}(g) \, dg \Big]\,.
\label{FisoLFappendice}
\end{equation}
The integral over $g$ in equation (\ref{FisoLFappendice}) is numerically evaluated to 
${\mathrm{2.162372}}$ finally giving the result quoted in equation (\ref{FLFdetail}).

\section{Transfer coefficients from interpolated isotropic kernels} \label{AppinterpolKernels}

\begin{figure}
\begin{center}
\includegraphics[width=0.45\hsize]{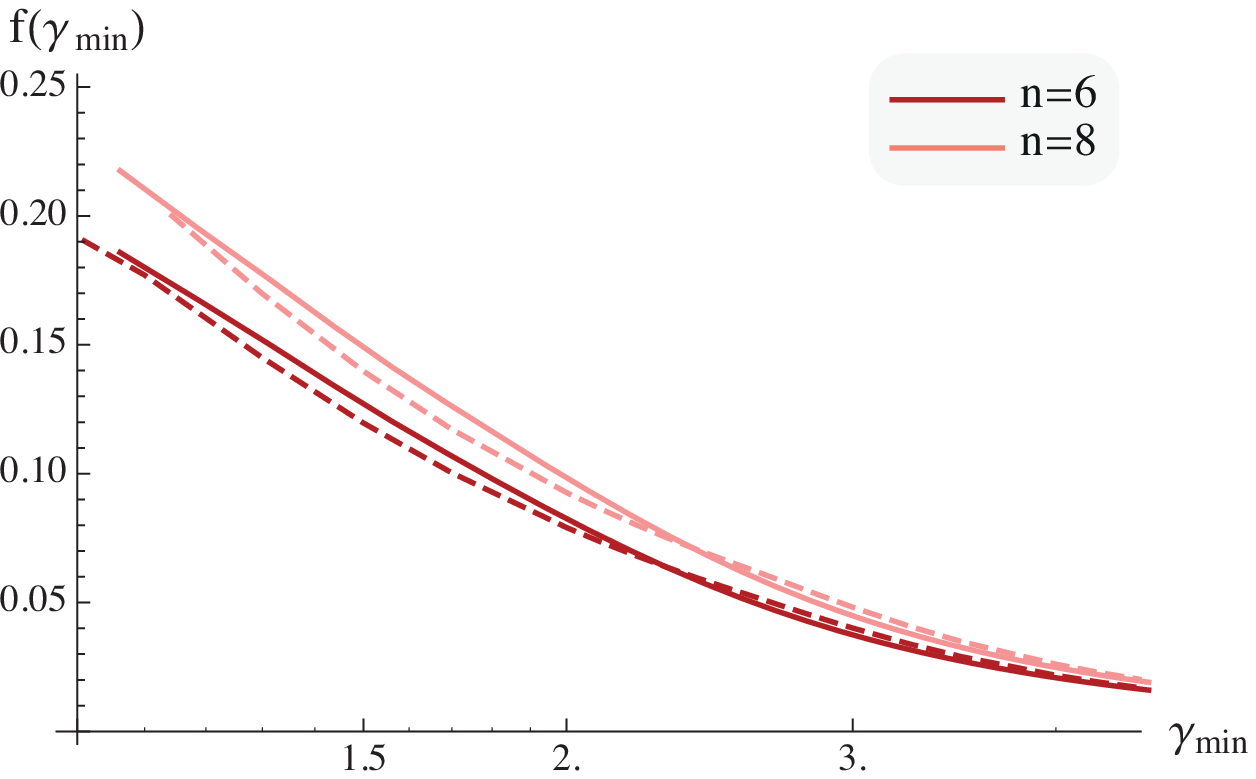}
\includegraphics[width=0.45\hsize]{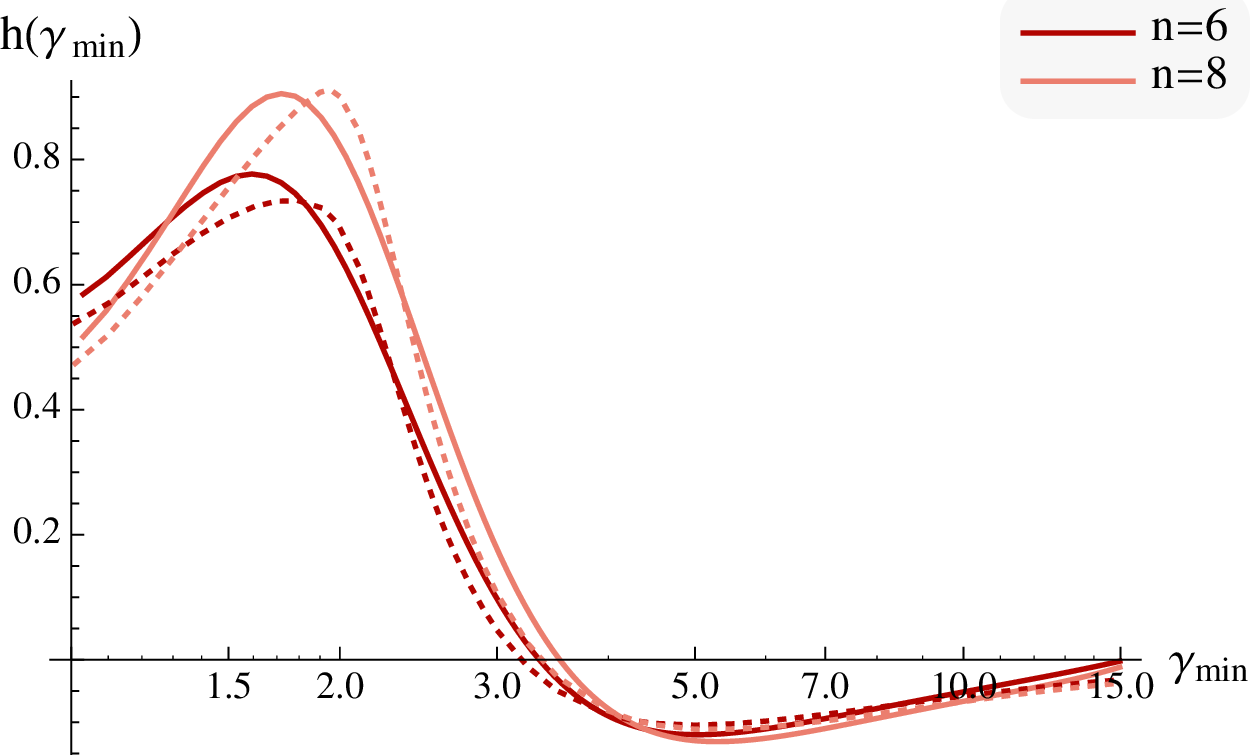}
\caption{ Left: quasi-exact values of the Faraday rotation coefficient $f$ for power-law distributions
obtained by numerically calculating the double integral in equation (\ref{fexacteanticip})
(full lines) compared to those obtained from the interpolating formula (\ref{Fisointerpol})
for the kernel $F^{\rm iso}(\gamma)$ (dashed lines), for a range of values of the low-energy cutoff $\gamma_{\rm min}$
and for two different exponents $n$, as labelled. 
Right: Similar plots for the Faraday conversion coefficient $h$, from equations (\ref{hexactanticip})
and (\ref{Hisointerpol}). The coefficient $f$ is normalized to
$-\pi s_q \cos \alpha (\omega_{\rm pr}^2\!\mid\!\!\Omega\!\!\mid\!/c \omega^2)$
and $h$ to $\omega_{\rm pr}^2 \Omega^2\!/c\omega^3$. The radiation's parameters are $\omega = 15 \mid\!\Omega\!\mid$ and $\alpha = \pi/4$.
}
\label{figcomparPLexactinterpol}
\end{center}
\end{figure}

\begin{figure}
\begin{center}
\includegraphics[width=0.45\hsize]{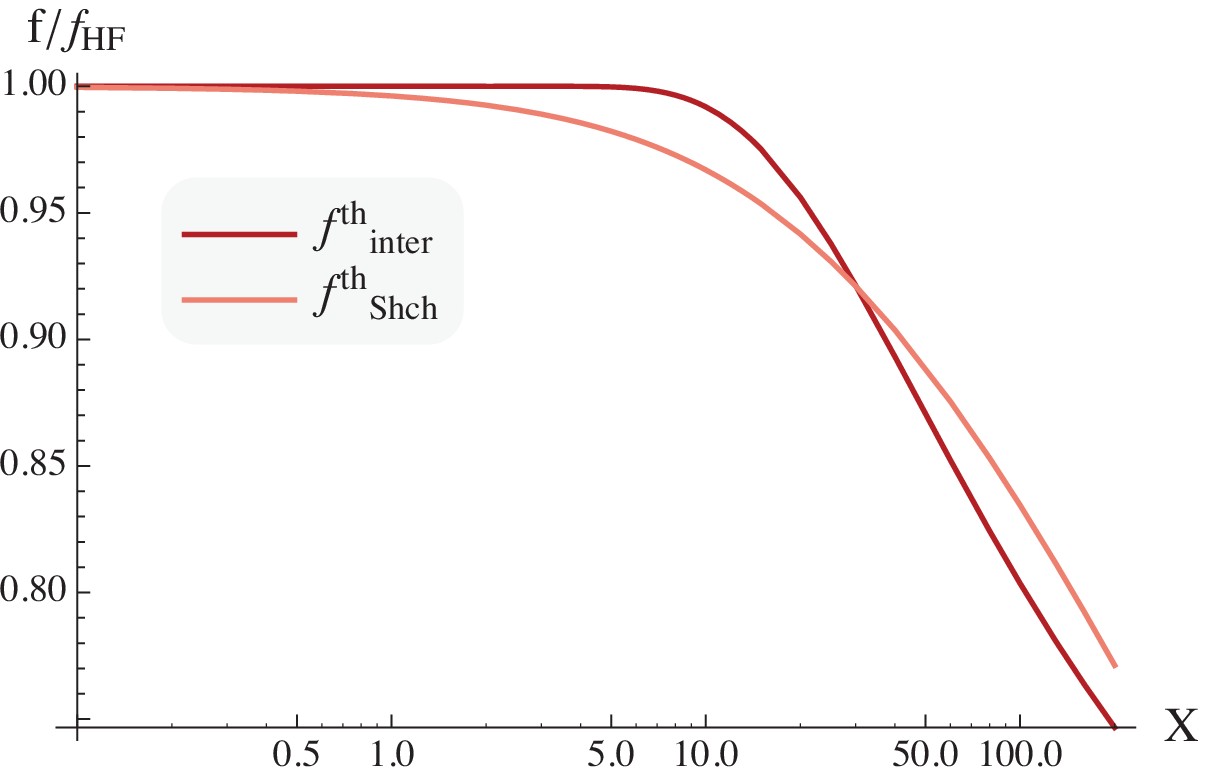}
\includegraphics[width=0.45\hsize]{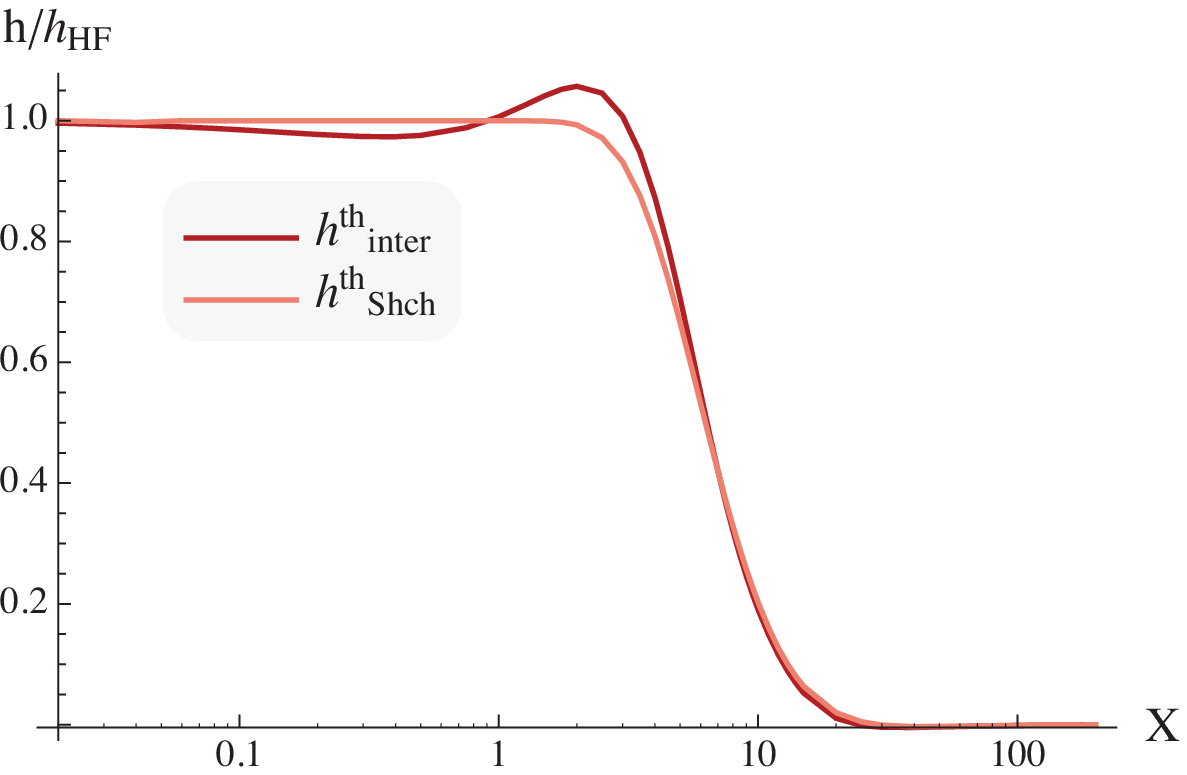}
\caption{ Left: The thick line represents the ratio of the Faraday 
rotation coefficient $f_{\rm inter}^{\rm th}$ for a thermal plasma, 
as deduced from the interpolating formula for the kernel $F_{\rm int}^{\rm iso}$  in equation (\ref{Fisointerpol}),
to the HF expression $f^{\rm th}_{\rm HF}$ deduced from 
equations (\ref{defnoyauxisotropes}) and (\ref{fisotfinal}). The 
thin line represents the fit by \citet{Cerbakov2008}
to the ratio of the exact value of $f$ to the HF formula. These quantities are plotted as a function
of the regime-change parameter $X$ defined by \citet{Cerbakov2008} and in equation (\ref{XCherbak}).
Right: Similar plots for the thermal 
Faraday conversion coefficient $h$, from equations (\ref{Hisointerpol}), (\ref{defnoyauxisotropes}) and
(\ref{hisotfinal}).
The radiation's parameters here are $\omega = 1000 \mid\!\Omega\!\mid$ and $\alpha = \pi/4$, allowing a comparison between
these figures  and
figures 4 and 5 of \citet{Cerbakov2008}.
}
\label{figcomparthermalCherbinterpol}
\end{center}
\end{figure}

The figures \ref{figcomparPLexactinterpol} and \ref{figcomparthermalCherbinterpol} illustrate how 
the interpolation formulae in equations (\ref{Fisointerpol}) and (\ref{Hisointerpol})  perform
in calculating the actual Faraday transfer coefficients in a plasma characterized by an isotropic  power-law energy distribution
with a low-energy cutoff (figure \ref{figcomparPLexactinterpol})
or in a thermal plasma (figure \ref{figcomparthermalCherbinterpol}). 
Figure \ref{figcomparPLexactinterpol} compares the quasi-exact values of the transfer coefficients 
in figure \ref{fig:powerlaw} to those derived from the interpolation formulae
(\ref{Fisointerpol}) and (\ref{Hisointerpol}) for the kernel or its derivative. 
The accuracy generally is of order 10 per cent for $h$ and somewhat better than 10 per cent for $f$.
Figure \ref{figcomparthermalCherbinterpol} displays the ratio, which \citet{Cerbakov2008} refers to
as the multiplier, of the Faraday transfer coefficients for a thermal plasma, calculated 
from the interpolated kernels (\ref{Fisointerpol}) and (\ref{Hisointerpol}), to their HF expression
deduced from equations (\ref{fisotfinal}), (\ref{hisotfinal}) and (\ref{defnoyauxisotropes}). In figures 4 and 5 of
his paper,
\citet{Cerbakov2008} presents similar ratios of numerically computed exact transfer
coefficients and corresponding HF expressions, which he displays as a function of the regime-change 
parameter $X$ defined by equation (\ref{XCherbak}). He proposes fitting formulae for these ratios 
(his equations (28) and (29)). The figures 4 and 5 of \citet{Cerbakov2008} can then 
be compared to our figures \ref{figcomparthermalCherbinterpol},
similar parameters having been adopted for the radiation. From this comparison it
appears that within the range of parameters represented here
our interpolation formula approaches the real value of the Faraday rotation coefficient
somewhat closer than his proposed fit,
while our interpolation also well represents the Faraday conversion coefficient, though slightly less accurately than his. 

\end{document}